\documentclass[a4paper,dvips,12pt]{article}
\usepackage{cite}
\usepackage{epsf}
\usepackage{multirow}
\usepackage{bm,bbm} 
\usepackage[pdftex]{graphicx}
\usepackage{tabularx}
\usepackage{latexsym}
\usepackage{amsmath,amssymb,exscale}
\usepackage{array,multicol}
\numberwithin{equation}{section}

\def\draftlabel#1{{\@bsphack\if@filesw {\let\thepage\relax
   \xdef\@gtempa{\write\@auxout{\string
      \newlabel{#1}{{\@currentlabel}{\thepage}}}}}\@gtempa
   \if@nobreak \ifvmode\nobreak\fi\fi\fi\@esphack}
        \gdef\@eqnlabel{#1}}
\def\@eqnlabel{}
\def\@vacuum{}
\def\draftmarginnote#1{\marginpar{\raggedright\scriptsize\tt#1}}
\def\draft{\oddsidemargin -.5truein
        \def\@oddfoot{\sl preliminary draft \hfil
        \rm\thepage\hfil\sl\today\quad\militarytime}
        \let\@evenfoot\@oddfoot \overfullrule 3pt
        \let\label=\draftlabel
        \let\marginnote=\draftmarginnote
   \def\@eqnnum{(\theequation)\rlap{\kern\marginparsep\tt\@eqnlabel}%
\global\let\@eqnlabel\@vacuum}  }

\newcommand{\PRL}[3]{\emph{ Phys.~Rev.~Lett.} \textbf{#1} (#2) #3}

\newcommand{\PR}[3]{\emph{ Phys.~Rep.} \textbf{#1} (#2) #3}

\def\ov{\overline}

\def\dalemb#1#2{{\vbox{\hrule height .#2pt
         \hbox{\vrule width.#2pt height#1pt \kern#1pt
                 \vrule width.#2pt}
         \hrule height.#2pt}}}

\def\half{{\textstyle{1\over2}}}
\let\a=\alpha    
    
    \let\p=\pi 
\let\s=\sigma     

      \let\G=\Gamma  
  \let\S=\Sigma   
\let\F=\Phi

 \def\bd{\begin{document}} \def\ed{\end{document}}
\def\ds{\documentstyle} \let\fr=\frac \let\bl=\bigl \let\br=\bigr
\let\Br=\Bigr \let\Bl=\Bigl
\let\bm=\bibitem
\let\na=\nabla
\let\pa=\partial
\let\ov=\overline
\def\ie{{\it i.e.\ }}
\def\tr{{\mbox{\rm tr}}}
\def\simlt{\mathrel{\lower2.5pt\vbox{\lineskip=0pt\baselineskip=0pt
            \hbox{$<$}\hbox{$\sim$}}}}
\def\simgt{\mathrel{\lower2.5pt\vbox{\lineskip=0pt\baselineskip=0pt
            \hbox{$>$}\hbox{$\sim$}}}}
\def\A{{\cal A}}
\def\a{{\mathcal a}}
\def\V{{\cal V}}
\def\F{{\cal F}}
\def\p{{\mathcal \phi}}
\def\L{{\mathcal L}}
\def\M{{\mathcal M}}
\def\bD{{\ov {\rm D}}}
\def\bO{{\ov {\rm O}}}
\def\bOp{{\ov {\rm O'}}}
\def\O{{ {\rm O}}}
\def\Dbar{D{\hspace{-7.5pt}\slash}}
\def\Om{\Omega}
\def\om{\omega}
\def\th{\theta}
\def\vt{\vartheta}
\def\vphi{\varphi}
\def\inte{{\bf Z}}
\def\pim{{\rm Im\,}}
\def\cn{{\cal N}}
\def\cpx{{\bf C}}
\def\real{{\bf R}}

\newcommand{\nsect}{\setcounter{equation}{0}
\def\theequation{\thesection.\arabic{equation}}\section}
\newcommand{\nappend}{\setcounter{equation}{0}
\def\theequation{\rm{A}.\arabic{equation}}\section*}
\newcommand{\appendixA}{\setcounter{equation}{0}
\def\theequation{\rm{A}.\arabic{equation}}\section*}
\newcommand{\appendixB}{\setcounter{equation}{0}
\def\theequation{\rm{B}.\arabic{equation}}\section*}
\newcommand{\appendixC}{\setcounter{equation}{0}
\def\theequation{\rm{C}.\arabic{equation}}\section*}
\newcommand{\appendixD}{\setcounter{equation}{0}
\def\theequation{\rm{D}.\arabic{equation}}\section*}
\newcommand{\appendixE}{\setcounter{equation}{0}
\def\theequation{\rm{E}.\arabic{equation}}\section*}
\newcommand{\appendixF}{\setcounter{equation}{0}
\def\theequation{\rm{F}.\arabic{equation}}\section*}
\newcommand{\appendixG}{\setcounter{equation}{0}
\def\theequation{\rm{G}.\arabic{equation}}\section*}
\def\baselinestretch{1.4}
\catcode`\@=11
\def\marginnote#1{}
\newcount\hour
\newcount\minute
\newtoks\amorpm
\hour=\time\divide\hour by60
\minute=\time{\multiply\hour by60 \global\advance\minute by-\hour}
\edef\standardtime{{\ifnum\hour<12 \global\amorpm={am}%
        \else\global\amorpm={pm}\advance\hour by-12 \fi
        \ifnum\hour=0 \hour=12 \fi
        \number\hour:\ifnum\minute<10 0\fi\number\minute\the\amorpm}}
\edef\militarytime{\number\hour:\ifnum\minute<10 0\fi\number\minute}
\def\draftlabel#1{{\@bsphack\if@filesw {\let\thepage\relax
   \xdef\@gtempa{\write\@auxout{\string
      \newlabel{#1}{{\@currentlabel}{\thepage}}}}}\@gtempa
   \if@nobreak \ifvmode\nobreak\fi\fi\fi\@esphack}
        \gdef\@eqnlabel{#1}}
\def\@eqnlabel{}
\def\@vacuum{}
\def\draftmarginnote#1{\marginpar{\raggedright\scriptsize\tt#1}}
\def\draft{\oddsidemargin -.5truein
        \def\@oddfoot{\sl preliminary draft \hfil
        \rm\thepage\hfil\sl\today\quad\militarytime}
        \let\@evenfoot\@oddfoot \overfullrule 3pt
        \let\label=\draftlabel
        \let\marginnote=\draftmarginnote
   \def\@eqnnum{(\theequation)\rlap{\kern\marginparsep\tt\@eqnlabel}%
\global\let\@eqnlabel\@vacuum}  }

\def\preprint{\twocolumn\sloppy\flushbottom\parindent 1em
        \leftmargini 2em\leftmarginv .5em\leftmarginvi .5em
        \oddsidemargin -.5in    \evensidemargin -.5in
        \columnsep 15mm \footheight 0pt
        \textwidth 250mmin      \topmargin  -.4in
        \headheight 12pt \topskip .4in
        \textheight 175mm
        \footskip 0pt
\def\@oddhead{\thepage\hfil\addtocounter{page}{1}\thepage}
        \let\@evenhead\@oddhead \def\@oddfoot{} \def\@evenfoot{} }

\def\titlepage{\@restonecolfalse\if@twocolumn\@restonecoltrue\onecolumn
     \else \newpage \fi \thispagestyle{empty}\c@page\z@ 
        \def\thefootnote{\fnsymbol{footnote}} }

\def\endtitlepage{\if@restonecol\twocolumn \else  \fi
        \def\thefootnote{\arabic{footnote}}
        \setcounter{footnote}{0}}  

\catcode`@=12
\relax
\def\abs#1{\left| #1\right|}
\def\bC{\mathop{\bf C}}
\def\bea{\begin{array}}
\def\bem{\begin{displaymath}}
\def\beq{\begin{equation}}
\def\beqa{\begin{eqnarray}}
\def\bR{\mathop{\bf R}}
\def\bra#1{\left\langle #1\right|}
\def\eea{\end{array}}
\def\eem{\end{displaymath}}
\def\eeq{\end{equation}}
\def\eeqa{\end{eqnarray}}
\def\eq{\beq\eeq}                          
\def\eqr#1{\beq\label#1\eeq}               
\def\half{\frac{1}{2}}
\def\Im{\mathop{\rm Im}}
\def\ket#1{\left| #1\right\rangle}
\def\sket#1{| #1 >}
\def\lie{\hbox{\it \$}}                          
\def\lineint{\oint \frac{d z}{2 \pi i}} 
\def\modsq#1{| #1 |^2}
\def\NP#1#2#3{Nucl. Phys. \underline{#1} (19#2) #3}
\def\ov{\overline}
\def\partder#1#2{{\partial #1\over\partial #2}}
\def\PL#1#2#3{Phys. Lett. \underline{#1} (19#2) #3}
\def\PR#1#2#3{Phys. Rev. \underline{#1} (19#2) #3}
\def\PRL#1#2#3{Phys. Rev. Lett. \underline{#1} (19#2) #3}
\def\Re{\mathop{\rm Re}}
\def\secder#1#2#3{{\partial^2 #1\over\partial #2 \partial #3}}
\def\s2w{\sin^2 \theta_W}
\def\Tr{\mathop{\rm Tr}}
\def\und{\underline}
\def\VEV#1{\left\langle #1\right\rangle} \let\vev\VEV
\def\mbf#1{\hbox{\boldmath $#1$}}
\relax
\def\dalpha{{\dot\alpha}}
\def\dbeta{{\dot\beta}}
\def\drho{{\dot\rho}}
\def\dsigma{{\dot\sigma}}
\def\crbig{\\\noalign{\vspace {3mm}}}
\def\bigint{{\displaystyle\int}}
\def\S{\Sigma}
\def\G{\Gamma}
\def\L{{\cal L}}
\def\SG{S_{\Gamma}}
\def\Fint{{\bigint d^2\theta\,}}
\def\Fbarint{{\bigint d^2\ov\theta\,}}
\def\Dint{{\bigint d^2\theta d^2\ov\theta\,}}

\title{ \vspace*{-1.8cm}
\begin{flushright}
\normalsize{CERN-PH-TH/2009-042}\\ 
\end{flushright}
\vspace{1cm}
\bf{\hspace{-1cm} Fermion Wavefunctions in Magnetized branes: 
\\Theta identities and Yukawa couplings } \vspace*{-0.3cm}}

\begin{document}
\author{\bf\large{
Ignatios Antoniadis$^{1}$\footnote{On leave from CPHT (UMR du CNRS 
7644), 
Ecole Polytechnique, F-91128 Palaiseau 
Cedex, France}
\footnote{Ignatios.Antoniadis@cern.ch}~,
Alok Kumar$^{2}$\footnote{kumar@iopb.res.in}~, 
Binata Panda$^{1,2}$\footnote{Binata.Panda@cern.ch,binata@iopb.res.in}}\\  
\\[-3mm]
\emph{\normalsize $^1$Department of Physics, CERN - Theory Division, 
CH-1211 Geneva 23, Switzerland }\\
\emph{\normalsize $^2$Institute of Physics, Bhubaneswar 751 005, India}\\
}

\date{\today}

\maketitle
\thispagestyle{empty}

\begin{abstract}
{
Computation of Yukawa couplings, determining superpotentials as well as
the K\"{a}hler metric, with oblique (non-commuting)
fluxes in magnetized brane  constructions 
is an interesting unresolved issue, in view of the importance of such fluxes
for obtaining phenomenologically viable models. In order to perform this
task, fermion (scalar) wavefunctions on toroidally compactified spaces
are presented for general fluxes,
parameterized by Hermitian matrices with eigenvalues of arbitrary signatures.
We also give explicit mappings among fermion wavefunctions, of different
internal chiralities on the tori, which interchange the role of the flux 
components with the complex structure of the torus.
By evaluating the overlap integral of the wavefunctions,
we give the expressions for Yukawa couplings among chiral multiplets
arising from an arbitrary set of branes (or their orientifold images). 
The method is based on constructing 
certain mathematical identities for general Riemann theta functions with
matrix valued modular parameter. We briefly discuss an 
application of the result, for the 
mass generation of non-chiral fermions, in the $SU(5)$ GUT model presented 
by us in arXiv: 0709.2799.}
\end{abstract}
\date

\vspace*{-0.8cm}
\newpage

\section{Introduction}\label{introduction}

Close form expressions for Yukawa couplings have been 
written down for string constructions involving branes at angles 
\cite{ibanez2,cvetic-yukawa} 
or those with magnetized branes 
\cite{bachas,AADS,ibanez,mag1,mag2,mag3,mag4,mag5,mag6,akm,mag7,mag8,mag9,mag10,mag11}. 
In the IIA picture, the interaction 
is described by the worldsheet instanton contributions from the 
sum of areas of various triangles that are formed by three $D6$ branes 
intersecting  at three vertices, forming a triangle. This is due to 
the fact that the intersection  of  branes relevant for Yukawa 
interactions are those which are point-like  
giving chiral multiplets. Line or surface like intersections, 
on the other hand, would give rise to  interactions of non-chiral matter.
In these discussions, the orientation of the branes themselves are 
parameterized by three angles in  the three orthogonal 2-planes, inside $T^6$.
These results have been further  generalized to include 
Euclidean $D2$ brane instanton contributions to the Yukawa couplings 
\cite{cvetic1,cvetic2,cvetic3,cvetic4,cvetic5,uranga1,uranga2,uranga3,uranga4}, 
generating up quark and right handed neutrino
masses through a Higgs mechanism, in a particular class of models. A limitation on the exercise 
performed in these papers comes from the factorized structure of the tori,
which arises from the orientations of the brane wrappings that are classified
by angles in three different $T^2$ planes, rather than their 
general orientations in the internal six dimensional space 
parameterized for instance  by the $SU(3)$ angles in  supersymmetric situations.

Similar results for perturbative Yukawa couplings 
have also been obtained in the magnetized brane picture, based
on their gauge theoretic representation \cite{ibanez}. 
In this case, the interactions are given  by the overlap integral of
three wavefunctions (contributing to  the interaction) along internal directions.
The wavefunctions correspond, in the ordinary field theory context, to those
belonging to two fermions and a scalar,  and
are given by  Jacobi theta functions, 
when fluxes are turned on along three diagonal 2-tori. The relationship
between the Yukawa interactions in the magnetized brane constructions and
those involving $D6$ branes, have also been established using
T-duality rules.  However, these exercises have once again 
been of limited scope due to 
the fact that explicit expressions are written down only
for  magnetized branes with fluxes that are diagonal along three $T^2$'s. 

Technically, the wavefunctions of  chiral fields participating in 
Yukawa interactions are defined in terms of  Jacobi theta functions, 
with a modular parameter identified as a product of the 
complex structure of the  $T^2$, with the flux that is turned on
along it. The Yukawa interactions are therefore computed for the case
when the six dimensional internal space is of a factorized form:
\beq
T^2 \times T^2 \times T^2 \in T^6.
\label{t2t6}
\eeq

For applications to  model building with moduli stabilization though, 
one in general 
needs to include both `diagonal' and `oblique' fluxes \cite{mag5,akm,mag7,akp}.
 Therefore it is 
imperative that we generalize previous results further and obtain 
interactions involving  branes with oblique fluxes. As stated, 
in the language of $D6$ branes such 
generalizations would amount to intersections of branes with  orientations 
given by $SU(3)$ rotation angles,  resulting to $N=1$ 
supersymmetry in $D=4$ with chiral matter. In view of the importance of 
such fluxes in obtaining realistic particle physics models with stabilized moduli, and to 
describe the interactions among the chiral  fields, we shall 
study the explicit construction of  fermion (and scalar) wavefunctions
on  compact toroidal spaces with arbitrary constant fluxes.

Scattered results on fermion wavefunctions in presence of constant
gauge fluxes, on tori of arbitrary dimensions, exist  already in the 
literature \cite{japan,ibanez}.
However, they are of limited use for our purpose. First, any wavefunction
obtained through a diagonalization process of the gauge 
fluxes \cite{japan}, is   not in general suitable for obtaining an 
overlap integral of wavefunctions. This is because the
flux matrices need not commute along different stacks of branes that
participate in the interaction through the chiral multiplets, arising 
from the strings that join these branes and 
  therefore they are not simultaneously diagonalizable.

In \cite{ibanez}, a set of wavefunctions was given 
for constant gauge fluxes. However, once again,  explicit results are 
 valid only for those fluxes which satisfy a set of 
`Riemann conditions', including a positivity criterion on the 
flux matrices. As the analysis in our paper will
clarify, the positivity restrictions on the fluxes is due to the fact that 
the given wavefunction in \cite{ibanez } corresponds to a specific
component of the $2^n$ dimensional Dirac spinor for a $2n$-dimensional 
torus $T^{2n}$. We will show that this restriction is relaxed, if one 
considers wavefunctions of various  chiralities, such that
all possible flux matrices are allowed, though in our case we 
restrict to only those fluxes that are consistent with the requirements of
space-time supersymmetry .

In fact, we give explicit solutions for the wavefunctions for arbitrary
fluxes, that are well defined globally on the toroidal space. We also 
give explicit mappings among the wavefunctions of different chiralities,
satisfying different consistency criterion. These mappings are 
shown to relate wavefunctions  corresponding to different fluxes and 
complex structures of the tori. We in fact reconfirm that our 
wavefunctions, as well as mappings are indeed correct, by showing that 
equations of motion also map into each other for the fermion
wavefunctions just described, corresponding to different 
internal chiralities.

Apart from the lack of enough knowledge about the fermion wavefunctions,
the limitations on available information about the Yukawa couplings 
for general gauge fluxes also arose from the technicalities in dealing
with  general Riemann theta functions that are used for defining the 
wavefunctions on  toroidal spaces.
Internal wavefunctions of chiral fermions participating in the interaction 
are given by a general Riemann theta function whose modular parameter
argument is determined in terms of the complex structure of 
$T^6$ as well as the `oblique' fluxes that we turn on.
Hence,  the limitations on available results for
Yukawa interactions in the literature, 
arise due to the intricacies involved in evaluating the 
overlap integrals of the trilinear product of general Riemann theta functions 
over the six dimensional internal space. 
In particular, even for  positive chirality wavefunctions
along the internal $T^6$ given in \cite{ibanez},
one finds that theta identities \cite{mumford}  need to be
further generalized, in order to 
compute the Yukawa interactions with oblique fluxes.
The task goes beyond the identity given in \cite{mumford}, since
one needs to evaluate the overlap integral of three wavefunctions, 
all having different modular parameter matrices as arguments, due 
to the presence of different fluxes along the three brane stacks 
involved in generating the Yukawa coupling.

In this paper, first, 
we generalize the identities used in \cite{ibanez}
(available from mathematical literature \cite{mumford}) for the known positive chirality wavefunctions 
to those with general Riemann theta functions representing the fermion wavefunctions.
This gives an  explicit answer for the Yukawa interaction
in a close form and generalizes the results of \cite{ibanez,ibanez2}.
In particular,  we generalize the result further for the positive chirality 
wavefunction, when general (hermitian)  fluxes with all nine parameters
are turned on rather than the six components considered before.

On the other hand, as already stated earlier, we give explicit construction
of the other $T^6$ spinor wavefunctions, as well. In these cases too, we
obtain the selection rules among chiral multiplets giving nonzero Yukawa 
couplings. Now, however, the final answer is left as 
a real finite integration of a theta function,
over three toroidal coordinate variables. The integration can be 
evaluated numerically for any given example.
Finally, in the paper, we also briefly discuss the issue of mass generation
for non-chiral multiplets given in the $SU(5)$ GUT  model, constructed in
\cite{akp}. Although the Yukawa couplings can be used for giving  
the precise masses in our model, where all  close string moduli 
are fixed, we avoid entering into these details in this work.

The plan of the rest of the paper is as follows.  In Section 2,
we start by discussing the general setup, including the gauge fluxes
that can be turned on, in a consistent manner. In Section 3, we
review  the known results on the
Jacobi theta identity  given in \cite{mumford} and present a proof 
of its validity. We also give an expression for the Yukawa interaction 
for factorized tori and `diagonal' fluxes using 
the theta identity. In Section 4, we  construct a similar identity, 
now for the general Riemann theta function.
We then use this new mathematical relation for writing down the expression for
the Yukawa interaction when oblique fluxes are present and satisfy the
`Riemann conditions' of \cite{ibanez}. Results are further 
generalized to include the most general flux matrices consistent with 
supersymmetry and `Riemann condition' requirements. In order to 
relax the later,
in Section 5, we present the generalizations to include the 
wavefunctions of the other internal
chiralities, in order to accommodate general fluxes consistent with 
supersymmetry restrictions. 
In Section 6, we briefly
present an independent analysis of the superpotential and D-terms 
for the model of \cite{akp} in order to show how masses for several non-chiral 
fermion multiplets can be generated, 
without evaluating explicitly  the superpotential coefficients, which would go beyond the scope of our present work.
Conclusions are presented in Section 7. In Appendix A we give the 
chiral fermion wavefunctions in the presence of constant fluxes.
Appendix B contains information on fluxes in terms of windings and 
Chern numbers, while Appendix C gives some details of our
model in \cite{akp}, needed for  the mass generation analysis of  Section 6.

\section{Fluxes}\label{fluxes}

We now start by describing the gauge fluxes that can be turned on
along internal tori. A general gauge flux, 
on $T^6$ with coordinates $X^I \equiv (x^i, y^i)$, 
$i=1, 2, 3$, has the form:
\beqa
	F \equiv p_{IJ} dX^I \wedge dX^J \hspace{1.0in} \cr 
  = p_{x^i x^j} dx^i \wedge dx^j + p_{y^i y^j} dy^i \wedge dy^j
	+ p_{x^i y^j} dx^i \wedge dy^j + p_{y^i x^j} dy^i \wedge dx^j \,\,.
\label{general-flux} 
\eeqa
Then using the definition of a general complex structure matrix $\Omega$:
\beq
     dz^i = dx^i + \Omega^i_j dy^j,\,\,\,
     d\bar{z^i} = dx^i + \bar{\Omega}^i_j dy^j,
\label{zi-barzi}
\eeq
we obtain:
\beq
	F = F_{z^i z^j} dz^i \wedge dz^j + 
	F_{z^i \bar{z}^j} (idz^i \wedge d\bar{z}^j)
	+ F_{\bar{z}^i \bar{z}^j} d\bar{z}^i \wedge d\bar{z}^j .
\label{general-flux2} 
\eeq
Choosing the basis $e^{i \bar{j}}$ of the cohomology $H^{1,1}$ to be of the form
\beq
e^{i \bar{j}} = i dz^i \wedge d\bar{z}^j ,
\label{chosen-basis}
\eeq 
we obtain:
\beq
       F_{z^i {z}^j} = {(\bar{\Omega} - \Omega)^{-1}}^T
	\left( \bar{\Omega}^T p_{xx} \bar{\Omega} - \bar{\Omega}^T p_{xy}
	+ p_{xy}^T \bar{\Omega} + p_{yy}\right){(\bar{\Omega} - \Omega)^{-1}}
\label{F(2,0)}
\eeq
and 
\beq
       F_{z^i \bar{z}^j} = (-i){(\bar{\Omega} - \Omega)^{-1}}^T
	\left( \bar{\Omega}^T p_{xx} {\Omega} - \bar{\Omega}^T p_{xy}
	+ p_{xy}^T {\Omega} + p_{yy}\right){(\bar{\Omega} - \Omega)^{-1}}.
\label{F(1,1)}
\eeq
In addition, $F_{\bar{z}^i \bar{z}^j}$  is complex conjugate to 
$F_{z^i z^j}$ and $F_{\bar{z}^i {z}^j} = - F_{z^j \bar{z}^i}$. 

Now, supersymmetry demands  all fluxes to be of $(1, 1)$ form which gives us
the condition:
\beq
	\left( \bar{\Omega}^T p_{xx} \bar{\Omega} - \bar{\Omega}^T p_{xy}
	+ p_{xy}^T \bar{\Omega} + p_{yy}\right) = 0,	
\label{(2,0)=0}
\eeq
or equivalently:
\beq
	\left( {\Omega}^T p_{xx} {\Omega} - {\Omega}^T p_{xy}
	+ p_{xy}^T {\Omega} + p_{yy}\right) = 0.
\label{(0,2)=0}
\eeq
Eqs. (\ref{(2,0)=0}) and (\ref{(0,2)=0}) together give
two real matrix equations. These equations can then be used to 
eliminate some of the variables and write the final 
$(1, 1)$ form in terms of certain independent variables only.

Using eq. (\ref{(0,2)=0}), eq. (\ref{F(1,1)}) reduces to the following form,
\beq
 F_{z^i \bar{z}^j} = -i \left( p_{xx} {\Omega}  -  p_{xy} \right) 
(\bar{\Omega} - \Omega)^{-1}  \,\,.
\label{newF(1,1)-1}
\eeq
On the other hand, use of eq. (\ref{(2,0)=0}) in eq. (\ref{F(1,1)}) gives,
\beq
F_{z^i \bar{z}^j} = -i{(\bar{\Omega} - \Omega)^{-1}}^T
\left( -\bar{\Omega}^T p_{xx} - p_{xy}^T\right)  \,\,.
\label{newF(1,1)-2}
\eeq
We also notice that the $(1, 1)$ form $F_{z^i \bar{z}^j}$ given in 
eq. (\ref{F(1,1)}) satisfies the hermiticity property: 
$F_{z^i \bar{z}^j} = F^{\dagger}_{z^i \bar{z}^j}$. 
To explicitly see that,  we use eqs. (\ref{newF(1,1)-1}), (\ref{newF(1,1)-2}).
\beqa \nonumber
F^{\dagger}_{z^i \bar{z}^j} = \left[ \left( -i \left( p_{xx} {\Omega}  -  p_{xy} \right) (\bar{\Omega} - \Omega)^{-1}\right) ^{*}\right] ^{T} \\
= -i{(\bar{\Omega} - \Omega)^{-1}}^T\left( -\bar{\Omega}^T p_{xx} - p_{xy}^T\right) 
= F_{z^i \bar{z}^j}
\label{hermiticity1} 
\eeqa
There are some  special cases, however, in which 
eqs. (\ref{(2,0)=0}) and (\ref{(0,2)=0}) simplify further and 
the resulting $F_{z^i \bar{z}^j}$ can be written more compactly. 
One such case arises when $p_{xx}$ and $p_{yy}$ components are
turned off. In such a situation $F_{(2, 0)} = 0$ condition (\ref{(0,2)=0}),
reduces to:
\beq
	 {\Omega}^T p_{xy} = p_{xy}^T {\Omega}.
\label{(0,2)=0-2}
\eeq

Thus far, we have concentrated on the spatial components of the gauge fluxes,
but ignored the gauge indices. In the magnetized $D$-brane construction,
gauge quantum numbers arise from the Chan-Paton factors associated with
the end points of the open strings for a given stack of branes. 
The simplest possibility is to consider 
fluxes with gauge indices given by an $n\times n$  identity matrix
for a stack of $D$-branes:
\beq
    F  = {m} I_n , 
\label{flux-u1(n)}
\eeq
with $m$ an arbitrary integer giving the 1st Chern number. All 
spatial indices of the gauge flux above have been suppressed, which are given
as in eq. (\ref{general-flux})  by the components :
$p_{x^iy^j}$, $p_{x^ix^j}$, $p_{y^iy^j}$, while  their quantized 
values are expressed in eq. (\ref{quantization}) for general wrapping 
of the branes. Actually,  eq. (\ref{flux-u1(n)}) corresponds to the 
situation when all the wrapping numbers are trivial: $n^{x^i} =  n^{ y^i} = 1$ 
in the denominators of eq. (\ref{quantization}).
 $F$, then represents a stack of $n$
magnetized $D$-branes with a $U(1)^n$ gauge flux.  The first Chern number 
for each of the $U(1)$ fluxes is equal to $m$. Moreover, $D$-brane wrapping
numbers on the internal directions, are all unity, 
given by a diagonal embedding of the brane in target space
and winding around each 1-cycle once. In most of the paper, we will 
consider fluxes of the above type. 

For multiple stacks of $n_i$ branes with respective 1st Chern numbers
$m_i$, the flux matrix is of block diagonal form:
\beqa
    F  = \begin{pmatrix} m_1 I_{n_1} & & & & \cr  
			& m_2 I_{n_2} & & &\cr
			& & . &  & \cr & & & . & \cr 
			& & & & m_{n_p}I_{n_p}\end{pmatrix}
\label{flux-blocks}
\eeqa
and corresponds to gauge fluxes in the diagonal $U(1)$'s of 
$U(n_1)\times U(n_2)\times \cdots$ gauge group.
Chiral fermion wavefunctions, on the other hand, belong to the 
bifundamental representations $(n_a, \, \bar{n}_b)$ etc. and are
determined by the field equations in terms of the difference of
fluxes in the respective stacks, parameterized by a matrix {\bf N}:
\beq
	{\bf N}^T = p_{xy}^a - p_{xy}^b \, .
\label{def-N}
\eeq
Using (\ref{(0,2)=0-2}), it follows that: 
\beq
	({\bf N} \Omega )^T = ({\bf N} \Omega ).
\label{def-N-symmetry}
\eeq


Moreover, further conditions are imposed on the matrix $\bf{N}$ 
in order to have well defined bifundamental wavefunctions. These are
the so-called Riemann conditions \cite{ibanez} and are written as:
\beqa
	{\bf{N}}_{\bar{i} j} \in {\bf{Z}} \,,\cr
	{(\bf{N}}. Im {\Omega})^T = {\bf{N}}. Im {\Omega} \,,\cr
	{\bf{N}} . Im {\Omega} > 0.
\label{eq-riemann-conditions}
\eeqa
The first condition 
in eq. (\ref{eq-riemann-conditions}) is the integrality of the elements
of $\bf{N}$, that we discuss later on, in the absence of any non-abelian 
Wilson lines \cite{ibanez}, following from the Dirac quantization of 
fluxes. To understand the last condition of eq. 
(\ref{eq-riemann-conditions}), one rewrites the $(1, 1)$ form 
$F_{z^i \bar{z}^j}$, for the case when $p_{xx} = p_{yy} = 0$.
Indeed using eq. (\ref{(0,2)=0-2}), one obtains:
\beq
       F_{z^i \bar{z}^j} =  - i p_{xy}{({\Omega} - \bar{\Omega})^{-1}},
\label{F(2,0)-n1}
\eeq
which matches with the expression for $H$ in eq. (4.73) of \cite{ibanez}
upon the identification $\bf{N}^T = p_{xy}$ and 
$ H = \frac{1}{2} {\bf N}^T . Im \Omega^{-1}$.
The positivity requirement on $H$ then arises from the condition that
the solutions of the Dirac equation, corresponding to  chiral 
wavefuntions, be normalizable. 

 Gauge fluxes on  branes with higher wrapping numbers
can also be given a gauge theoretic interpretation.
The method, as stated earlier, is based on a representation of the 
magnetized brane constructions\cite{ibanez} in terms of fluxes along 
internal directions in a compactified gauge theory. In this picture,
the effect of windings of  branes around $T^6$
 is simulated by the rank of the gauge group. 
In particular, due to the Dirac quantization condition on fluxes,
a $U(n)$ flux on, say $T^2$: 
\beq
    F  = \frac{m}{n} I_n , 
\label{flux-u(n)}
\eeq
with $I_n$ being the $n$-dimensional identity matrix, and $(n, m)$ relatively
prime, represents a single brane wound $n$ times around $T^2$ with flux quantum
$m$ and resulting gauge symmetry being only $U(1)$. On the other hand, if $m$
is an integer multiple of $n$ such that $m = p n$, then each of the entries in 
the identity matrix represents a well defined $U(1)$ flux of quantum $p$ and
the gauge symmetry is  $U(n)$, given by a stack of $n$ such magnetized branes,
as described in the last paragraph.
It turns out that explicit realization of fluxes with $(n, m)$ 
relatively prime, needs gauge configurations with non-abelian Wilson lines.

The wavefunctions of the chiral fermion bifundamentals, with both 
abelian and non-abelian Wilson lines,
involved in Yukawa computations, are given in \cite{ibanez} for the case
of the factorized tori, eq. (\ref{t2t6}), and diagonal fluxes. 
For oblique fluxes, we postpone the discussion of 
non-abelian Wilson lines and rational fluxes to the last part
of the paper (Section \ref{conclusion}) 
and for the moment we consider the case of 
integral fluxes only. This restriction, nevertheless, allows for  a rich structure of 
phenomenological value, since semi-realistic models with three
generations of chiral fermions and stabilized moduli can be built  even in the context of such integral fluxes, by turning on NS-NS antisymmetric 
tensor background. For example, a three generation $SU(5)$ GUT with stabilized 
moduli given in \cite{akp} was constructed with all winding numbers, $n=1$,
for different stacks of branes.
Also, the presence of a half-integral NS-NS antisymmetric 
tensor does not modify any of our results, since all the relevant chiral 
fermion wavefunctions depend on the difference of fluxes along pairs of brane stacks
which is always integral.

\section{Yukawa computation on factorized tori}\label{yukawa-factorized}

\subsection{Wavefunction}\label{factorized-wavefunction}

A detail discussion of the chiral fermion wavefunctions in the presence of
constant gauge fluxes is given in Appendix \ref{wavefunction} for general 
tori and fluxes. In the case of factorized tori, eq. (\ref{t2t6}), 
the six dimensional chiral/anti-chiral wavefunctions are written as 
a product of wavefunctions on $T^2$. 
To show this explicitly, we present (as in Appendix \ref{wavefunction})
the case of $T^4$ as an example,  with $T^6$ case working out in a similar fashion.  
More precisely, 
considering that on two $T^2$'s, fermion wavefunctions
\beqa
\psi^{(1)} =  \begin{pmatrix} \psi_+^{(1)} \cr 
			\psi_-^{(1)} \end{pmatrix},\,\,\,\,
\psi^{(2)} =  \begin{pmatrix} \psi_+^{(2)} \cr 
			\psi_-^{(2)} \end{pmatrix},
\label{2Dwavefunction}
\eeqa
with their internal $U(n_1)\times U(n_2)$ structure being represented in a 
manner as in eq. (\ref{bi-wavefunction}), satisfy the equations:
\beqa
\bar{\partial}_1 \psi_+^{(1)}  +
	(A^1 - A^2)_{\bar{z_1}} \psi_+^{(1)}  = 0, \cr
{\partial}_1 \psi_-^{(1)}  +
	(A^1 - A^2)_{{z_1}} \psi_-^{(1)}  = 0, \cr
\bar{\partial}_2 \psi_+^{(2)} + 
	(A^1 - A^2)_{\bar{z_2}} \psi_+^{(2)}  = 0, \cr
{\partial}_2 \psi_-^{(2)} + 
	(A^1 - A^2)_{{z_2}} \psi_-^{(2)}  = 0 .
\label{Dirac-equation3-2d}
\eeqa
$T^4$ fermion wavefunctions are then constructed through a direct product
of $\psi^1$ and $\psi^2$ (in the notations of Appendix \ref{wavefunction}):
\beqa
   \begin{pmatrix} \Psi_+^1 \cr \Psi_-^2 \cr \Psi_-^1 \cr 
		\Psi_+^2 \end{pmatrix} \equiv 
\begin{pmatrix} \psi_+^{(1)} \cr \psi_-^{(1)} \end{pmatrix} \otimes
\begin{pmatrix} \psi_+^{(2)} \cr \psi_-^{(2)} \end{pmatrix} .
\label{Dwavefunction-product}
\eeqa
In particular, 
\beq
	\Psi_+^1 \equiv \psi_+^{(1)} \otimes \psi_+^{(2)}
\label{chiral-product}
\eeq
satisfies precisely the equations (\ref{Dirac-equation3}) for  
chiral fermions on $T^4$. We can further extend these results to show
that $T^6$ chiral wavefunctions can also be written as a product of 
the chiral wavefunctions on three $T^2$'s in the decomposition 
(\ref{t2t6}).

Yukawa interaction on $T^6$ is then also given by an expression which is a 
direct product of the interaction terms for the three $T^2$'s. 
Wavefunctions for the chiral fermions on a  $T^2$ (with 
coordinates $x$, $y$) are 
expressed in terms of the basis wavefunctions 
$\psi^{j,N}$ \cite{ibanez}:
\beq
 \psi^{j, N} (\tau, z) = \cn \cdot e^{i\pi N z \pim z/\pim \tau} 
\cdot 
\vt
\left[
\begin{array}{c}
\frac{j}{N} \\ 0
\end{array}
\right]
(N z, N \tau), \quad \quad j=0,\dots,N-1\,\, , 
\label{basis1}
\eeq
with $N$ denoting the difference of the $U(n_a)$ and $U(n_b)$ 
magnetic gauge fluxes as given in eq. (\ref{def-N}),
turned on along the Cartan generators, 
representing stacks of $n_a$ and $n_b$
branes respectively and gives the degeneracy of the chiral fermions:
\beq
	N = m_a - m_b \equiv I_{ab},
\label{Nab}
\eeq
with $m_a$ and $m_b$ being the 1st Chern number of fluxes
along stacks $a$ and $b$, with unit windings, as defined through eq. (\ref{quantization}).

Using such a basis, the chiral and anti-chiral (left and right handed 
fermions) basis wavefunctions:
\beqa
\psi^j =  \begin{pmatrix} \psi_+^j \cr \psi_-^j \end{pmatrix},\,\,\,\,
\label{2Dwavefunction-1}
\eeqa
are given by:
\beqa
	\psi_+^j = \psi^{j, N}(\tau, z +\zeta),\,\,
	(\psi_+^j)^* = \psi^{-j, -N}(\bar{\tau}, \bar{z} +\bar{\zeta}),\cr
	\cr
	\psi_-^j = \psi^{j, N}(\bar{\tau}, \bar{z} +\bar{\zeta}),\,\,
	(\psi_-^j)^* = \psi^{-j, -N}(\tau, z +\zeta),
\label{chiral-solutions}
\eeqa
and satisfy the equations:
\beqa
D\psi_+^j \equiv 
(\bar{\partial} + \frac{\pi N}{2 Im \tau (z +\zeta)}) \psi_+^j = 0,\cr
D^{\dagger}(\psi_+^j)^* \equiv 
({\partial} - \frac{\pi N}{2 Im \tau (z +\zeta)})(\psi_+^j)^* = 0,\cr
D^{\dagger}\psi_-^j \equiv 
({\partial} - \frac{\pi N}{2 Im \tau (z +\zeta)})\psi_-^j = 0,\cr
D(\psi_-^j)^* \equiv 
(\bar{\partial} + \frac{\pi N}{2 Im \tau (z +\zeta)})(\psi_-^j)^* = 0,
\label{chiral-equations}
\eeqa
with $\zeta$ representing the Wilson lines. In the following 
we set the Wilson lines $\zeta =0$.
Furthermore, expressions of the chiral and anti-chiral solutions, as given in  
eqs. (\ref{chiral-solutions}) and (\ref{basis1}), are well defined provided 
$N > 0$ for the wavefunctions $\psi_+^j$ and 
$N < 0$ for the wavefunctions $\psi_-^j$. In these cases, for $\psi_+^j$ and
$\psi_-^j$ to be properly normalized:
\beq
	\int_{T^2} dz d\bar{z}\, \psi_{\pm}^j (\psi_{\pm}^k)^* = \delta_{jk},
\label{normalizable}
\eeq
an additional factor
\beq
 {\cal{N}}_j = \left(\frac{2 Im\tau |N|}{{\cal{A}}^2}   \right)^{\frac 14}
\label{normalization-constant}
\eeq
needs to be introduced, with $\cal{A}$ being the area of the $T^2$.

In fact, the basis functions (\ref{basis1}) are also  eigenfunctions
of the Laplacian. We elaborate on this point more in Section \ref{laplace}
and now proceed to make use of these fermion and boson basis functions to 
determine the Yukawa interaction in the case of factorized tori
and `diagonal' fluxes.


\subsection{Interaction for factorized tori}\label{interaction-factorized}

We now summarize the basic results of \cite{ibanez} regarding the
computations of Yukawa interactions.  Such four dimensional interaction
terms were obtained  through a dimensional reduction 
of the $D=10$, $N=1$ super-Yang-Mills theory to four dimensions in the
presence of constant magnetic fluxes. The Yukawa coupling is given by
\beq
	Y_{ijk} = \int_{\cal{M}} \psi_i^{a\dagger} \Gamma^m \psi_j^b \phi_{k, m}^c
			f_{abc},
\label{yukawa}
\eeq
where $\cal{M}$ is the internal space on which the gauge theory has been 
compactified and $\psi$ and $\phi$ are the internal zero mode fluctuations 
of the gaugino and Yang-Mills fields with $f_{abc}$ being the structure constants of the
higher dimensional gauge group. For the torus compactification that we are 
discussing, the internal wavefunctions are factorized
into those depending on the coordinates of three $T^2$'s.
In turn, these
involve the evaluation of  terms of the type:
\beq
	\int_{T^2} dz d\bar{z} Tr\{\psi_+ .[\phi_-, \psi_+]\}\quad
	\mathrm{and}\quad
	\int_{T^2} dz d\bar{z} Tr\{\psi_- .[\phi_+, \psi_-]\},
\label{integral}
\eeq
with $\phi_{\pm}$ being the wavefunctions of the bosonic fluctuations
of the ten dimensional gauge fields with helicity $\pm 1$ along the
particular $T^2$ direction. Similarly $\psi_{\pm}$ denotes the
spinor fluctuations with helicities $\pm \frac12$. Therefore, In the factorized case
of eq. (\ref{t2t6}), the full interaction term is computed
as a product of three such integrals. 
To evaluate these integrals, one uses the wavefunctions  (\ref{2Dwavefunction})
and basis functions as given in eq. (\ref{basis1}).

In the language of string construction with magnetized branes, 
$N \equiv I_{a b}$ corresponds to the intersection number for the 
string starting at a stack $a$ and ending on another one $b$.
The Yukawa interaction then reads:
\beq
    Y_{ijk} = g\sigma_{a b c}  \int_{T^2} dz d\bar{z}\, 
	\psi^{i, I_{ab}}(\tau, z).\psi^{j, I_{ca}}(\tau, z).
	(\psi^{k, I_{cb}}(\tau, z))^*  
\label{yukawa-t2-abc}
\eeq
with $I_{bc} < 0$, corresponding to the fact that when the 
intersection numbers $I_{ab}$ and $I_{ca}$ are positive, then 
$I_{bc}$ has to be negative, since  
$I_{ab} + I_{bc} + I_{ca} = 0$. 
A similar expression exists for $I_{bc} > 0$ as well.
To evaluate this integral,  one uses an identity, satisfied by
the theta functions appearing in the definition of the basis functions
 (\ref{basis1}). The aim of this relation is to establish a connection 
between the wavefunctions with 
intersection numbers $N_1$ and $N_2$ for bifundamental states
in brane intersections $ab$ and $ca$ with the one in the intersection $bc$
with $N_3 = N_1 + N_2$. However, in view of the further 
generalization to the oblique flux case, we establish this identity 
explicitly in the next subsection and generalize it further in 
Section {\ref{general-tori}}.

\subsection{Jacobi theta function identities}\label{jacobi-theta}

We now explicitly prove the following theta function identity\cite{mumford} used in 
\cite {ibanez} for computing the Yukawa couplings:
\beqa 
&&\hskip -0.7cm \vt
\left[
\begin{array}{c}
\frac{r}{N_1} \\ 0
\end{array}
\right]
(z_1, \tau N_1)
\cdot
\vt
\left[
\begin{array}{c}
\frac{s}{N_2} \\ 0
\end{array}
\right]
(z_2, \tau N_2)   = 
\sum_{m \in \inte_{N_1 + N_2}}
\vt
\left[
\begin{array}{c}
\frac{r + s + N_1m}{N_1+ N_2} \\ 0
\end{array}
\right]
(z_1+ z_2, \tau (N_1 + N_2))   \cr
&&\hskip 3cm\times 
\vt
\left[
\begin{array}{c}
\frac{N_2 r - N_1 s + N_1N_2 m }{N_1 N_2(N_1+ N_2)} \\ 0
\end{array}
\right]
(z_1N_2 - z_2 N_1, \tau N_1 N_2 (N_1 + N_2)), 
{}
\label{identity}
\eeqa
 where $\vt$ is the  Jacobi theta-function:
\beq
\vt \left[
\begin{array}{c}
a \\ b
\end{array}
\right] (\nu,\tau) =  \sum_{l \in \inte} 
e^{\pi i (a + l)^2 \tau} \ e^{2\pi i (a + l)(\nu + b)}.
\label{theta}
\eeq

To proceed with the proof of the above identity, 
we write its LHS explicitly as: 
\beqa 
\vt
\left[
\begin{array}{c}
\frac{r}{N_1} \\ 0
\end{array}
\right]
(z_1, \tau N_1)
\cdot
\vt
\left[
\begin{array}{c}
\frac{s}{N_2} \\ 0
\end{array}
\right]
(z_2, \tau N_2)  & = &
\sum_{l_1 \in \inte} \sum_{l_2 \in \inte} 
e^{\pi i (\frac{r}{N_1} + l_1)^2 \tau N_1} \ e^{2\pi i (\frac{r}{N_1} + l_1)z_1} \cr
& \cdot & 
e^{\pi i (\frac{s}{N_2} + l_2)^2 \tau N_2} \ e^{2\pi i (\frac{s}{N_2} + l_2)z_2}.
\label{lhs}
\eeqa

Similarly the RHS of the identity (\ref{identity} ) can be written as:
\beqa 
&&\hskip -1.7cm \sum_{m \in \inte_{N_1 + N_2}}
\vt
\left[
\begin{array}{c}
\frac{r + s + N_1m}{N_1+ N_2} \\ 0
\end{array}
\right]
(z_1+ z_2, \tau (N_1 + N_2))\cr 
&&\times \vt
\left[
\begin{array}{c}
\frac{N_2 r - N_1 s + N_1N_2 m }{N_1 N_2(N_1+ N_2)} \\ 0
\end{array}
\right]
(z_1N_2 - z_2 N_1, \tau N_1 N_2 (N_1 + N_2)) 
\cr
&&\hskip -1.5cm = \sum_{m \in \inte_{N_1 + N_2}}
\sum_{l_3 \in \inte} \sum_{l_4 \in \inte}
e^{\pi i (\frac {r + s + N_1m}{N_1+ N_2} + l_3)^2 \tau (N_1 + N_2)} \ 
e^{2\pi i (\frac {r + s + N_1m}{N_1+ N_2} + l_3)(z_1+ z_2)}\cr  
&&\times e^{\pi i (\frac {N_2 r - N_1 s + N_1N_2 m }
{N_1 N_2(N_1+ N_2)}+ l_4)^2 \tau N_1 N_2 (N_1 + N_2)} 
\ e^{2\pi i (\frac {N_2 r - N_1 s + N_1N_2 m }{N_1 N_2(N_1+ N_2)}+ l_4)
(z_1N_2 - z_2 N_1)}. 
\label{rhs}
\eeqa
Now, to match the $ z_1,z_2 $ terms in both sides of eq. (\ref{identity}), we 
first note the identity:
\beq
\left( \frac {r + s}{N_1+ N_2}\right)\left(z_1+ z_2\right) + 
\left(\frac {N_2 r - N_1 s}{N_1 N_2(N_1+ N_2)}\right)
\left(z_1N_2 - z_2 N_1\right) = 
\left(\frac{r}{N_1}z_1 + \frac{s}{N_2}z_2 \right),
\hspace{1.0in}
\label{co11}
\eeq
and find coefficients $ p_1, p_2, q_1, q_2 $ such that, 
\beq
\left( p_1l_1  + p_2l_2\right)\left(z_1+ z_2\right) + 
\left(q_1l_1 + q_2l_2\right)
\left(z_1N_2 - z_2 N_1\right) =  
\left(l_1z_1 + l_2 z_2\right).
\hspace{1.0in}
\label{co12}
\eeq
Eq. (\ref{co12}) leads to the following values for $ p_1, p_2, q_1, q_2 $ :
\beqa
p_1 = \frac {N_1} {N_1+ N_2},\,\,\,
p_2 = \frac {N_2} {N_1+ N_2},\,\,\,\cr
q_1 = \frac {1} {N_1+ N_2},\,\,\,
q_2 = \frac {-1} {N_1+ N_2}.
\label {val1}
\eeqa
Then  the two terms, containing $ z_1,z_2 $, in the 
RHS of eq. (\ref{lhs}) can be rewritten as:
\beq
 e^{2\pi i (\frac{r}{N_1} + l_1)z_1} \ e^{2\pi i (\frac{s}{N_2} + l_2)z_2} = 
e^{2\pi i (\frac {r + s}{N_1+ N_2} + \frac {N_1 l_1} {N_1+ N_2} + 
\frac {N_2 l_2} {N_1+ N_2})(z_1+ z_2)} 
\  e^{2\pi i (\frac {N_2 r - N_1 s}{N_1 N_2(N_1+ N_2)} + 
\frac {l_1 - l_2} {N_1+ N_2})(z_1N_2 - z_2 N_1)}.
\label{z-simplify}
\eeq
Similarly, coefficients $ p, q$ satisfying identity:
\beqa
p \left[\frac {r + s}{N_1+ N_2} + \frac {N_1 l_1} {N_1+ N_2} + 
\frac {N_2 l_2} {N_1+ N_2} \right]^2 +
q \left[ \frac {N_2 r - N_1 s}{N_1 N_2(N_1+ N_2)} + 
\frac {l_1 - l_2} {N_1+ N_2} \right ] ^2 =  \\ \nonumber
\left[\frac{r}{N_1} + l_1\right]^2  N_1 + \left[\frac{s}{N_2} + l_2\right]^2  N_2 , 
\label {co2}
\eeqa 
are given as:
\beqa
p = N_1 + N _2 ,\,\,\,
q = N_1 N_2(N_1+ N_2) .
\label {val2}
\eeqa
Using eqs. (\ref{co11}), (\ref{co12}), (\ref{z-simplify}) and (\ref{co2}), 
the RHS of eq. (\ref{lhs}) (appearing in the LHS of eq.
(\ref {identity}) ) can be re-written :
\beqa
\sum_{l_1 \in \inte} \sum_{l_2 \in \inte} 
e^{\pi i (\frac{r}{N_1} + l_1)^2 \tau N_1} \ e^{2\pi i (\frac{r}{N_1} + l_1)z_1} 
 \cdot 
e^{\pi i (\frac{s}{N_2} + l_2)^2 \tau N_2} \ e^{2\pi i (\frac{s}{N_2} + l_2)z_2} =  
\cr
\sum_{l_1 \in \inte} \sum_{l_2 \in \inte} 
e^{\pi i (\frac {r + s}{N_1+ N_2} + \frac {N_1 l_1} {N_1+ N_2} + 
\frac {N_2 l_2} {N_1+ N_2})^2 \tau (N_1 + N_2)} 
\ e^{2\pi i (\frac {r + s}{N_1+ N_2} + \frac {N_1 l_1} {N_1+ N_2} + 
\frac {N_2 l_2} {N_1+ N_2})(z_1+ z_2)} 
 \cdot  \cr
\ e^{\pi i (\frac {N_2 r - N_1 s}{N_1 N_2(N_1+ N_2)} + 
\frac {l_1 - l_2} {N_1+ N_2})^2 \tau N_1 N_2(N_1+ N_2)}
\  e^{2\pi i (\frac {N_2 r - N_1 s}{N_1 N_2(N_1+ N_2)} + 
\frac {l_1 - l_2} {N_1+ N_2})(z_1N_2 - z_2 N_1)}.
\label {relhs}
\eeqa

Proving the identity, eq. (\ref{identity}), now amounts to showing that the RHS
of eq. (\ref{rhs}) matches precisely with that of eq. (\ref{relhs}) with
$m$ in eq. (\ref{rhs}) taking value as
$ m = 0,1,.......,(N_1 + N_2 -1) $. We note:
\begin {enumerate}
\item When  $ l_1 = l_2 $ in eq. (\ref {relhs}), the terms in the RHS are
identical to those  in the RHS of 
eq. (\ref {rhs}), with $ m=0, l_4 = 0$, 
if we identify $l_2$ with $l_3$. \\
When $l_1 = l_2 + 1 $, the terms in eq. (\ref {relhs}) exactly match with 
those in eq. (\ref {rhs}) obtained for the values $ m = 1, l_4 = 0$ with the 
identification of $ l_2 $ with $l_3 $. \\
This goes on up to $l_1 = l_2 + (N_1 + N_2 - 1) $  which corresponds to the case 
for $l_3 ( = l_2 ), m = (N_1 + N_2 - 1)$ and  $ l_4 = 0 $. 
\item  The terms obtained in eq. (\ref {relhs}) for $ l_1 = l_2 + ( N_1 + N_2 ) $  
corresponds to  
$ m=0, l_4 = 1 $ and $l_2 + N_1 $ identified with  $l_3 $ in eq. (\ref {rhs}). \\
When $ l_1 = l_2 +( N_1 + N_2 ) + 1 $ the terms correspond to the case  
$ m=1 , l_4 = 1 $ and $l_2 + N_1 $ identified with  $l_3 $ in eq. (\ref {rhs}).  \\
This goes on till $ l_1 = l_2 +( N_1 + N_2 ) + (N_1 + N_2 - 1) $ when they correspond
to $ m = (N_1 + N_2 -1), l_4 = 1 $ and $l_2 + N_1 $ identified with  
$l_3 $ in eq. (\ref {rhs}). 

\item Similarly  the terms for $ l_1 = l_2 + 2 ( N_1 + N_2 ) $ correspond 
to  the terms for $ m = 0, l_4 =2 $ and $ l_3 = (l_2 + 2 N_1) $ .
And so on....
\end {enumerate}

We have therefore shown a one-to-one correspondence 
between the terms in the RHS of eqs. (\ref{rhs}) and (\ref{relhs}).
The identity eq. (\ref{identity} ) has thus been proved explicitly.

\subsection{Application to Yukawa computation for factorized tori}

We now make use of the above Jacobi theta identity as well as of the 
explicit forms of the fermion and scalar wavefunctions,
defined in terms of the basis functions in eq. (\ref{basis1}) to write the 
expression for the Yukawa interaction term. More precisely, in order to evaluate the
Yukawa coupling  given in eq. (\ref{yukawa-t2-abc}), one uses
the theta identity of eq. (\ref{identity}) and the basis function in 
eq. (\ref{basis1}) and proceeds by  writing down:
\beqa
\psi^{i, I_{ab}}(\tau, z).\psi^{j, I_{ca}}(\tau, z) = 
\left(\frac{2 Im\tau }{{\cal{A}}^2}   \right)^{\frac 12} 
(I_{ab} I_{ca})^{\frac 14}
e^{i\pi (N_1 + N_2) z \pim z/\pim \tau} \times \hspace{1.5in}\cr
\times
\vt
\left[
\begin{array}{c}
\frac{i}{N_1} \\ 0
\end{array}
\right]
( N_1 z, N_1 \tau) \cdot
\vt
\left[
\begin{array}{c}
\frac{j}{N_2} \\ 0
\end{array}
\right]
( N_2 z, N_2 \tau), 
\,\,i=0,\dots, N_1-1,\,\, j=0,\dots, N_2-1. \cr
\label{yukawa-basis-expansion}
\eeqa
where we have also made use of the normalization factor, $\cal{N}$ 
given in eq. (\ref{normalization-constant}), and  identified 
for a $T^2$ compactification:
\beq
	N_1 = I_{ab},\,\,\,\,\,N_2 = I_{ca},
\label{N1N2}
\eeq
with 
\beq
	I_{ab} = m_a - m_b,\,\,\,\, etc.
\label{Iab}
\eeq
giving
\beq
	N_3 = (N_1 + N_2) = I_{cb} .
\label{N_3}
\eeq

Now, using the theta identity (\ref{identity}), 
eq. (\ref{yukawa-basis-expansion}) can be rewritten in the form:
\beqa
\psi^{i, I_{ab}}(\tau, z).\psi^{j, I_{ca}}(\tau, z) = 
\left(\frac{2 Im\tau }{{\cal{A}}^2}   \right)^{\frac 14} 
\left(\frac{I_{ab} I_{ca}}{I_{cb}}\right)^{\frac 14}
\sum_{m \in \inte_{I_{cb}}}
\psi^{i+j+ I_{ab} m, I_{cb}}(\tau, z) \times \cr
\times 
\vt
\left[
\begin{array}{c}
\frac{I_{ca} i - I_{ab} j + I_{ab} I_{ca} m }{I_{ab} I_{ca} I_{cb}} \\ 0
\end{array}
\right] 
(0, \tau I_{ab} I_{ca} I_{cb}). 
\label{yukawa-basis-expansion-2}
\eeqa
The Yukawa interaction (\ref{yukawa-t2-abc}), is then evaluated using the
orthogonality property of the wavefunctions given in eq. (\ref{normalizable})
and reads:
\beq
	Y_{ijk} = \sigma_{abc} g
\left(\frac{2 Im\tau }{{\cal{A}}^2}   \right)^{\frac 14} 
\left(\frac{I_{ab} I_{ca}}{I_{cb}}\right)^{\frac 14}
\sum_{m \in \inte_{I_{cb}}} \delta_{k, i + j + I_{ab} m} \cdot
\vt
\left[
\begin{array}{c}
\frac{I_{ca} i - I_{ab} j + I_{ab} I_{ca} m }{I_{ab} I_{ca} I_{cb}} \\ 0
\end{array}
\right] 
(0, \tau I_{ab} I_{ca} I_{cb}). 
\label{yukawa-series}
\eeq
After imposing the Kronecker delta constraint, we obtain:
\beq
	Y_{ijk} = \sigma_{abc} g
\left(\frac{2 Im\tau }{{\cal{A}}^2}   \right)^{\frac 14} 
\left(\frac{I_{ab} I_{ca}}{I_{cb}}\right)^{\frac 14}
\vt
\left[
\begin{array}{c}
- \left( \frac{j}{I_{ca}} + \frac{k}{I_{bc}} \right) / {I_{ab}} \\ 0
\end{array}
\right] 
(0, \tau I_{ab} I_{ca} I_{cb}). 
\label{yukawa-simple}
\eeq
The final answer can be expressed as :
\beq
	Y_{ijk} = \sigma_{abc} g
\left(\frac{2 Im\tau }{{\cal{A}}^2}   \right)^{\frac 14} 
\left(\frac{I_{ab} I_{ca}}{I_{cb}}\right)^{\frac 14}
\vt
\left[
\begin{array}{c}
\delta_{ijk} \\ 0
\end{array}
\right] 
(0, \tau |I_{ab} I_{bc} I_{ca}|),
\label{yukawa-answer-t2}
\eeq
with 
\beq
   \delta_{ijk} = \frac{i}{I_{ab}} + \frac{j}{I_{ca}} + \frac{k}{I_{bc}} . 
\label{delta}
\eeq
The result can be easily extended to the case of factorized $T^6$  (\ref{t2t6}) and the interaction is then written in 
 terms of the products of theta functions of the type appearing in 
eq. (\ref{yukawa-answer-t2}). We refer the reader to \cite{ibanez}
for the details and now go on to the generalization when fluxes 
of both oblique and diagonal forms are present. Such magnetic fluxes
do not respect the factorization and hence involve the wavefunctions 
written in terms of the general Riemann theta functions. 

\section{General tori and `oblique' fluxes}\label{general-tori}

Let us now consider the more general case where the $2n$-dimensional torus is not 
necessarily factorizable.
A generic flat $2n$-dimensional torus, $T^{2n} \simeq \cpx^n/\Lambda$, inherits 
a complex structure from the covering space $\cpx^n$. Its geometry can hence be 
described in terms of a K\"ahler metric and complex structure as
\beq \begin{array}{rcl} \vspace*{.1cm}
ds^2 & = & h_{\mu\bar{\nu}} dz^{\mu} d\bar{z}^{\bar{\nu}} \\
dz^\mu & = & dx^\mu + \tau^{\mu}_{\nu} dy^\nu
\end{array}
\label{complexn}
\eeq
where $x^{\mu}, y^{\mu} \in (0,1)$, $\mu=1,\dots,n$,  parametrize the $2n$ vectors 
of the lattice $\Lambda$. The natural generalization of the Jacobi theta 
function (\ref{theta}) to this higher-dimensional tori is known as Riemann 
$\vt$-functions, as defined in eq. (\ref{general-theta}): 
\beq
\vt
\left[\begin{array}{c}
{\vec{a}} \\ {\vec{b}}
\end{array}\right]
(\vec{\nu}|{{\Omega}}) = \sum_{\vec{l} \in {\bf{Z}}^n}
	e^{i\pi (\vec{l} + \vec{a}) . {{\Omega}} . (\vec{l} + \vec{a})}
	e^{2 \pi i (\vec{l} + \vec{a}) . (\vec{\nu} + \vec{b})}	.
\label{general-theta-text}
\eeq

As already elaborated upon earlier, in our case, although the geometry 
itself may be such that $T^6$ is factorizable as in eq. (\ref{t2t6}),
 the fluxes turned on, may violate in general  the factorizable 
structure of the tori. Indeed, the general wavefunctions for bifundamentals
given in terms of basis functions (\ref{general-basis}){\footnote{See
Appendix A for more details on the properties of
the wavefunctions and Section \ref{fluxes}, as well as
Appendix  \ref{Appendix-A} for discussion on fluxes.}}:
\beqa \nonumber
  \psi^{\vec{j}, {\bf{N}}} (\vec{z}, {{\Omega}}) & = &
\cn \cdot e^{\{i\pi [{\bf{N}}.\vec{z}]. ({\bf{N}}.Im {{\Omega}})^{-1} 
Im [{\bf{N}}.\vec{z}]\}} 
\cdot 
\vt
\left[
\begin{array}{c}
{\vec{j}} \\ 0
\end{array}
\right]
({\bf{N}}.\vec{z} | {\bf{N}}. {{\Omega}}), \\
& = & 
\cn \cdot e^{i\pi [{\bf{N}} \cdot \vec{z}] \cdot (\pim {{\Omega}})^{-1}  \cdot \pim \vec{z} }
\cdot 
\vt
\left[
\begin{array}{c}
\vec{j} \\ 0
\end{array}
\right]
\left({\bf{N}} \cdot \vec{z}\ | {\bf{N}} \cdot {{\Omega}} \right),
\label{general-basis1}	
\eeqa
with ${\bf{N}}$'s being the intersection matrices,
 depend on such fluxes explicitly in terms of its modular parameter 
argument: 
${\bf{N}} \Omega$; this breaks in general  the factorized structure, even if the complex structure $\Omega$ is diagonal.
The explicit form of the normalization factor ${\cal N}$ appearing in 
eq. (\ref{general-basis1}) is given eq. (\ref{general-normalization-constant}).
One  needs to obtain an overlap integral of three basis
functions of the type  (\ref{general-basis1}), in order to generalize the results
of Yukawa computations given in eqs. (\ref{yukawa-t2-abc}), 
(\ref{yukawa-basis-expansion-2}) - (\ref{yukawa-answer-t2}).

\subsection{ Riemann theta function identity} \label{Riemann-Theta-Function-Identity}

We now generalize eq. (\ref{identity}) to the case of general Riemann
theta functions given in eq. (\ref{general-theta-text}). Explicitly, 
we consider the LHS of our identity to be given by an expression:
\beq
\vt \left[
\begin{array}{c}
\vec{j_1} \\  0
\end{array}
\right] (\vec{z_1} |  {\bf{N_1}} \cdot \Omega)  \cdot 
\vt \left[
\begin{array}{c}
\vec{j_2} \\  0
\end{array}
\right] (\vec{z_2} |  {\bf{N_2}} \cdot \Omega) 
\label{riemann1} 
\eeq
where $\Omega$ is an $n \times n$ complex matrix and $ {\bf{N_1}} $,
$ {\bf{N_2}} $  are $n \times n $ integer-valued symmetric matrices 
satisfying the constraints (\ref{eq-riemann-conditions}).  These 
constraints, in turn, follow from the convergence of 
theta series expansion, as well as from 
the holomorphicity of fluxes: for instance, eq. (\ref{(0,2)=0-2}) when
$p_{xx}$ and $p_{yy}$ components of
fluxes are zero, with $x^i, y^i$, 
($i=1, 2, 3$) denoting the coordinates of three $T^2$'s in the decomposition
(\ref{t2t6}) and (\ref{complexn}).
Generalization to the case when
$p_{xx}$ and $p_{yy}$ flux components
are also present is discussed in Section \ref{fluxes}, as well as later on in  
subsection \ref{hermitian}, and is 
relevant for evaluating the Yukawa couplings in  models with moduli stabilization, such as the 
one of \cite{akp}.

Initially, we also restrict ourselves to the case when
$ \Omega = \tau I_n $ with $I_n$ being a 
$n\times n$ identity matrix, implying that the geometric structure is
factorized as in eq. (\ref{t2t6}). However, in 
Section \ref{general-complex-structure}, we generalize the results
further to the case when $\Omega$ is an arbitrary matrix satisfying
the $F_{(2, 0)}=0$ supersymmetry condition, as given in eqs. (\ref{(2,0)=0}) 
and (\ref{(0,2)=0}).
Then, using the definition of Riemann $\vt $-functions 
 (\ref{general-theta-text}),
the expression in eq. (\ref{riemann1}) can be expanded as:
\beqa
\vt \left[
\begin{array}{c}
\vec{j_1} \\  0
\end{array}
\right] (\vec{z_1} |  {\bf{N_1}}\tau )  \cdot 
\vt \left[
\begin{array}{c}
\vec{j_2} \\  0
\end{array}
\right] (\vec{z_2} | {\bf {N_2}}\tau  )   = 
\sum_{\vec{l_1}, \vec{l_2}\in \inte^n} e^{\pi i (\vec{j_1} + \vec{l_1}) 
\cdot {\bf{N_1}} \tau \cdot (\vec{j_1} + \vec{l_1})}
e^{2\pi i (\vec{j_1} + \vec{l_1}) \cdot \vec{z_1} } \cdot \cr
e^{\pi i (\vec{j_2} + \vec{l_2}) \cdot {\bf{N_2}} \tau \cdot (\vec{j_2} + \vec{l_2})}
e^{2\pi i (\vec{j_2} + \vec{l_2}) \cdot \vec{z_2} }.
\label{ri-identity}
\eeqa
Now, by defining $2n$-dimensional vectors:
\beqa
(\vec{\textbf{j}} + \vec{\textbf{l}}) = \left( \begin{array}{c}
\vec{j_1} + \vec{l_1}  \\ \vec{j_2} + \vec{l_2} 
\end{array} \right),\,\,\,
\vec{\textbf{z}} = \left( \begin{array}{c} \vec{z_1} \\ \vec{z_2} \end{array} \right),\,\,\,
\label{vectors}
\eeqa
and the $2n\times 2n$ dimensional matrix:
\beqa
\textbf{Q} = \left(
\begin{array}{cc} {\bf{N_1}}\tau & 0 \\ 0 & {\bf{N_2}}\tau \end{array} \right),
\label{matrix}
\eeqa
eq. (\ref{ri-identity}) can be re-written as:
\beq
\vt \left[
\begin{array}{c}
\vec{j_1} \\  0
\end{array}
\right] (\vec{z_1} |  {\bf{N_1}} \tau)  \cdot 
\vt \left[
\begin{array}{c}
\vec{j_2} \\  0
\end{array}
\right] (\vec{z_2} | {\bf {N_2}} \tau ) = 
\sum_{\vec{l}\in {Z}^{2n}}e^{\pi i(\vec{\textbf{j}} + \vec{\textbf{l}})^{T} \cdot \textbf{Q} 
\cdot (\vec{\textbf{j}}+\vec{\textbf{l}})} 
 e^{2\pi i (\vec{\textbf{j}} + \vec{\textbf{l}})^{T} \cdot \vec{\textbf{z}}}.
\label{riemann3}
\eeq

Our aim in combining the terms into $2n$ dimensional vectors and matrices is to 
generalize the procedure outlined in \cite{mumford} to our situation, namely when
two theta functions appearing in the LHS of the identity (that we propose below)
carry independent modular parameter matrices ${\bf{N_1}}\tau$ and 
${\bf{N_2}}\tau$, which generally may not commute. Note that the results of 
\cite{mumford} are insufficient
to give such an identity as they involve theta functions whose modular
parameter matrices are proportional to each other and therefore commute. 
In order to proceed, we note that using a transformation matrix:
\beq
\label{transmatrix}
T = 
\left(
\begin{array}{cc}
1 & 1 \\
\alpha {\bf{N_1}} ^ {-1} & -\alpha {\bf{N_2}}^{-1}
\end{array}
\right),
\eeq
\beq
\label{transmatrixT}
T^ {T} = 
\left(
\begin{array}{cc}
1 &  {\bf{N_1}} ^ {-1} \alpha ^ {T}\\
1  & - {\bf{N_2}}^{-1}\alpha ^ {T}
\end{array}
\right),
\eeq 
and
\beq
\label{transmatrixI}
T^ {-1} = ({\bf{N_1}} ^ {-1} + {\bf{N_2}} ^ {-1} ) ^ {-1}
\left(
\begin{array}{cc}
{\bf{N_2}} ^ {-1} &   \alpha ^ {-1}\\
{\bf{N_1}}^{-1} &  -\alpha ^ {-1}
\end{array} 
\right),
\eeq 
with $\alpha $ being an arbitrary matrix (to be determined below) and 
$ {\bf {N_1}}, {\bf {N_2}}$ being real symmetric matrices,
due to the condition (\ref{eq-riemann-conditions}) (for $\Omega = \tau I_n$),
one obtains:
\beqa
{\textbf{Q}}^{\prime} \equiv T  \cdot \textbf{Q} \cdot T ^ {T} =  \left(
\begin{array}{cr}
 ({\bf{N_1}} + {\bf{N_2}})\tau  & 0  \\
0  &  \alpha {({\bf{N_1}}^{-1} + {\bf{N_2}}^{-1})\tau  } \alpha ^ {T}
\end{array}
\right).
\label{trans4}
\eeqa
In the following we  also make use of the identities:
\beq
	({\bf{N_1}}^{-1} + {\bf{N_2}}^{-1}) = 
	{\bf{N_1}}^{-1} ({\bf{N_1}} + {\bf{N_2}}){\bf{N_2}}^{-1}
	= {\bf{N_2}}^{-1} ({\bf{N_1}} + {\bf{N_2}}){\bf{N_1}}^{-1} 
\label{N1N2-identity1}
\eeq
and 
\beq
	({\bf{N_1}}^{-1} + {\bf{N_2}}^{-1})^{-1} = 
	{\bf{N_1}} ({\bf{N_1}} + {\bf{N_2}})^{-1}{\bf{N_2}}
	= {\bf{N_2}} ({\bf{N_1}} + {\bf{N_2}})^{-1}{\bf{N_1}} 
\label{N1N2-identity2}
\eeq
in simplifying certain expressions. 

The transformation matrix $T$ defined above is used to transform the product of 
theta functions in the LHS of eq. (\ref{riemann3}), in terms of a finite
sum over another product of theta's, now with modular parameter 
matrices:
$({\bf{N_1}} + {\bf{N_2}})\tau$ and 
$\alpha {({\bf{N_1}}^{-1} + {\bf{N_2}}^{-1})\tau  } \alpha^{T}$.
Explicitly, we can write the terms appearing in the exponents
in the RHS of eq. (\ref{riemann3}) as:
\beq
(\vec{\textbf{j}} + \vec{\textbf{l}})^{T} \cdot \textbf{Q} 
\cdot (\vec{\textbf{j}} +\vec{\textbf{l}})
=(\vec{\textbf{j}} + \vec{\textbf{l}})^{T} \cdot  (T ^{-1} T ) \cdot \textbf{Q} \cdot
(T ^ {T} (T ^{-1})^{T} ) 
\cdot (\vec{\textbf{j}}+ \vec{\textbf{l}}) 
\label{riemann4}
\eeq
\beq
 (\vec{\textbf{j}} + \vec{\textbf{l}})^{T} 
\cdot \vec{\textbf{z}}=
    (\vec{\textbf{j}} + 
\vec{\textbf{l}})^{T} (T ^ {-1} T ) \cdot \vec{\textbf{z}}.
\label{riemann5}
\eeq
Then using:
\beqa
T  \cdot \vec{\textbf{z}} = \left( \begin{array}{c} \vec{z_1} + \vec{z_2} \\
\alpha {\bf{N_1}}^ {-1} \vec{z_1} - \alpha {\bf{N_2}}^{-1} \vec{z_2}
\end{array} \right),
\label{trans1}
\eeqa
\beqa
(\vec{\textbf{j}} + \vec{\textbf{l}})^{T} T^{-1} = 
\left( \begin{array}{c}
(\vec{j_1} + \vec{l_1})({\bf{N_1}}^{-1} + {\bf{N}}_2^{-1}) ^ {-1} {\bf{N}}_2^{-1} + 
(\vec{j_2} + \vec{l_2})({\bf{N}}_1^{-1} + {\bf{N}}_2^{-1}) ^ {-1} {\bf{N}}_1^{-1} \\ 
((\vec{j_1} + \vec{l_1}) - (\vec{j_2} + \vec{l_2})) 
({\bf{N}}_1^{-1} + {\bf{N}}_2^{-1} ) ^ {-1} \alpha ^ {-1}
\end{array} \right)^{T},
\label{trans2}
\eeqa
and
\beqa
(T^{-1})^{T} (\vec{\textbf{j}} + \vec{\textbf{l}}) =\left( \begin{array}{c}
{\bf{N}}_2^{-1}({\bf{N}}_1^{-1} + {\bf{N}}_2^{-1} )^{-1}(\vec{j_1} + \vec{l_1}) + 
{\bf{N}}_1^{-1}({\bf{N}}_1 ^ {-1} + {\bf{N}}_2 ^ {-1} ) ^ {-1}(\vec{j_2} + \vec{l_2}) \\
(\alpha ^ {-1})^{T}({\bf{N}}_1 ^ {-1} + {\bf{N}}_2 ^ {-1} ) ^ {-1}[(\vec{j_1} + 
\vec{l_1}) - (\vec{j_2} + \vec{l_2})]
\end{array} \right)
\label{trans3}
\eeqa
we can re-write eq. (\ref{ri-identity}) as,
\beqa \nonumber
&&\vt \left[
\begin{array}{c}
\vec{j_1} \\  0
\end{array}
\right] (\vec{z_1} |  N_1 \tau)  \cdot 
\vt \left[
\begin{array}{c}
\vec{j_2} \\  0
\end{array}
\right] (\vec{z_2} |  N_2 \tau)    =    \\ \nonumber 
&& \hspace{-0.5in}
\sum_{\vec{l_1}, \vec{l_2}\in \inte^n} 
e^ {\pi i \left[ \{((\vec{j_1} +\vec{l_1} ) {\bf{N_1}} +
 (\vec{j_2} + \vec{l_2}) {\bf{N_2}}) 
({\bf{N}}_1 + {\bf{N}}_2 ) ^ {-1} \}\cdot  
( {\bf{N}}_1 + {\bf{N}}_2)\tau \cdot
\{({\bf{N}}_1 + {\bf{N}}_2 ) ^ {-1} ({\bf{N}}_1(\vec{j_1} +\vec{l_1}) + 
{\bf{N}}_2(\vec{j_2} + \vec{l_2})) \}\right] }  \\ \nonumber
&&
\times
e^ {2\pi i \{ [((\vec{j_1} +\vec{l_1}){\bf{N}}_1 + (\vec{j_2} +\vec{l_2}){\bf{N}}_2)
  (\bf{N}_1 + \bf{N}_2 )^{-1}] \cdot [\vec{z_1} + \vec{z_2}]\}} \times \\ \nonumber
&& \hspace{-0.5in}
e^ {\pi i \{ [((\vec{j_1} - \vec{j_2}) + (\vec{l_1} - \vec{l_2})) 
{\bf{N}}_1 ({\bf{N}}_1 + {\bf{N}}_2 ) ^ {-1}{\bf{N}}_2  \alpha ^ {-1}] \cdot
[\alpha ({{\bf{N}}_1}^{-1}{({\bf{N}}_1 + {\bf{N}}_2){{\bf{N}}_2}^{-1} })
\tau \alpha ^ {T}]
\cdot
[(\alpha ^ {-1})^{T} {\bf{N}}_2 ({\bf{N}}_1 + {\bf{N}}_2 ) ^ {-1}{\bf{N}}_1
((\vec{j_1} - \vec{j_2}) + (\vec{l_1} - \vec{l_2}))] \}} \\ 
&&
\times
e^ {2\pi i \{[((\vec{j_1} - \vec{j_2}) + (\vec{l_1} - \vec{l_2}))   
{\bf{N}}_1 ({\bf{N}}_1 + {\bf{N}}_2 ) ^ {-1}{\bf{N}}_2
\alpha ^ {-1}] \cdot
[\alpha N_1 ^ {-1} \vec{z_1} - \alpha N_2^{-1} \vec{z_2}]\}}. 
\label{transflhs}
\eeqa

Now, to reexpress the above series expansion in terms of a sum over theta functions
with modular parameter matrices: ${\bf{N_1}} + {\bf{N_2}}$
and $\alpha {({\bf{N_1}}^{-1} + {\bf{N_2}}^{-1})  } \alpha ^ {T}$, 
we rearrange the series in eq. (\ref{transflhs})
in terms of new summation variables 
$ \vec{l_3}, \vec{l_4} ,\vec{m} $, whose values and ranges will be assigned later.
In the course of going from eq. (\ref{transflhs}) to (\ref{exprhs}) below, 
however,
one needs to make sure that such redefined variables are  integers.
This requirement
constrains the matrix $\alpha$ whose `minimal' solution will be taken to 
be 
\beq
	\alpha = (\det {\bf{N}_1} \det {\bf{N}_2}) I.
\label{def-alpha}
\eeq
We will later on discuss also the possibility of choosing other forms of 
$\alpha$ and show that such choices lead to the cyclicity of the superpotential
coefficients, as in eqs. (\ref{yukawa-answer-t2}), (\ref{delta}).
Using eq. (\ref{def-alpha}), the RHS of eq. (\ref{transflhs}) takes the form:
\beqa \nonumber
&& \hskip -0.8cm \sum_{\vec{l_3}, \vec{l_4}\in \inte^n}\sum_{\vec{m}} 
e^ {\pi i [ (\vec{j_1} {\bf{N}}_1 + \vec{j_2} {\bf{N}}_2   
+ {\vec{m}}{\bf{N}}_1)
({\bf{N}}_1 + {\bf{N}}_2 ) ^ {-1} + \vec{l_3}] 
\cdot ( {\bf{N}}_1 + {\bf{N}}_2) \tau \cdot
[ ({\bf{N}}_1  + {\bf{N}}_2  ) ^ {-1} ( {\bf{N}}_1\vec{j_1} +  {\bf{N}}_2 \vec{j_2}  
+{\bf{N}}_1 {\vec{m}})
 + \vec{l_3}]} \\ \nonumber
&& \hskip -0.8cm \cdot e^ {2\pi i [ (\vec{j_1} {\bf{N}}_1 + \vec{j_2} {\bf{N}}_2   
+ {\vec{m}}{\bf{N}}_1)
({\bf{N}}_1 + {\bf{N}}_2 ) ^ {-1} + \vec{l_3}]\cdot 
[\vec{z_1} + \vec{z_2}]} \times  \\ \nonumber
&& \hskip -0.8cm e^ {\pi i [(\vec{j_1} - \vec{j_2} + \vec{m}) 
\frac{{\bf{N}}_1 ({\bf{N}}_1 + {\bf{N}}_2 ) ^ {-1}{\bf{N}}_2} 
{\det {\bf{N}_1} \det {\bf{N}_2} } + \vec{l_4}]\cdot
[(\det {\bf{N}_1} \det {\bf{N}_2})^2  {\bf{N}}_1^{-1}({\bf{N}}_1 + 
{\bf{N}}_2) {\bf{N}}_2^{-1}]\tau \cdot
[\frac{{\bf{N}}_2 ({\bf{N}}_1 + {\bf{N}}_2 ) ^ {-1}{\bf{N}}_1} 
{\det {\bf{N}_1} \det {\bf{N}_2} } 
(\vec{j_1} - \vec{j_2} + \vec{m}) + \vec{l_4}]} \\ 
&& \hskip -0.8cm \cdot  e^ {2\pi i [(\vec{j_1} - \vec{j_2} + \vec{m}) 
\frac{{\bf{N}}_1 ({\bf{N}}_1 + {\bf{N}}_2 ) ^ {-1}{\bf{N}}_2} 
{\det {\bf{N}_1} \det {\bf{N}_2} } + \vec{l_4}]\cdot
\det {\bf{N}_1} \det {\bf{N}_2}
[{\bf{N}}_1 ^ {-1} \vec{z_1} - {\bf{N}}_2^{-1} \vec{z_2}]}\, .
\label{exprhs}
\eeqa
This series can now be reexpressed in terms of a finite sum over product of 
generalized theta functions given in eq. (\ref{general-theta-text}), leading to 
 a generalization of the identity (\ref{identity}) to:

\beqa \nonumber
& &
\hspace{-.35in}\vt
\left[
\begin{array}{c}
\vec{j_1} \\ 0
\end{array}
\right]
(\vec{z_1}| {\bf{N}_1}\tau)
\cdot
\vt
\left[
\begin{array}{c}
\vec{j_2} \\ 0
\end{array}
\right]
(\vec{z_2}|  {\bf{N}_2}\tau)   =  \cr
& & \hspace{-0.2in}
\sum_{\vec{m}}
\vt
\left[
\begin{array}{c}
(\vec{j_1}{\bf{N}_1} +\vec{j_2}{\bf{N}_2} +\vec{ m}.{\bf{N}_1} )({\bf{N}_1} 
+ {\bf{N}_2})^{-1} \\ 0
\end{array}
\right]
(\vec{z_1}+ \vec{z_2}| ({\bf{N}_1} + {\bf{N}_2})\tau)\times   \cr
& &  
\hspace{-0.2in}\vt
\left[
\begin{array}{c}
[(\vec{j_1} - \vec{j_2})+ \vec{m}]\frac{{\bf{N}_1}({\bf{N}_1} + {\bf{N}_2})^{-1} {\bf{N}_2}}
{\det{\bf{N}_1} \det{\bf{N}_2}}\\ 0
\end{array}
\right]   \cr
&  & \hspace{-0.7in}
 \hspace{1in}((\det{\bf{N}_1} \det{\bf{N}_2}) ({\bf{N}_1}^{-1}
\vec{z_1} - {\bf{N}_2}^{-1}\vec{z_2}) |
(\det{\bf{N}_1} \det{\bf{N}_2}) ^2({\bf{N}_1} ^{-1} 
({\bf{N}_1} + {\bf{N}_2}){\bf{N}_2} ^{-1})\tau), \\
\label{general-identity}
\eeqa
where $\vec{m} = \sum_{i} m_i \vec{e_i} $  are all vectors generated by the basis 
vectors $\vec{e_i}$: 
\beqa
  \begin{pmatrix} 1 \cr 0 \cr . \cr .\cr 0 \end{pmatrix}, \,\,\,
	\begin{pmatrix} 0 \cr 1 \cr . \cr . \cr 0 \end{pmatrix} \,\,\, {\mathrm etc.},
\label{basis-e}
\eeqa
and lied within the unit-cell defined by the new basis vectors:
\beq
\vec{e'} = \vec{e}(\det{\bf{N}_1} \det{\bf{N}_2})({\bf{N}_1} ^{-1} 
({\bf{N}_1} + {\bf{N}_2}){\bf{N}_2} ^{-1}) . 
\label{unit-cell}
\eeq  

The above identity already assumes the form $\Omega = \tau I_n$ for the 
complex structure 
of $T^{2n}$. As mentioned already,  in subsection  {\ref{general-complex-structure}} below, we
make further generalization to include arbitrary complex structure $\Omega$ as well.
Also, note that, due to the identities (\ref{N1N2-identity1}) and 
(\ref{N1N2-identity2}), the theta functions appearing in the RHS of eq. 
(\ref{general-identity}) satisfy the constraint (\ref{j-integer}) with 
respect to their own arguments.

\subsection{Proof of the identity} \label{proof}

We now show the equality of the series expansions (\ref{transflhs}) 
and (\ref{exprhs}) to establish the identity eq. (\ref{general-identity}).
We also show that matrix $\alpha$ needs to be chosen as in eq. (\ref{def-alpha})
for showing the equality of eqs. (\ref{transflhs}) 
and (\ref{exprhs}) for the case when $\det {\bf{N}_1}$ and $\det {\bf{N}_2}$
are relatively prime. In other cases $\alpha$ can be chosen as the least common 
multiple of $\det {\bf{N}_1}$ and $\det {\bf{N}_2}$. Here we assume them to 
be relatively prime, while the remaining cases can  be worked out in a 
similar fashion.

We now follow an exercise similar to the one in Section \ref{jacobi-theta},
to show that 
series in eqs. 
(\ref{transflhs}) and (\ref{exprhs}) precisely match with $\vec{m}$ restricted
to be an integer, provided $\alpha$ is given by eq. (\ref{def-alpha}). 
\begin{enumerate}
 \item  When $ \vec{l_1} =\vec{l_2} $ in eq. (\ref{transflhs}), we have:
\beq
(\vec{l_1}{\bf{N}_1} + \vec{l_2} {\bf{N}_2})  ({\bf{N}_1} + {\bf{N}_2} ) ^ {-1} 
=\vec{l_2}
\eeq
and
\beq
(\vec{l_1} - \vec{l_2}) {\bf{N}}_1 ({\bf{N}}_1 + {\bf{N}}_2 ) ^ {-1}{\bf{N}}_2
\alpha ^ {-1} =0
\eeq
These terms are exactly same if we consider the series given in eq. (\ref{exprhs})
for the values $ \vec{l_3} ( \equiv \vec{l_2} ), \vec{l_4} = 0$ and $\vec{m}= 0 $, 
irrespective of the choice for the matrix $\alpha$. 

\item In order to see the restriction on the matrix $\alpha$, one needs to understand how 
the nonzero integers $\vec{l_4} \neq 0$ in eq. (\ref{exprhs}) 
are generated from the terms in eq. (\ref{transflhs}). In other words, 
one needs to make sure that 
\beq
(\vec{l_1} - \vec{l_2}) {\bf{N}}_1 ({\bf{N}}_1 + {\bf{N}}_2 ) ^ {-1}{\bf{N}}_2
\alpha ^ {-1} \equiv \vec{l_4}
\label{restriction-alpha0}
\eeq
 is an integer. This in turn is possible only if $\vec{l_1}$ is
of the form:
\beq
	\vec{l_1} =  \vec{l_2} + \vec{l_4} \alpha 
{\bf{N}}_2^{-1} ({\bf{N}}_1 + {\bf{N}}_2 ){\bf{N}}_1^{-1}.
\label{restriction-alpha1}
\eeq
However, since $\vec{l_4}$, ${\bf{N}}_1$, ${\bf{N}}_2$, take integral values, the 
RHS in eq. (\ref{restriction-alpha1}) is an integer only if 
$ \alpha ({\bf{N}}_1^{-1} + {\bf{N}}_2^{-1} )$
is an integer. In other words, for $\det {\bf{N}_1}$ and $\det {\bf{N}_2}$
relatively prime, $\alpha$ needs to be of the form:
\beq
	\alpha = (\det {\bf{N}_1} \det {\bf{N}_2}) P.
\label{def-alpha2}
\eeq
with $P$ being an arbitrary invertible integer matrix. `Minimal' choice also demands
$\det P = 1$, otherwise $\vec{l_4}$ will not span over {\it all} integers.  
Then, since  $P$ is invertible, it is fixed to be the identity matrix.
We have therefore established the restriction on $\alpha$ as in 
eq. (\ref{def-alpha}). At the same time, we have also proved that the series 
in eqs. (\ref{transflhs}) and (\ref{exprhs}) precisely match for $\vec{m} = 0$
provided $\vec{l_2} + \det {\bf{N}_1} \det {\bf{N}_2} \vec{l_4}{\bf{N}}_2^{-1}$
is identified with $\vec{l_3}$ in eq.(\ref{exprhs}). Note  that
$(\det {\bf{N}_1} \det {\bf{N}_2}) {\bf{N}}_2^{-1}$ is also integer valued and ensures that
such an identification with $\vec{l_3}$ holds. 

\item On the other hand, 
When $ \vec{l_1} =\vec{l_2} + \vec{m} $ in eq. (\ref{transflhs}), 
we end up with terms like:
\beq
(\vec{l_1}  {\bf{N_1}} +  \vec{l_2} {\bf{N_2}}) 
({\bf{N}}_1 + {\bf{N}}_2 ) ^ {-1}  = 
\vec{l_2} +   \vec{ m}.{\bf{N}_1} ({\bf{N}_1} 
+ {\bf{N}_2})^{-1}
\label{l1l2m}
\eeq
and
\beq
(\vec{l_1} - \vec{l_2})\frac{{\bf{N}_1}({\bf{N}_1} + {\bf{N}_2})^{-1} {\bf{N}_2}}
{\det{\bf{N}_1} \det{\bf{N}_2}}
=\vec{m}\frac{{\bf{N}_1}({\bf{N}_1} + {\bf{N}_2})^{-1} {\bf{N}_2}}
{\det{\bf{N}_1} \det{\bf{N}_2}}
\label{l1l2m2}
\eeq
These terms can also be obtained in the series (\ref{exprhs}), for the 
following values of the variables:  $\vec{l_3} ( \equiv \vec{l_2} )$, $\vec{l_4} = 0$,
$\vec{m}$ arbitrary. 
However, as seen above in eqs. (\ref{restriction-alpha0}),
(\ref{restriction-alpha1}),  the sum over $\vec{m}$ is  finite due to the fact
that
\beq
	\vec{l_1} - \vec{l_2} = 
 \vec{m} = \vec{L}  \det {\bf{N}_1} \det {\bf{N}_2}
{\bf{N}}_2^{-1} ({\bf{N}}_1 + {\bf{N}}_2 ){\bf{N}}_1^{-1},
\label{restriction-m}
\eeq
for $\vec{L}$ arbitrary integers, contributes to the values of $\vec{l_4}$ in 
the RHS of eq. (\ref{exprhs}) by an amount $\vec{L}$,
while setting $\vec{m}$ to zero,  $\vec{l_3}$ is identified with 
$\vec{l_2} + \det {\bf{N}_1} \det {\bf{N}_2} \vec{L}{\bf{N}}_2^{-1}$. 
In other words, we have shown that the sum 
over $\vec{m}$ in (\ref{exprhs}) is over all integrally defined vectors in the 
unit cell generated by the basis elements:
\beq
	\vec{e'} = \vec{e}  \det {\bf{N}_1} \det {\bf{N}_2}
{\bf{N}}_2^{-1} ({\bf{N}}_1 + {\bf{N}}_2 ){\bf{N}}_1^{-1}
\label{basis-e'}
\eeq
with $\vec{e}$ being the elements of the canonical basis (\ref{basis-e}). 

\end{enumerate}

We have therefore proved that identity eq. (\ref{general-identity}) holds by 
explicitly showing a one to one correspondence between the series in 
eqs. (\ref{transflhs}) and (\ref{exprhs}).

\subsection{Yukawa expressions for oblique fluxes}\label{yukawa-general}

We now use the wavefunctions given in eqs. (\ref{general-basis1}) 
and (\ref{general-theta-text}),
to obtain the expression of Yukawa interactions when oblique fluxes, 
specified by intersection matrices
\beq
	{\bf{N}_1} = F_a - F_b,\,\,
	{\bf{N}_2} = F_c - F_a,\,\,
	{\bf{N}_3} = F_c - F_b.
\label{flux-abc}
\eeq 
are turned on along branes $a$, $b$ and $c$. As already mentioned,
in eq. (\ref{def-N}), ${\bf N_1}$, ${\bf N_2}$ and
${\bf N_3}$ are all real symmetric matrices 
(in the absence of components $p_{xx}$,
$p_{yy}$) and in addition the complex structure matrix is 
chosen to be proportional to the identity: $\tau I_n$, with $\tau$ complex.  
We then have:
\beqa
&&\hskip -1cm  \psi^{\vec{i}, {\bf{N}_1}} (\vec{z}, {\bf{\Omega}} = \tau I_n) \cdot
  \psi^{\vec{j}, {\bf{N}_2}} (\vec{z}, {\bf{\Omega}}= \tau I_n) = 
\left(2^{\frac n2}  \right)
\left(Vol(T^{2n})\right)^{- 1} 
\left(|\det {\bf{N}_1}|.|\det {\bf{N}_2}| (Im \tau)^6\right)^{\frac 14}
\cr && \hskip 2cm
\times e^{i\pi {\bf{N}_3}. \vec{z} \pim \vec{z}/\pim \tau} 
\vt \left[
\begin{array}{c}
\vec{i} \\  0
\end{array}
\right] ( {\bf{N_1}}\cdot\vec{z} |  {\bf{N_1}} \cdot \tau)  \cdot 
\vt \left[
\begin{array}{c}
\vec{j} \\  0
\end{array}
\right] ( {\bf{N_2}}\cdot\vec{z} |  {\bf{N_2}} \cdot \tau).  
\label{general-prod-wavefunction} 
\eeqa
Using the Riemann theta identity derived earlier in 
eq. (\ref{general-identity}),  
eq. (\ref{general-prod-wavefunction}) can be rewritten as:
\beqa
 &&\hspace{-1.7cm} \psi^{\vec{i}, {\bf{N}_1}} (\vec{z}) \cdot
  \psi^{\vec{j}, {\bf{N}_2}} (\vec{z}) = 
\sum_{\vec{m}}
\left(2^{\frac n2}  \right)^{\frac 12}
\left(Vol(T^{2n})\right)^{- {\frac 12}} 
\left[\frac{\left(|\det {\bf{N}_1}|.|\det {\bf{N}_2}| (Im \tau)^3\right)}
{|\det {\bf{N}_3}|}\right]^{\frac 14}\times 
\cr
&&\hspace{-1.7cm}\psi^{(\vec{i}{\bf{N}_1} + \vec{j}{\bf{N}_2} + \vec{m}{\bf{N}_1}).
{\bf{N}_3}^{-1}, {\bf{N}_3}} (\vec{z})
\cdot
\vt
\left[
\begin{array}{c}
[(\vec{i} - \vec{j})+ \vec{m}]\frac{{\bf{N}_1}{\bf{N}_3}^{-1} {\bf{N}_2}}
{\det{\bf{N}_1} \det{\bf{N}_2}}\\ 0
\end{array}
\right]   
\cr
&&\hspace{5.0cm}({0} |
(\det{\bf{N}_1} \det{\bf{N}_2}) ^2({\bf{N}_1} ^{-1} 
{\bf{N}_3}{\bf{N}_2} ^{-1})\tau) . 
\label{general-prod-wavefunction2}
\eeqa

Note that the integrality condition (\ref{j-integer}) is maintained by 
$\psi^{(\vec{i}{\bf{N}_1} + 
\vec{j}{\bf{N}_2} +\vec{m}{\bf{N}_1}){\bf{N}_3}^{-1}, {\bf{N}_3}} (\vec{z})$
appearing in the RHS of the above equation, since the expression
\beq
	\left[(\vec{i}{\bf{N}_1} +
	\vec{j}{\bf{N}_2} + \vec{m}{\bf{N}_1}){\bf{N}_3}^{-1}\right]\cdot {\bf{N}_3}
\label{vec-k-integer}
\eeq
is always an integer. On the other hand,
the  sum $\vec{m}$ in eq. (\ref{general-prod-wavefunction2})
is over the integers inside the cell generated
by the lattice vectors in eq. (\ref{basis-e'}) and total number of them
is given by the volume of this compact space. The size of the cell, i.e.,
its volume matches with those in eq. (\ref{N_3}) and 
(\ref{yukawa-basis-expansion-2}) for the $T^2$ case which is just the 
number, $N_3 = I_{cb}$ in eq. (\ref{N_3}), of chiral states for 
brane intersection $bc$.
However, the situation is different for $T^{2n}$, $n>1$. This becomes
clear by observing that the size of the cell given in eq. (\ref{basis-e'})
is bigger than the number of states ($\vec{k}$) 
in the intersection ${\bf N_3}$ between the branes $b$ and $c$
by a factor $det(\det {\bf{N}_1} \det {\bf{N}_2}
{\bf{N}}_2^{-1} {\bf{N}}_1^{-1})$.  This factor,
on the other hand, for $T^2$ is unity. We therefore notice that 
the sum $\vec{m}$ is over many more terms\footnote{We thank R. Russo and 
S. Sciuto for invaluable suggestions on this point.}
than the actual number of 
states ($\vec{k}$) in the intersection ${\bf N_3}$ between the branes
$b$ and $c$.

The extra factor of terms appearing in eq. (\ref{general-prod-wavefunction2})
can be explained by noticing that the sum over 
terms in eqs. (\ref{general-prod-wavefunction2}) and 
(\ref{general-yukawa-series}) is over the states 
$\psi^{(\vec{i}{\bf{N}_1} + \vec{j}{\bf{N}_2} + \vec{m}{\bf{N}_1}).
{\bf{N}_3}^{-1}, {\bf{N}_3}} (\vec{z})$ that are inside the cell
in eq. (\ref{basis-e'}) and contribute to the Yukawa coupling  
by the orthogonality relation
eq. (\ref{normalizable-general}). As any state (with more details 
given in the subsection-\ref{summation}) $\vec{k}$, satisfying
integrality conditions such as (\ref{j-integer}) is defined only upto the 
integer lattice shifts, one therefore has appearance of the same states
inside the volume of lattice (\ref{basis-e'}), multiple times. 
In other words, for any given state,
in the RHS of eqs. 
(\ref{general-prod-wavefunction2}), all those integer vector ($\vec{m}$)
shifts also contribute to the sum which satisfy the integrality condition for 
$\vec{m} {\bf N_1} {\bf N_3}^{-1}$ inside the cell (\ref{basis-e'}). 
Explicit solution of this condition is presented later on 
in section \ref{summation} in eq. (\ref{integer-m-in-cell}).

Then, as in the $T^2$ case, orthonormality of wavefunctions
(\ref{normalizable-general}), implies that the Yukawa coupling,
whose explicit form is given in section \ref{summation}, 
can be `formally' written in a form :
\beqa
&&\hskip -0.7cm	Y_{ijk} = g\sigma_{abc} 
		\left(2^{\frac n2}  \right)^{\frac 12}
\left(Vol(T^{2n})\right)^{- {\frac 12}} 
\left[\frac{\left(|\det {\bf{N}_1}|.|\det {\bf{N}_2}| (Im \tau)^3 \right)}
{|\det {\bf{N}_3}|}\right]^{\frac 14}\hskip -0.2cm\times \hskip -0.2cm
\sum_{\vec{m} \in \{\vec{e'}\}} \delta_{\vec{k}, {\bf{N}_3}^{-1}
({\bf{N}_1}\vec{i} + {\bf{N}_2}\vec{j} + 
{\bf{N}_1} \vec{m})} \cr 
&&\hskip 1cm\times\vt
\left[
\begin{array}{c}
[(\vec{i} - \vec{j})+ \vec{m}]\frac{{\bf{N}_1}{\bf{N}_3}^{-1} {\bf{N}_2}}
{\det{\bf{N}_1} \det{\bf{N}_2}}\\ 0
\end{array}
\right]   
({0} |
(\det{\bf{N}_1} \det{\bf{N}_2}) ^2({\bf{N}_1} ^{-1} 
{\bf{N}_3}{\bf{N}_2} ^{-1})\tau), 
\label{general-yukawa-series}
\eeqa
where by the summation index $\vec{m} \in \{\vec{e'}\}$, one means to 
sum over all integer points inside the lattice generated by 
$\vec{e_1^{\prime}}, \vec{e_2^{\prime}}\cdots 
\vec{e_n^{\prime}}$ in eq. (\ref{basis-e'}) and the Kronecker delta
is to identify all the states $\vec{k}$
upto integer shifts.

The above expression  reduces in the case of $T^2$ flux compactification to 
eq. (\ref{yukawa-simple}), since the Kronecker delta constraint 
has a unique solution in such a situation. 
To compare the two expressions, note that the
indices $i,j,k$ in the factorized case are scaled with respect to 
the one of general tori, by  the factors ${\frac{1}{N_1}}$,
${\frac{1}{N_2}}$ and ${\frac{1}{N_3}}$, respectively. Then, 
the Kronecker delta constraint
in eq. (\ref{general-yukawa-series})  precisely matches with the 
one in eq. (\ref{yukawa-series}). In the case of general tori, however, 
the constraint implies that the interaction terms involve 
the states  which satisfy the equation
\beq
	{\bf{N}_3}\vec{k} =  
({\bf{N}_1}\vec{i} + {\bf{N}_2}\vec{j} + 
{\bf{N}_1} \vec{m})
\label{integer-constraint}
\eeq
among the vectors ${\bf{N}_1}\vec{i}$, ${\bf{N}_2}\vec{j}$, 
${\bf{N}_3}\vec{k}$ for $\vec{m}$ integers 
inside the unit cell given in eq. (\ref{unit-cell}) and corresponding 
states $\vec{k}$ are only defined upto integer lattice shifts.
We now find all such solutions of the lattice shifts in the next subsection
and present the explicit answer for the Yukawa coupling for general tori.

\subsection{Explict Yukawa coupling expressions}\label{summation}

In this subsection we now present the set of terms that contribute
to eqs. (\ref{general-prod-wavefunction2}) and 
(\ref{general-yukawa-series}). In order to clarify the situation 
we analyze the correspondence between the chiral multiplet families
of states such as the ones appearing in eq. (\ref{vec-k-integer})
and  the fluxes along the branes. Our discussion is restricted to 
${\bf N}$ being real symmetric matrices, due to the imposition
of the Riemann conditions (\ref{eq-riemann-conditions}) for the special 
complex structure $\Omega = \tau I_n $ under discussion.


For a given pair of brane-stacks with intersection matrix ${\bf N}$,
the condition eq. (\ref{j-integer}) that a 
state $\hat{i}$ needs to satisfy is 
$N . {\hat{i}} = integer$. The solution of this condition is: 
${\hat{i}} = {\bf N}^{-1} \vec{e}$,
with $\vec{e}$ being the integer basis vectors in an $n$-dimensional 
space as given in eq. (\ref{basis-e}). 
The states are therefore generated by the set of $n$ vectors:
${\hat{i}}_i = \vec{e_i}{\bf N}^{-1} $, with subscript 
$i=1,2\cdots n$ and are  $det({\bf N})$ in number, namely those
which are inside the cell generated by $\vec{e_i}$'s. Here and in following
we also keep in mind that all the chiral multiplet states 
that we are discussing,
are defined only upto the shift by integer lattice vectors $\vec{e_i}$'s.

To give an example: for $n=2$ (corresponding to $T^4$), with
\beqa
{\bf N} = \begin{pmatrix}\alpha & \gamma \cr \gamma  & \beta 
\end{pmatrix},
\label{n-example}
\eeqa
we have the basis vectors for generating the states:
\beqa
\hat{i}_1 =
{1\over (\alpha \beta - \gamma^2)} 
\begin{pmatrix} \beta \cr - \gamma 
\end{pmatrix},\,\,\,\,
{\hat{i}_2} =
{1\over (\alpha \beta - \gamma^2)} 
\begin{pmatrix}- \gamma \cr  \alpha 
\end{pmatrix}.
\label{i1i2}
\eeqa

To obtain the degeneracy count, we note that for the above example
we have:
\beqa
\vec{e_1} = \alpha \vec{i_1} + \gamma \vec{i_2},\cr
\vec{e_2} = \gamma \vec{i_1} + \beta \vec{i_2} .
\label{e1e2-i1i2}
\eeqa
The number of independent states inside the cell with lattice vectors
$\vec{e_1}$ and $\vec{e_2}$ is then the determinant of the above
transformation  which is $det {\bf N}$.
A generic state appearing in eq. (\ref{integer-constraint}) then has a form:
\beq
\vec{i} = m_1 \vec{i_1} + m_2 \vec{i_2},\,\,\,
\vec{j} = n_1 \vec{j_1} + n_2 \vec{j_2}, \,\,\,
\vec{k} = p_1 \vec{k_1} + p_2 \vec{k_2} .
\label{state-generic}
\eeq
with $\vec{j_i}$, $\vec{k_i}$ defined in a similar way as in 
eq. (\ref{i1i2}) with respect to the corresponding intersection matrices. 
Also, integers $m_i, n_i, p_i$ label the states of a chiral 
family in a given brane stack.

We now go on to give explicit solution for the vector $\vec{m}$
that contribute to the sum of terms in Yukawa coupling expressions
(\ref{general-prod-wavefunction2}) and (\ref{general-yukawa-series}),
namely those inside the cell defined in 
eq. (\ref{basis-e'}). 
The size of the cell, namely the number of states
that it contains is equal to $det(\det {\bf{N}_1} \det {\bf{N}_2}
{\bf{N}}_2^{-1} ({\bf{N}}_1 + {\bf{N}}_2 ){\bf{N}}_1^{-1})$, 
as stated earlier. In a situation with
$2\times 2$ matrices, for example, it is  
$det{\bf N_1} det{\bf N_2} det{\bf N_3}$.
For illustration purposes we restrict ourselves to the 
discussion with $2\times 2$ matrices. However, all the results we
write below are valid for other situations as well.

Now, restricting to this $2\times 2$ case for the simplicity of
discussion, we write all possible solutions for $\vec{m}$ that provide
integer solutions for $\vec{m} {\bf N_1} {\bf N_3}^{-1}$,
as appearing in the definition of states in 
eqs. (\ref{general-prod-wavefunction2}), (\ref{vec-k-integer}),
and show that they are 
$det{\bf N_1} det{\bf N_2}$ in number. So that the degeneracy of
the state matches with 
$det{\bf N_1} det{\bf N_2} det{\bf N_3}$ given in the last
paragraph. To compare, note that for a diagonal flux situation,
as in section-\ref{yukawa-factorized}, we have $m = n_3$ as a 
single solution of an analogous condition
$m n_1 n_3^{-1} = integer$, corresponding to the state degeneracy 
which is $n_3$.

The integer solutions for 
$\vec{m} {\bf N_1} {\bf N_3}^{-1}$ are:
\beq
\vec{m} = \vec{p} 
det{\bf N_1} {\bf N_3} {\bf N_1}^{-1} + 
\vec{\tilde{p}} det{\bf N_2} {\bf N_3} {\bf N_2}^{-1}, 
\label{integer-m-in-cell}
\eeq
where $\vec{p}$ is all integer vectors within a cell generated by
$\vec{e} det{\bf N_2} {\bf N_2}^{-1}$
and $\vec{\tilde{p}}$ is all integer vectors within a cell generated
by $\vec{e} det{\bf N_1} {\bf N_1}^{-1}$. It is easy to see that $\vec{m}$
satisfies $\vec{m} {\bf N_1} {\bf N_3}^{-1} = integer$ 
(by making use of ${\bf N_1} = {\bf N_3} - {\bf N_2}$).
Together, for every solution of the first term in $\vec{m}$ we have 
$det{\bf N_1}$
solution for the second term and this goes on for 
$det{\bf N_2}$ number
of terms from the first term. So that total degeneracy of such
$\vec{m}$ is $det{\bf N_1} det{\bf N_2}$, as stated earlier.

About the states: $\vec{m}$ given in eq. (\ref{integer-m-in-cell}) 
defines a periodic set, in the same way as 
for the $T^2$ case $m = n_3$  defines the periodic set of states
in the RHS of eqs. (\ref{yukawa-basis-expansion-2}) and 
(\ref{yukawa-series}). There the states are explicitly given as  
$k= (0), (n_1/n_3), (2n_1/n_3), 
\cdots [(n_3-1)n_1/n_3]$ with a periodicity $n_3$ 
for this series. Various states inside the cell (\ref{basis-e'}) 
can also be found using eq. (\ref{integer-constraint})
and making use of the condition: 
${\bf N_1} = {\bf N_3} - {\bf N_2}$ as: (also the fact that
any state is defined upto integer vectors). The states are:
\beq
\vec{k} = 
\vec{p} det{\bf N_1} {\bf N_3}^{-1} 
+ \vec{\tilde{p}} det{\bf N_2} {\bf N_3}^{-1}\,\,\, etc.
\label{states-detn1-detn2}
\eeq
and the state degeneracy is 
$det{\bf N_1} det{\bf N_2} det{\bf N_3}$.

The Yukawa coupling can now be written in an explicit form 
given by a sum of $ det{\bf N_1} det{\bf N_2}$
number of terms, which can be read off from eq.
(\ref{general-prod-wavefunction2}) directly, with
$\vec{m}$ replaced by
\beq
\vec{\tilde m} + \vec{p} det{\bf N_1} {\bf N_3} {\bf N_1}^{-1} + 
\vec{\tilde{p}} det{\bf N_2} {\bf N_3} {\bf N_2}^{-1}
\label{replace}
\eeq
and now such $\vec{\tilde m}$  are the unique solutions of  eq. 
(\ref{integer-constraint}) where all other solutions defined upto
the shifts in 
$\vec{\tilde m}$
by $\vec{p} detN_1 N_3 N_1^{-1} + \vec{\tilde{p}} detN_2 N_3 N_2^{-1}$
have been identified.


Eq. (\ref{general-yukawa-series}) now reads as:
\beqa
&&\hskip -0.7cm	Y_{ijk} = g\sigma_{abc} 
		\left(2^{\frac n2}  \right)^{\frac 12}
\left(Vol(T^{2n})\right)^{- {\frac 12}} 
\left[\frac{\left(|\det {\bf{N}_1}|.|\det {\bf{N}_2}| (Im \tau)^3 \right)}
{|\det {\bf{N}_3}|}\right]^{\frac 14}\hskip -0.2cm\times \hskip -0.2cm
\sum_{\vec{p}, \vec{\tilde{p}} }
\cr
&&\times\vt
\left[
\begin{array}{c}
[\{(\vec{i} - \vec{j})+ (\vec{k} {\bf N_3} - \vec{i} {\bf N_1}
		- \vec{j}{\bf N_2}) {\bf N_1}^{-1}\}
\frac{{\bf{N}_1}{\bf{N}_3}^{-1} {\bf{N}_2}}
{\det{\bf{N}_1\det{\bf{N}_2}}} + 
(\vec{p}\frac{\bf N_2}{det {\bf N_2}} + 
\vec{\tilde{p}}\frac{\bf N_1}{det {\bf N_1}}) ] \cr 0
\end{array}
\right]  \cr 
&&\hspace{2.0in}({0} |
(\det{\bf{N}_1}\det{\bf{N}_2} ) ^2({\bf N_1}^{-1} {\bf{N}_3}{\bf{N}_2} ^{-1} 
\tau), 
\label{general-yukawa-series-sum}
\eeqa
or equivalently:
\beqa
&&\hskip -0.7cm	Y_{ijk} = g\sigma_{abc} 
		\left(2^{\frac n2}  \right)^{\frac 12}
\left(Vol(T^{2n})\right)^{- {\frac 12}} 
\left[\frac{\left(|\det {\bf{N}_1}|.|\det {\bf{N}_2}| (Im \tau)^3 \right)}
{|\det {\bf{N}_3}|}\right]^{\frac 14}\hskip -0.2cm\times \hskip -0.2cm
\sum_{\vec{p}, \vec{\tilde{p}} }
\cr
&&\times\vt
\left[
\begin{array}{c}
[(- \vec{j}+ \vec{k} )
\frac{{\bf{N}_2}}{\det{\bf{N}_1\det{\bf{N}_2}}} + 
(\vec{p}\frac{\bf N_2}{det {\bf N_2}} + 
\vec{\tilde{p}}\frac{\bf N_1}{det {\bf N_1}}) ] \cr 0
\end{array}
\right]  \cr 
&&\hspace{2.0in}({0} |
(\det{\bf{N}_1}\det{\bf{N}_2} ) ^2({\bf N_1}^{-1} {\bf{N}_3}{\bf{N}_2} ^{-1} 
\tau). 
\label{general-yukawa-simple}
\eeqa

Note that the sum over $\vec{m}$ is now broken 
into sum over $\vec{p}$ and $\vec{\tilde{p}}$. We end this discussion 
by reminding ourselves once again that
$\vec{p}$ runs over all the states inside the cell generated by
$\vec{e_1} det{\bf N_2} {\bf N_2}^{-1}$ 
and $\vec{e_2} det{\bf N_2} {\bf N_2}^{-1}$.
Similarly $\vec{\tilde{p}}$ runs over all the states inside the 
cell generated by
$\vec{e_1} det{\bf N_1} {\bf N_1}^{-1}$ 
and $\vec{e_2} det{\bf N_1} {\bf N_1}^{-1}$.


We now present two explicit examples, one for the oblique situation and the 
other for the commuting diagonal fluxes. We show that our answer for the 
diagonal flux is identical to the one for the diagonal yukawa coupling 
expression given in \cite{ibanez} for $T^{2n}$. In fact this holds for 
any set of fluxes  with ${\bf N_1}$, ${\bf N_2}$, ${\bf N_3}$
diagonal. On the other hand, we also show that
the set of terms given above in 
eqs. (\ref{general-yukawa-series-sum}) and 
(\ref{general-yukawa-simple})
can also be summed up in a number of cases, for the oblique cases as well.

\vskip .6cm

\noindent {\bf Example : Oblique flux}
\vskip .6cm

For the oblique case,  by taking two noncommuting matrices:
\beqa
{\bf N_1} = \begin{pmatrix} 2 & 1 \cr 1  & 2
\end{pmatrix},\,\,\,
{\bf N_2} = \begin{pmatrix} 1 &  \cr   & 2
\end{pmatrix},
\label{example-oblique-n1-n2}
\eeqa
we have:
\beqa
(det {\bf N_1}) {\bf N_1}^{-1} = \begin{pmatrix} 2 & - 1 \cr - 1  & 2
\end{pmatrix},\,\,\,
(det {\bf N_2}) {\bf N_2}^{-1} = \begin{pmatrix} 2 &  \cr   & 1
\end{pmatrix}.
\eeqa
The set of integer points inside the cell generated by
$\vec{e_1} det{\bf N_2} {\bf N_2}^{-1} = (2, 0)$ and 
$\vec{e_2} det{\bf N_2} {\bf N_2}^{-1} = (0, 1)$,
are: $(0, 0)$ and $(1, 0)$, as 
$det (det{\bf N_2} {\bf N_2}^{-1}) = 2$. 
The set of integer points inside the cell generated by
$\vec{e_1} det{\bf N_1} {\bf N_1}^{-1} = (2, -1)$ 
and $\vec{e_2} det{\bf N_1} {\bf N_1}^{-1} = (-1, 2)$,
are : $(0, 0)$, $(1, 0)$ and $(0, 1)$, as 
$det (det{\bf N_1} {\bf N_1}^{-1}) = 3$.\footnote{Another example 
with mixed eigenvalues for the matrix ${\bf N_1}$ can be constructed 
by exchanging the off-diagonal and diagonal entries in 
eq. (\ref{example-oblique-n1-n2}) for ${\bf N_1}$. Such an example 
will be relevant for the situtation discussed in later sections
where intersection matrices
with both positive and negative eigenvalues are discussed.}

Now, to illustrate our method,
we concentrate on finding a particular Yukawa interaction 
among states: $\vec{i} = \vec{j} = \vec{k} = (0, 0)$.
This particular Yukawa now has the form, making use of
Eq. (\ref{general-yukawa-series-sum}) as:
\beqa
&&\hskip -0.7cm	Y_{000} = g\sigma_{000} 
		\left(2^{\frac n2}  \right)^{\frac 12}
\left(Vol(T^{2n})\right)^{- {\frac 12}} 
\left[\frac{\left(|\det {\bf{N}_1}|.|\det {\bf{N}_2}| (Im \tau)^3 \right)}
{|\det {\bf{N}_3}|}\right]^{\frac 14}\hskip -0.2cm\times \hskip -0.2cm
\sum_{\vec{p}, \vec{\tilde{p}} } \cr
&&\hspace{1.5in}\vt
\left[
\begin{array}{c}
[
(\vec{p}\frac{\bf N_2}{det {\bf N_2}} + 
\vec{\tilde{p}}\frac{\bf N_1}{det {\bf N_1}}) ] \cr 0
\end{array}
\right]
({0} | (\det{\bf{N}_1}\det{\bf{N}_2} ) ^2
({\bf N_1}^{-1} {\bf{N}_3}{\bf{N}_2} ^{-1}  
\tau)), 
\label{general-yukawa-series2}
\nonumber
\eeqa

To see what terms in $\vec{p}$ and $\vec{\tilde{p}}$ dependent arguements
appear in the sum, we write down all the possibilities that arise from 
the combinations:
\beqa
  (\vec{p}\frac{\bf N_2}{det {\bf N_2}} + 
\vec{\tilde{p}}\frac{\bf N_1}{det {\bf N_1}})
	= \vec{p} \begin{pmatrix} \frac{1}{2} &  \cr   & 1 \end{pmatrix}
	+ \vec{\tilde{p}} \frac{1}{3}
\begin{pmatrix} 2 & 1 \cr 1  & 2 \end{pmatrix}
\eeqa 
with $\vec{p} = (0, 0), (1, 0)$ and 
$\vec{\tilde{p}} = (0,0), (0, 1), (1, 0)$.
All the six possibilities then imply that in Theta function we get the 
following explicit sum:
\beqa
 &&
\left(
\vt
\left[
\begin{array}{c}
[(0, 0) ] \cr 0
\end{array}
\right] + 
\vt
\left[
\begin{array}{c}
[(\frac{1}{2}, 0) ] \cr 0
\end{array}
\right]+
\vt
\left[
\begin{array}{c}
[(\frac{2}{3}, \frac{1}{3}) ] \cr 0
\end{array}
\right] +
\vt
\left[
\begin{array}{c}
[(\frac{1}{3}, \frac{2}{3}) ] \cr 0
\end{array}
\right] +
\vt
\left[
\begin{array}{c}
[(\frac{1}{6}, \frac{1}{3}) ] \cr 0
\end{array}
\right] +  \right. \hspace{0.0in}\cr
&&\left. \hspace{1.0in}\vt
\left[
\begin{array}{c}
[(\frac{5}{6}, \frac{2}{3}) ] \cr 0
\end{array}
\right] \right)
({0} |
(\det{\bf{N}_1}\det{\bf{N}_2} ) ^2({\bf N_1}^{-1} {\bf{N}_3}{\bf{N}_2} ^{-1} 
\tau)) 
\label{six-terms}
\eeqa
where a common modular parameter arguement of the all the 
six Theta terms have been written outside
of the bracket for saving space.
The integer sums of the six terms over integer $\vec{l}$ are 
of the forms: 
\beq
	\sum_{\vec{l}} e^{[\vec{l} + (q_1, q_2)]
(\det{\bf{N}_1}\det{\bf{N}_2} )^2({\bf N_1}^{-1} {\bf{N}_3}{\bf{N}_2} ^{-1} 
\tau)[\vec{l} + (q_1, q_2)]	}
\label{sum-illustration}
\eeq
with $\vec{l} + (q_1, q_2)$ given explicitly as:
\beq
	\vec{l} + (0, 0),\,\,
	\vec{l} + (\frac{1}{2}, 0),\,\,
	\vec{l} + (\frac{2}{3}, \frac{1}{3}),\,\,
	\vec{l} + (\frac{1}{3}, \frac{2}{3}),\,\,
	\vec{l} + (\frac{1}{6}, \frac{1}{3}),\,\,
	\vec{l} + (\frac{5}{6}, \frac{2}{3}),
\label{l-q1q2}
\eeq
for the six terms in eq. (\ref{six-terms}). It can also be seen that
we can write them as:
\beqa
	\begin{pmatrix} l_1 \cr l_2   \end{pmatrix}
    +   \begin{pmatrix} \frac{m}{2} + \frac{2n}{3} 
			\cr \frac{n}{3}   \end{pmatrix}	
\equiv		\begin{pmatrix} l_1 \cr l_2   \end{pmatrix}
    +   \frac{1}{6}
\begin{pmatrix} 3 & 4 \cr 0  & 2 \end{pmatrix}
\begin{pmatrix} {m}  \cr {n}   \end{pmatrix}	
\label{l-shift} 
\eeqa
with $m = 0, 1$ and $n = 0, 1, 2$.
Now, using the inverse of the matrix 
\beqa
P = \frac{1}{6}\begin{pmatrix} 3 & 4 \cr 0  & 2 \end{pmatrix},
\label{P-matrix}
\eeqa
appearing in eq. (\ref{l-shift}): 
\beqa
P^{-1} = \begin{pmatrix} 2 & -4 \cr 0  & 3 \end{pmatrix} ,
\label{P-inverse-matrix}
\eeqa
we can write eq. (\ref{l-shift}) as:
\beqa
\frac{1}{6}
\begin{pmatrix} 3 & 4 \cr 0  & 2 \end{pmatrix}
\left[	
\begin{pmatrix} 2l_1 - 4 l_2 \cr 3 l_2   \end{pmatrix}
    +   \begin{pmatrix} m  \cr n   \end{pmatrix} \right]
\label{sum-l-n}
\eeqa
with $m = 0, 1$ and $n = 0, 1, 2$.

It can now be seen that as $l_1, l_2$ vary over all integers, and
$m = 0, 1$ and $n = 0, 1, 2$, then the combination of terms in the
big square bracket in eq. (\ref{sum-l-n}) also span over  ALL integers. 
As a result we are able to take the factor of matrix P
out by summing over all the six terms, while reducing the six terms
in eq.(\ref{six-terms}) to one. The net result is then the arguement
of theta function modifies by the factor: 
\beq
(\det{\bf{N}_1}\det{\bf{N}_2} ) ^2({\bf N_1}^{-1} {\bf{N}_3}{\bf{N}_2} ^{-1} 
\tau) \rightarrow  
 P^T (\det{\bf{N}_1}\det{\bf{N}_2} ) ^2({\bf N_1}^{-1} {\bf{N}_3}{\bf{N}_2} ^{-1} 
\tau) P
\eeq
and final answer for Yukawa coupling is:
\beqa
&&\hskip -0.7cm	Y_{000} = g\sigma_{000} 
		\left(2^{\frac n2}  \right)^{\frac 12}
\left(Vol(T^{2n})\right)^{- {\frac 12}} 
\left[\frac{\left(|\det {\bf{N}_1}|.|\det {\bf{N}_2}| (Im \tau)^3 \right)}
{|\det {\bf{N}_3}|}\right]^{\frac 14}\hskip -0.2cm\times \hskip -0.2cm  \cr
&&\hspace{1.5in}\vt
\left[
\begin{array}{c}
0
 \cr 0
\end{array}
\right]
({0} | P^T (\det{\bf{N}_1}\det{\bf{N}_2} ) ^2
({\bf N_1}^{-1} {\bf{N}_3}{\bf{N}_2} ^{-1} 
\tau) P ). 
\label{general-yukawa-series2-final-sum}
\nonumber
\eeqa
We can similarly take care of other nonzero values 
$\vec{i}, \vec{j}, \vec{k}$ etc. as well, but details are being left.
\vskip 0.6cm

\noindent{\bf Example : Diagonal Flux}
\vskip .6cm

We take another example, now with diagonal fluxes :
\beqa
{\bf N_1} = \begin{pmatrix} 2 &  \cr   & 3
\end{pmatrix},\,\,\,
{\bf N_2} = \begin{pmatrix} 5 &  \cr   & 2
\end{pmatrix} .
\eeqa
Then:
\beqa
(det {\bf N_1}) {\bf N_1}^{-1} = \begin{pmatrix} 3 &  \cr   & 2
\end{pmatrix},\,\,\,\,
(det {\bf N_2}) {\bf N_2}^{-1} = \begin{pmatrix} 2 &  \cr   & 5
\end{pmatrix} .
\eeqa
Set of integer points inside the cell generated by
$\vec{e_1} det{\bf N_2} {\bf N_2}^{-1} = (2, 0)$ and 
$\vec{e_2} det{\bf N_2} {\bf N_2}^{-1} = (0, 5)$,
are: $(0, 0)$, $(0, 1)$, $(0, 2)$, $(0, 3)$, $(0, 4)$, $(1, 0)$,
$(1, 1)$, $(1, 2)$, $(1, 3)$, $(1, 4)$,  
as $det (det{\bf N_2} {\bf N_2}^{-1}) = 10$. 
On the other hand, set of integer points inside the cell generated by
$\vec{e_1} det{\bf N_1} {\bf N_1}^{-1} = (3, 0)$ 
and $\vec{e_2} det{\bf N_1} {\bf N_1}^{-1} = (0, 2)$,
are: $(0, 0)$, $(1, 0)$ $(0, 1)$, $(1, 1)$, $(2, 0)$, $(2, 1)$, 
as $det (det{\bf N_1} {\bf N_1}^{-1}) = 6$. 

We now have:
\beqa
 \vec{l} + (\vec{p}\frac{\bf N_2}{det {\bf N_2}} + 
\vec{\tilde{p}}\frac{\bf N_1}{det {\bf N_1}})
	= \vec{l} + \vec{p} 
\begin{pmatrix} \frac{1}{2} &  \cr   & \frac{1}{5} \end{pmatrix}
	+ \vec{\tilde{p}}
\begin{pmatrix} \frac{1}{3} &  \cr   & \frac{1}{2} \end{pmatrix} , 
\label{vec-l-diagonal}
\eeqa 
which can also be written as:
\beqa
 \vec{l} + (\vec{p}\frac{\bf N_2}{det {\bf N_2}} + 
\vec{\tilde{p}}\frac{\bf N_1}{det {\bf N_1}})
	= \begin{pmatrix} l_1 \cr l_2  \end{pmatrix} +  
\begin{pmatrix} \frac{1}{2} &  \cr   & \frac{1}{5} \end{pmatrix}
\begin{pmatrix} p_1 \cr p_2  \end{pmatrix}
	+ \begin{pmatrix} \frac{1}{3} &  \cr   & \frac{1}{2} \end{pmatrix}
	\begin{pmatrix} \tilde{p}_1 
	\cr \tilde{p}_2  \end{pmatrix} ,
\label{vec-l-diagonal2}
\eeqa 
with $p_1 = 0, 1$, $p_2 = 0, 1, 2, 3, 4$,
$\tilde{p}_1 = 0, 1, 2$, $\tilde{p}_2 = 0, 1$.
  
By taking a  factor of 
$\frac{{\bf N_1} {\bf N_2}}{det{\bf N_1} det{\bf N_2}}$ out, the above
equation can also be rewritten as:
\beqa
\frac{{\bf N_1} {\bf N_2}}{det{\bf N_1} det{\bf N_2}}
 [\vec{l} + (\vec{p}\frac{\bf N_2}{det {\bf N_2}} + 
\vec{\tilde{p}}\frac{\bf N_1}{det {\bf N_1}})]
	= 
\begin{pmatrix} \frac{1}{6} &  \cr   & \frac{1}{10}
\end{pmatrix}
\left[\begin{pmatrix} 6 l_1 \cr 10 l_2  \end{pmatrix} +  
\begin{pmatrix} 3 p_1 \cr 2 p_2  \end{pmatrix} +
	\begin{pmatrix} 2 \tilde{p}_1 
	\cr 5 \tilde{p}_2  \end{pmatrix} \right]
\eeqa 
with $p_1 = 0, 1$, $p_2 = 0, 1, 2, 3, 4$,
$\tilde{p}_1 = 0, 1, 2$, $\tilde{p}_2 = 0, 1$.
It can again be checked explicitly that it leads to 
ALL integer variables inside the square bracket. The net result of summing 
over different terms in the diagonal case therefore is the appearance of 
the matrix outside the square bracket:
$\frac{{\bf N_1} {\bf N_2}}{det{\bf N_1} det{\bf N_2}}$. 
When multiplying the modular parameter
arguement as appearing in eq. (\ref{general-yukawa-series-sum}), from both
left and the right, this precisely reproduces 
a modified modular parameter which
matches with the known diagonal flux solultion for Yukawa
coupling in \cite{ibanez}. This holds for the diagonal flux in general, 
not restricted to the example above.


\subsection{arbitrary-$\alpha$}\label{alpha}

The results, obtained so far in this  section, are derived for a 
particular  choice of $\alpha$ given in the eq. (\ref{def-alpha}). 
However, all the results can be re-derived for arbitrary  $\alpha$,
appearing in eq. (\ref{transmatrix}) etc.. 
For the factorized case, we saw in that the Yukawa coupling expression (\ref{yukawa-simple}) can be recast into a symmetric form 
in eq. (\ref{yukawa-answer-t2}) (apart from the prefactor), where 
the arguments of the Jacobi theta functions are  
invariant under a cyclic change: $a \rightarrow b \rightarrow c$. 
This is due to the cyclic property of the superpotential coefficients obtained by
a third derivative of the superpotential $W_{ijk}$. The prefactor does not
obey in general this symmetry, since it depends on the wave function normalizations
(K\"ahler metric).
Here, we  show a similar cyclic property in the non-factorized case, given 
above in the Yukawa coupling expression (\ref{general-yukawa-simple}),
by making different choices of the matrix $\alpha$ in eq. (\ref{def-alpha}).
Note that different choices of this matrix provide equivalent expressions
for the wavefunctions, and in turn Yukawa couplings, since they are related
though a change of variables inside the theta sum. The 
$\alpha$ matrix can be chosen appropriately so that the redefined variables in
eqs. (\ref{restriction-alpha1}) and (\ref{restriction-m})  are well 
defined integers.
Below we present a few examples with different choices of $\alpha$, to 
demonstrate the cyclicity mentioned above.

Eq. (\ref{exprhs}), for arbitrary $\alpha$, can be written as:
\beqa 
\hspace{-0.35in} \sum_{\vec{l_3}, \vec{l_4}\in \inte^n}\sum_{\vec{m}} 
\left( \right.
e^ {\pi i [ (\vec{j_1} {\bf{N}}_1 + \vec{j_2} {\bf{N}}_2  
+ {\vec{m}}{\bf{N}}_1)
({\bf{N}}_1 + {\bf{N}}_2 ) ^ {-1} + \vec{l_3}] 
\cdot ( {\bf{N}}_1 + {\bf{N}}_2) \tau \cdot
[ ({\bf{N}}_1  + {\bf{N}}_2  ) ^ {-1} ( {\bf{N}}_1\vec{j_1} +  {\bf{N}}_2 \vec{j_2} 
+{\bf{N}}_1 {\vec{m}})
 + \vec{l_3}]} 
 \hspace{1.0in} \cr
\times e^ {2\pi i [ (\vec{j_1} {\bf{N}}_1 + \vec{j_2} {\bf{N}}_2  
+ {\vec{m}}{\bf{N}}_1)
({\bf{N}}_1 + {\bf{N}}_2 ) ^ {-1} + \vec{l_3}]\cdot 
[\vec{z_1} + \vec{z_2}]} 
\left. \right)  \hspace{3in} \cr
\times \left( \right.
e^ {\pi i [(\vec{j_1} - \vec{j_2} + \vec{m}) 
{\bf{N}}_1 ({\bf{N}}_1 + {\bf{N}}_2 ) ^ {-1}{\bf{N}}_2 
\alpha^{-1} + \vec{l_4}]\cdot
[\alpha  {\bf{N}}_1^{-1}({\bf{N}}_1 + 
{\bf{N}}_2) {\bf{N}}_2^{-1} \tau ] \alpha^T\cdot
[(\alpha^{-1})^T{\bf{N}}_2 ({\bf{N}}_1 + {\bf{N}}_2 ) ^ {-1}{\bf{N}}_1 
 (\vec{j_1} - \vec{j_2} + \vec{m}) + \vec{l_4}]} 
\hspace{1.0in}  \cr
\times e^ {2\pi i [(\vec{j_1} - \vec{j_2} + \vec{m}) 
{\bf{N}}_1 ({\bf{N}}_1 + {\bf{N}}_2 ) ^ {-1}{\bf{N}}_2 
\alpha^{-1} + \vec{l_4}]\cdot 
[\alpha {\bf{N}}_1 ^ {-1} \vec{z_1} - {\bf{N}}_2^{-1} \vec{z_2}]}
\left. \right), 
\hspace{3in} \nonumber
\eeqa\vskip -1.5cm
\beq
\label{exprhs-arbitrary-alpha}
\eeq
provided $\vec{l_4} $, defined  in eq. (\ref{restriction-alpha0}), is an integer vector, and so  is $\vec{m}$ given in eq. (\ref{restriction-m}). In addition the unit-cell, within which $\vec{m}$ lie, is now defined by the basis vectors :
\beq
\vec{e'} = \vec{e} \alpha ({\bf{N}_1} ^{-1} 
({\bf{N}_1} + {\bf{N}_2}){\bf{N}_2} ^{-1}) . 
\label{unit-cell-arbitrary-alpha}
\eeq 
Moreover, eq. (\ref{general-identity}) takes the form:
\beqa 
& &
\hspace{-0.8in}\vt
\left[
\begin{array}{c}
\vec{j_1} \\ 0
\end{array}
\right]
(\vec{z_1}| {\bf{N}_1}\tau)
\cdot
\vt
\left[
\begin{array}{c}
\vec{j_2} \\ 0
\end{array}
\right]
(\vec{z_2}|  {\bf{N}_2}\tau)   =  \cr
& &\hskip -0.8cm
\sum_{\vec{m}}
\vt
\left[
\begin{array}{c}
(\vec{j_1}{\bf{N}_1} +\vec{j_2}{\bf{N}_2} +\vec{ m}.{\bf{N}_1} )({\bf{N}_1} 
+ {\bf{N}_2})^{-1} \\ 0
\end{array}
\right]
(\vec{z_1}+ \vec{z_2}| ({\bf{N}_1} + {\bf{N}_2})\tau)   \cr
&&  \times
\vt
\left[
\begin{array}{c}
[(\vec{j_1} - \vec{j_2})+ \vec{m}]{\bf{N}_1}({\bf{N}_1} + {\bf{N}_2})^{-1} {\bf{N}_2}
\alpha^{-1}\\ 0
\end{array}
\right]   \cr &  & \hspace{1in}
 \left(\alpha ({\bf{N}_1}^{-1}
\vec{z_1} - {\bf{N}_2}^{-1}\vec{z_2}) |
\alpha ({\bf{N}_1} ^{-1} 
({\bf{N}_1} + {\bf{N}_2}){\bf{N}_2} ^{-1}\tau )\alpha^T\right). 
\label{general-identity-alpha}
\eeqa
It is then easy to see, all  equations from (\ref{general-prod-wavefunction}) 
to (\ref{general-yukawa-series}) go through for arbitrary $\alpha$, giving 
the following expression for the Yukawa couplings:
\beqa
&&	Y_{ijk} = g \sigma_{abc} 
		\left(2^{\frac n2}  \right)^{\frac 12}
\left(Vol(T^{2n})\right)^{- {\frac 12}} 
\left[\frac{\left(|\det {\bf{N}_1}|.|\det {\bf{N}_2}| (Im \tau)^3\right)}
{|\det {\bf{N}_3}|}\right]^{\frac 14}  
\times \sum_{\vec{m}} \cr
&&\hskip 1cm\vt
\left[
\begin{array}{c}
(- \vec{j}+ \vec{k}){\bf{N}_2} \alpha^{-1} + \vec{m} {\bf{N}_1}{\bf{N}_3}^{-1} 
{\bf{N}_2} \alpha^{-1}\\ 0
\end{array}
\right]   
({0} |
\alpha ({\bf{N}_1} ^{-1} 
{\bf{N}_3}{\bf{N}_2} ^{-1}\tau) \alpha^T) . \hspace{1.0in}
\label{general-yukawa-arbi-alpha}
\eeqa  
where the sum $\vec{m}$ is now over all the integer solutions of 
$\vec{m} {\bf N_1} {\bf N_3}^{-1}$ in the cell given in 
eq. (\ref{unit-cell-arbitrary-alpha}). Explicit contributions to 
this sum, of course, will depend on the exact form of $\alpha$.
In subsection \ref{summation}, we have presented the case of 
$\alpha = det {\bf N_1} det {\bf N_2}$.

We now study how the above expression 
(\ref{general-yukawa-arbi-alpha}) reduces for another 
choice of  $\alpha$, such as:
\beq
\alpha = {\bf{N}_3} ^{-1}{\bf{N}_1}(\det {\bf{N}_2}.\det {\bf{N}_3}).
\label{choice-alpha}
\eeq
Note, for this choice of $\alpha$, that the degeneracy of states in the cell 
given in eq. (\ref{unit-cell-arbitrary-alpha}) is 
$det (det {\bf N_3} det {\bf N_2} {\bf N_2}^{-1})$. As a result, for the 
case of $2\times 2$ matrices for example, one now expects the sum over
$\vec{m}$ to run over $det{\bf N_2} det{\bf N_3}$ values. Explicit solutions are
now given as:
\beq
\vec{m} = \vec{p} det{\bf N_2} {\bf N_3} {\bf N_2}^{-1}
+\vec{\tilde{p}} det{\bf N_3} , 
\label{integer-m-in-cell-alpha}
\eeq
where $\vec{p}$ is all integer vectors within a cell generated by
$\vec{e} det{\bf N_3} {\bf N_3}^{-1}$
and $\vec{\tilde{p}}$ is all integer vectors within a cell generated
by $\vec{e} det{\bf N_2} {\bf N_2}^{-1}$. It is again easy to see that $\vec{m}$
satisfies $\vec{m} {\bf N_1} {\bf N_3}^{-1} = integer$ 
(by making use of ${\bf N_1} = {\bf N_3} - {\bf N_2}$).

The characteristic of the $\vt$-function in 
eq. (\ref {general-yukawa-arbi-alpha}), becomes:
\beqa \nonumber
(- \vec{j}+ \vec{k}){\bf{N}_2}
\alpha^{-1} &=& (-\vec{k} {\bf{N}_1} + \vec{i} {\bf{N}_1} + 
\vec{m} {\bf{N}_1}) \frac{{\bf{N}_1} ^{-1}{\bf{N}_3}}{(\det {\bf{N}_2}.\det {\bf{N}_3})} \\
 &= & \frac{(- \vec{k}+ \vec{i}){\bf{N}_3}}{(\det {\bf{N}_2}.\det {\bf{N}_3})},
\eeqa
where in the first equality we have made use of eq. (\ref{integer-constraint}).
Also we have,
\beqa \nonumber\hskip -0.8cm
\alpha ({\bf{N}_1} ^{-1} 
{\bf{N}_3}{\bf{N}_2} ^{-1}\tau) \alpha^T & = & 
 ({\bf{N}_3} ^{-1}{\bf{N}_1})({\bf{N}_1} ^{-1} 
{\bf{N}_3}{\bf{N}_2} ^{-1}) ({\bf{N}_1}{\bf{N}_3} ^{-1})  
\tau  (\det {\bf{N}_2}.\det {\bf{N}_3})^2 \\
&=&  ({\bf{N}_2} ^{-1} 
{\bf{N}_1}{\bf{N}_3} ^{-1}\tau) (\det {\bf{N}_2}.\det {\bf{N}_3})^2 .
\eeqa
The Yukawa couplings then read (following the exercise performed in 
subsection \ref{summation}):
\beqa
&&Y_{ijk} = g \sigma_{abc} 
		\left(2^{\frac n2}  \right)^{\frac 12}
\left(Vol(T^{2n})\right)^{- {\frac 12}} 
\left[\frac{\left(|\det {\bf{N}_1}|.|\det {\bf{N}_2}| (Im \tau)^3\right)}
{|\det {\bf{N}_3}|}\right]^{\frac 14}\times \cr 
&& \hskip -1cm \sum_{\vec{p}, \vec{\tilde{p}}}\vt
\left[
\begin{array}{c}
(- \vec{k}+ \vec{i})\frac{{\bf{N}_3}}
{\det{\bf{N}_2} \det{\bf{N}_3}} + 
(\vec{p}\frac{\bf N_3}{det {\bf N_3}} + 
\vec{\tilde{p}}\frac{\bf N_2}{det {\bf N_2}})
\\ 0
\end{array}
\right]  
({0} |
(\det{\bf{N}_2} \det{\bf{N}_3}) ^2({\bf{N}_2} ^{-1} 
{\bf{N}_1}{\bf{N}_3} ^{-1})\tau), 
\cr
&&
\label{general-yukawa-choi-alpha}
\eeqa
where the summation over indices $\vec{p}$ and $\vec{\tilde{p}}$ 
is explained earlier after eq. (\ref{integer-m-in-cell-alpha}).
We can also explicitly obtain the sums, as done for various examples
in the last subsection.

Now, a comparison of eqs. (\ref{general-yukawa-simple}) and 
(\ref{general-yukawa-choi-alpha}) shows a symmetry between the $\vt$-function 
characteristics in these cases, including the summation variables 
$\vec{p}$ and $\vec{\tilde{p}}$. 
It is obvious that the replacement 
 $\vec{i} \rightarrow \vec{j}, \vec{j} 
\rightarrow \vec{k}, \vec{k} \rightarrow \vec{i} $ 
and ${\bf{N}_1} \rightarrow{\bf{N}_2}, 
{\bf{N}_2} \rightarrow{\bf{N}_3}, {\bf{N}_3} \rightarrow{\bf{N}_1}$ 
in  eq. (\ref{general-yukawa-simple}) 
results  eq. (\ref{general-yukawa-choi-alpha}). We have thus established
that just as in the factorized case, for oblique fluxes too, one can 
show the cyclicity property of the Yukawa superpotential coefficients, as naively expected.

\subsection{General complex structure}\label{general-complex-structure}

In the previous subsections \ref{Riemann-Theta-Function-Identity} 
- \ref{yukawa-general}, 
we have confined ourselves to the 
complex structure matrix $\Omega = \tau I_n$ for a $2n$ dimensional 
torus. This implies the restriction to orthogonal tori, a solution 
which is already used in many phenomenologically interesting models. 
However, the results are easily generalized to 
complex structure with arbitrary $\Omega$. More precisely, to write 
down an identity generalizing eq. (\ref{general-identity}) 
one starts
with the product expression given in eq. (\ref{riemann1}) and rescales
${\bf{N}_1}$, ${\bf{N}_2}$ in eqs. (\ref{matrix}) - (\ref{unit-cell}) 
to ${\bf{N}_1}\Omega/{\tau}$, ${\bf{N}_2}\Omega/{\tau}$. At the same time,
the matrix $\alpha$ in eq. (\ref{def-alpha}) is also rescaled :
\beq
	\alpha \rightarrow \tilde{\alpha} 
	= \det {\bf{N}_1} \det {\bf{N}_2} \Omega/\tau = \frac{\alpha \Omega} {\tau}.
\label{def-alpha3}
\eeq
Moreover, one  needs to take into account that in relations such as 
(\ref{transmatrixT}) earlier, we have made use of the property
${\bf{N}}^T = {\bf{N}}$,
which is true for the complex structure of the form: $\tau I_n$.
Replacements: ${\bf{N}}\tau \rightarrow {\bf{N}}\Omega$ are, however, to be 
done in the original expression.

Explicitly, under the changes mentioned, the transformation matrix $T$ 
in eq. (\ref{transmatrix}) remains unchanged, while its transposition in 
eq. (\ref{transmatrixT}) is now written as:
\beq
\label{transmatrixT'}
T^ {T} = 
\left(
\begin{array}{cc}
1 &  {{\bf{N_1}} ^ {-1}}^T \alpha ^ {T}\\
1  & - {{\bf{N_2}}^{-1}}^T \alpha ^ {T}
\end{array}
\right).
\eeq 
Also, (\ref{transmatrixI}) is unchanged, whereas 
${\textbf{Q}}^{\prime}$ in eq. (\ref{trans4}) goes over to 
\beqa
{\textbf{Q}}^{\prime} \equiv T  \cdot \textbf{Q} \cdot T ^ {T} =  \left(
\begin{array}{cr}
 ({\bf{N_1}} + {\bf{N_2}})\Omega  & 0  \\
0  &  \alpha {({\bf{N_1}}^{-1} + {\bf{N_2}}^{-1})\Omega^T  } \alpha ^ {T}
\end{array}
\right)\, ,
\label{trans4-2}
\eeqa
where we have made use of the fact that
both $({\bf N_1 + N_2 }) \Omega$ and
$({\bf N_1}^{-1} + {\bf N_2 }^{-1}) \Omega^T$ are symmetric matrices, due 
to the condition (\ref{(0,2)=0-2}), with ${\bf N}$ defined in eq. (\ref{def-N}).
Then expressions  (\ref{trans1}) and (\ref{trans2}) 
remain unchanged, while (\ref{trans3}) is modified to:
\beqa
(T^{-1})^{T}\! (\vec{\textbf{j}} + \vec{\textbf{l}}) \!=\!\! \left( \!\!\!\begin{array}{c}
{{\bf{N}}_2^{-1}}^T({{\bf{N}}_1^{-1}}^T \!\!+\! 
{{\bf{N}}_2^{-1}}^T )^{-1}(\vec{j_1} + \vec{l_1}) + 
{{\bf{N}}_1^{-1}}^T({{\bf{N}}_1 ^ {-1}}^T 
\!\!+\! {{\bf{N}}_2 ^ {-1}}^T ) ^ {-1}(\vec{j_2} + \vec{l_2}) \\
(\alpha ^ {-1})^{T}({{\bf{N}}_1 ^ {-1}}^T \!\!+\! {{\bf{N}}_2 ^ {-1}}^T ) ^ {-1}
[(\vec{j_1} + \vec{l_1}) - (\vec{j_2} + \vec{l_2})]
\end{array}\!\!\!\! \right) 
\label{trans3'}
\eeqa

The identity (\ref{general-identity}) then takes the form:
\beqa 
& &
\hspace{-.2in}\vt
\left[
\begin{array}{c}
\vec{j_1} \\ 0
\end{array}
\right]
(\vec{z_1}| {\bf{N}_1}\Omega)
\cdot
\vt
\left[
\begin{array}{c}
\vec{j_2} \\ 0
\end{array}
\right]
(\vec{z_2}|  {\bf{N}_2}\Omega)   =  \\ \nonumber
& &
\sum_{\vec{m}}
\vt
\left[
\begin{array}{c}
(\vec{j_1}{\bf{N}_1} +\vec{j_2}{\bf{N}_2} +\vec{ m}.{\bf{N}_1} )({\bf{N}_1} 
+ {\bf{N}_2})^{-1} \\ 0
\end{array}
\right]
(\vec{z_1}+ \vec{z_2}| ({\bf{N}_1} + {\bf{N}_2})\Omega)\times   \\ \nonumber
&&
\hspace{-.1in}\vt
\left[
\begin{array}{c}
[(\vec{j_1} - \vec{j_2})+ \vec{m}]\frac{{\bf{N}_1}
({\bf{N}_1} + {\bf{N}_2})^{-1} {\bf{N}_2}}
{\det{\bf{N}_1} \det{\bf{N}_2}}\\ 0
\end{array}
\right]   \\ \nonumber
&  & 
 \hspace{-.0in}((\det{\bf{N}_1} \det{\bf{N}_2}) ({\bf{N}_1}^{-1}
\vec{z_1} - {\bf{N}_2}^{-1}\vec{z_2}) |
(\det{\bf{N}_1} \det{\bf{N}_2})^2 ( {\bf{N}_1} ^{-1} 
({\bf{N}_1} + {\bf{N}_2}){\bf{N}_2} ^{-1}\Omega^T)),
\label{general-identity2}
\eeqa
leading to the expression for the Yukawa interaction:
\beqa
&&Y_{ijk} = \sigma_{abc} g
		\left(2^{\frac n2}  \right)^{\frac 12}
\left(Vol(T^{2n})\right)^{- {\frac 12}} 
\left[\frac{\left(|\det {\bf{N}_1}|.|\det {\bf{N}_2}|
|\det \Omega| \right)}
{|\det {\bf{N}_3}|}\right]^{\frac 14} \times 
\sum_{\vec{p}, \vec{\tilde{p}}}\cr 
&&\hskip -1cm \vt
\left[
\begin{array}{c}
(- \vec{j}+ \vec{k})\frac{{\bf{N}_2}}
{\det{\bf{N}_1} \det{\bf{N}_2}} + 
(\vec{p}\frac{\bf N_2}{det {\bf N_2}} + 
\vec{\tilde{p}}\frac{\bf N_1}{det {\bf N_1}})\\ 0
\end{array}
\right]   
({0} |
(\det{\bf{N}_1} \det{\bf{N}_2}) ^2( {\bf{N}_1} ^{-1} 
{\bf{N}_3}{\bf{N}_2} ^{-1}\Omega^T)). 
\cr
&&
\label{general-yukawa-simple2}
\eeqa
We leave the rest of the details, which readers can work out.

\subsection{Hermitian intersection matrices}\label{hermitian}

In subsections \ref{Riemann-Theta-Function-Identity},
\ref{proof}, \ref{yukawa-general},
we have assumed that intersection matrices
${\bf{N}_1}, {\bf{N}_2}$ etc. are real symmetric.  
As explained, this restriction originates from the case when 
fluxes $p_{xx}$, $p_{yy}$ are zero and the intersection matrix 
${\bf N}$ is represented 
by the real matrix $p_{xy}$ in eq. (\ref{def-N}), which is symmetric 
whenever the  complex structure is of the canonical form: $\Omega = i I_d$.
Moreover, the Yukawa coupling expression was generalized nicely in the last
subsection to the case of arbitrary complex structure, as well.

In this subsection we discuss the case
when fluxes  $p_{xx}$ and $p_{yy}$ are also present, in addition to those of the type $p_{xy}$ and $p_{yx}$. Furthermore, all these 
fluxes are constrained by the conditions (\ref{(2,0)=0})
and (\ref{(0,2)=0}) giving a resulting $(1, 1)$ - form flux which 
can be represented by the Hermitian matrix (\ref{newF(1,1)-1}),
(\ref{newF(1,1)-2}). 
We explicitly present the case of $\Omega = iI_d$ solution 
($I_d$ : $d$-dimensional Identity matrix), which
is particularly simple, since in this case due to constraints
(\ref{constraint-tau=i}), the Hermitian flux has the simple final form
of eq. (\ref{11-tau=i}). 
The generalization to arbitrary complex structure $\Omega$ can also be 
done, but is left as an exercise to the reader.

Wavefunctions on $T^6$, as given in eq. (\ref{general-basis1}), satisfy 
the following field 
equations (\ref{Dirac-equation4}) and (\ref{difference-mapping}): 
\footnote{See Appendix A for details.}
\beq
\bar{\partial}_i \chi_+^{ab}  +
	(A^1 - A^2)_{\bar{z_i}} \chi_+^{ab}  = 0,\;\;\;\;(i=1,2,3). 
\label{diraceqnhermitian}
\eeq
We now show that the solution for the above equation, together with proper 
periodicity requirements on $T^6$, is given by the basis elements:
\beqa \nonumber
\psi^{\vec{j},\bf{N}}(\vec{z}) &= & \cn_{\vec{j}} \cdot  f (z,\bar{z})  
\cdot \hat{\varTheta} (z,\bar{z}) \\
& = & 
\cn_{\vec{j}} \cdot e^{i\pi [({\bf{N_R}} - i {\bf{N_I}}) \cdot \vec{z}]   
\cdot \pim \vec{z} }
\cdot 
\vt
\left[
\begin{array}{c}
\vec{j} \\ 0
\end{array}
\right]
\left({\bf{N_R}} \cdot \vec{z}\ |{\bf{N_R}} \cdot iI_3 \right)
\label{wavefunction-hermitian}
\eeqa
where ${\bf{N_R}}$ is a real, symmetric matrix.

The  wavefunction given in eq. (\ref{wavefunction-hermitian}) satisfies 
the Dirac equations  (\ref{diraceqnhermitian}) for the following 
gauge potentials.
\beq
(A^1 - A^2)_{\bar{z_j}} = \left( \frac{\pi}{2}\right)  z_i ({\bf{N_R}} -i {\bf{N_I}})_{i \bar{j}},
\eeq
which exactly matches with eq. (\ref{difference-mapping}) for the complex structure $\Omega = iI_3$.
The intersection matrix is therefore given by : 
\beq
{\bf{N}} =  {\bf{N_R}} - i {\bf{N_I}},
\label{NNRNI}
\eeq
where we identify, 
\beqa
  {\bf{N_R}} = p^a_{xy} - p^b_{xy},\qquad 
  {\bf{N_I}}=  p^a_{xx} - p^b_{xx}.
\label{NRNI}
\eeqa
The wavefunction described in eq. (\ref{wavefunction-hermitian}) can be re-written 
in terms of the real coordinates $\vec{x}$ and $\vec{y}$ as well as 
matrices ${\bf N_R}$, ${\bf N_I}$. By a slight abuse of notation, below, only 
for this subsection, we use ${\bf{N_R}} = p_{xy}$, 
$ {\bf{N_I}} =  p_{xx}$, by setting $p^b$'s to zero in eq. (\ref{NRNI})
and suppressing the superscript $a$ in $p^a$. Such a notational change,
 helps to make comparison of the transformation rules
we derive for the wavefunction written above in 
eq. (\ref{wavefunction-hermitian}) with general transition functions,
consistent with the gauge transformations along the $2n$ non-contractible
cycles of $T^n$, given in \cite{ibanez}. These transition functions 
are written in equations (4.40), (4.41) of \cite{ibanez}
for the fields that transform in fundamental representation 
rather than as bifundamentals. Hence, the notation changes above are meant to 
make the expressions consistent with the ones of \cite{ibanez}.

The wavefunction (\ref{wavefunction-hermitian}), in the real coordinates
$\vec{x}$ and $\vec{y}$, then reads:
\beqa \nonumber
\psi^{\vec{j},\bf{N}}(\vec{z}) &= &  \cn_{\vec{j}} \cdot
 e^{i\pi \left[ (x^i \cdot p_{x^iy^j} \cdot y^j ) +i (-x^i \cdot p_{x^ix^j} 
\cdot y^j + y^i \cdot p_{x^iy^j} \cdot y^j ) \right] } \\
& \cdot &  \sum_{l_i \in \inte^n} e^{i\pi (i) \left[ (l_i +j_i) 
\cdot p_{x^iy^j} \cdot (l_j +j_j)  \right] } 
e^{2i\pi \left[ (l_i +j_i) \cdot p_{x^iy^j} \cdot (x^j +i y^j)  \right] }. 
\label{wavefn-hermitian-xy}
\eeqa
This expression 
in  terms of real coordinates is 
useful in comparing  the  transformation properties of the wavefunction 
over $T^6$ with the one in \cite{ibanez}. 
The transformation properties, as derived from 
eq. (\ref{wavefunction-hermitian}), are given by,
\beqa 
 \begin{array}{rcl} \vspace{.2cm}
\psi^{\vec{j},{\bf{N}}}(\vec{z} + \vec{n})
& = &
e^{i\pi ([{\bf{N}} \cdot \vec{n}] \cdot  \pim \vec{z})}
\cdot \psi^{\vec{j},{\bf{N}}}(\vec{z} ),
\\ 
\psi^{\vec{j},{\bf{N}} }(\vec{z} + i \vec{n})
& = &
e^{-i\pi ([{\bf{N}}^t \cdot \vec{n}] \cdot  Re \vec{z})}
\cdot \psi^{\vec{j},{\bf{N}}}(\vec{z} ),
\end{array}
\label{trans1-hermitian-wvfn}
\eeqa
provided that
\begin{itemize}
\item $({\bf{N_R}})_{i\bar{j}} \equiv p_{x^iy^j} \in \inte$,  
i.e ${\bf{N_R}}$ is integrally quantized, 
\item $\vec{j}$ satisfies  $\vec{j} 
\cdot {\bf{N_R}} \in \inte^n$.     
\end{itemize}
We therefore notice that the integer quantization is imposed 
only on the  symmetric part ${\bf{N_R}}$ of the  intersection matrix 
from the periodicity of the wavefunction as well. However, Dirac 
quantization already imposes both $p_{xy}$ and $p_{xx}$ to be integral
for unit windings, as discussed in Section \ref{fluxes}.

 Using eq. (\ref{wavefn-hermitian-xy}), the expressions (\ref{trans1-hermitian-wvfn}) can be re-written in terms 
of real coordinates as:
\beqa 
\psi^{\vec{j},{\bf{N}}}(\vec{x} + \vec{n} + i \vec{y})
& = &
e^{i\pi [ n_i (p_{x^iy^j} - i p_{x^ix^j} ) y^j]}
\cdot \psi^{\vec{j},{\bf{N}}}(\vec{x} + i \vec{y} ),
\label{trans-herm-wvfn-x}
\eeqa
\beqa
\psi^{\vec{j},{\bf{N}}}(\vec{x}  + i [\vec{y}+ \vec{n}])
& = &
e^{-i\pi [ n_i (p_{x^jy^i} -i p_{x^jx^i} ) x^j ]} 
\cdot \psi^{\vec{j},{\bf{N}}}(\vec{x} + i \vec{y} ).
\label{trans-herm-wvfn-y}
\eeqa
In order to see that eqs. (\ref{trans-herm-wvfn-x}) and (\ref{trans-herm-wvfn-y})
 are the proper transformation properties 
of the fermion wavefunction over $T^6$,
let us compare them with the
the transition functions eq. (4.41) of \cite{ibanez} given for 
a fundamental representation in six real coordinates $X_I$, 
$I = 1, \cdots , 6$, as used in our eq. (\ref{general-flux}) as well. 
After changing variables first to the coordinates $x^i, y^i$, $i=1, 2, 3$ and then making
coordinate transformation to $z^i, i\bar{z^i}$, as described in Section  \ref{fluxes},
the general transition function is given by,
\beq
 \chi(x_i,y_i) = e^{i \pi [(m_i+ i n_i). F_{i \bar{j}} (y^j+ i x^j) + 
(im_i+ n_i). F_{ \bar{i} j} (x^j+ i y^j)]} .
\label{transitionfn}
\eeq
In correspondence to  the transformation along the 1-cycles, 
the integer parameters on $x_i$ and $y_i$ are denoted as 
$m_i$ and $n_i$ respectively. One then has two cases:\\
Case -I : When $n_i= 0$, i.e $\vec{x} \longrightarrow (\vec{x} + \vec{m})$, eq. (\ref{transitionfn}) reduces to
\beqa \nonumber
\chi(x_i,y_i) &= &e^{i \pi \{[ m_i.F_{i \bar{j}}. y^j - m_i.F_{ \bar{i} j}. y^j ] 
+ i [m_i.F_{i \bar{j}}. x^j + m_i.F_{ \bar{i} j}. x^j  ]\} }, \\
&=& e^{2i \pi ( m_i.F_{i \bar{j}}. y^j )}, 
\label{transitionfn-x}
\eeqa
where we used the hermiticity property of $F$. Using the expression 
(\ref{11-tau=i} ) in eq. (\ref{transitionfn-x}), we recover 
the transformation  given in eq. ( \ref{trans-herm-wvfn-x}).\\
Case -II : When $m_i= 0$ i.e $\vec{y} \longrightarrow (\vec{y} + \vec{n})$, 
eq. (\ref{transitionfn}) takes the form,
\beqa \nonumber
\chi(x_i,y_i) &= &  e^{i \pi \{ [- n_i  F_{i \bar{j}} x^j 
+ n_i F_{ \bar{i} j} x^j] + i [ n_i  F_{i \bar{j}} y^j  
+ n_i  F_{ \bar{i} j} y^j] \} }, \\
&=& e^{-2i \pi [ n_i.F_{i \bar{j}}. x^j ]} .
\label{transitionfn-y}
\eeqa
Again, using eq. (\ref{11-tau=i} ) in eq. (\ref{transitionfn-y}), we reproduce 
the transformation (\ref{trans-herm-wvfn-y}).

It can also be easily seen that the basis wavefunctions given in 
eqs. (\ref{wavefunction-hermitian}) and (\ref{wavefn-hermitian-xy}) satisfy 
the orthonormality condition
\beq
\int_{T^{2n}} (\psi^{\vec{k},{\bf{N}}})^{\dagger}
\psi^{\vec{j},{\bf{N}}} = \delta_{\vec{j},\vec{k}}\, ,  
\label{herm-wvfn-normn}
\eeq 
by fixing the normalization constant to
\beq
\cn_{\vec{j}} = \left( 2^n | {\rm det} {\bf{N_R}} 
|  \right)^{1/4} \cdot {\rm Vol} (T^{2n})^{-1/2},\quad \forall j \,\,.
\label{herm-wvfn-normn2}
\eeq
We have therefore confirmed that the wavefunction written in 
(\ref{wavefunction-hermitian}) is not only a solution of the 
field equation, but also has the correct periodicity properties on the
torus under the gauge transformation.
Now, regarding the Yukawa interaction, 
since only ${\bf{N_R}}$, which is real symmetric matrix, appears in 
the  $\hat{\varTheta} (z,\bar{z})$ part of the wavefunction (\ref{wavefunction-hermitian}), all the theta function identities 
described in Sections  \ref{Riemann-Theta-Function-Identity}, 
\ref{proof}
hold  for this new  $\hat{\varTheta} (z,\bar{z})$.
Similarly, as in the expression (\ref{general-yukawa-simple}), the 
Yukawa coupling $Y_{ijk}$ now has the following form,
\beqa
Y_{ijk} = g \sigma_{abc} 
		\left(2^{\frac n2}  \right)^{\frac 12}
\left(Vol(T^{2n})\right)^{- {\frac 12}} 
\left[\frac{\left(|\det {\bf{N}^1_R}|.|\det {\bf{N}^2_R}| \right)}
{|\det {\bf{N}^3_R}|}\right]^{\frac 14}\times 
\sum_{\vec{p}, \vec{\tilde{p}} }\hspace{1.5in}\cr
\vt
\left[
\begin{array}{c}
(- \vec{j}+ \vec{k})\frac{{\bf{N}^2_R}}
{\det{\bf{N}^1_R} \det{\bf{N}^2_R}}+ 
(\vec{p}\frac{\bf N^2_R}{det {\bf N^2_R}} + 
\vec{\tilde{p}}\frac{\bf N^1_R}{det {\bf N^1_R}})\\ 0
\end{array}
\right]  
({0} |
(\det{\bf{N}^1_R} \det{\bf{N}^2_R}) ^2({\bf{N}^1_R} ^{-1} 
{\bf{N}^3_R}{\bf{N}^2_R} ^{-1})\tau) 
\cr
&&
\hspace{-1cm}
\label{hermitian-yukawa-simple}
\eeqa
with $\vec{p}$ running over all the states inside the cell generated by
$\vec{e_1} det{\bf N^2_R} {\bf N^2_R}^{-1}$ 
and \\
 $\vec{e_2} det{\bf N^2_R} {\bf N^2_R}^{-1}$.
Similarly $\vec{\tilde{p}}$ runs over all the states inside the 
cell generated by
$\vec{e_1} det{\bf N^1_R} {\bf N^1_R}^{-1}$ 
and $\vec{e_2} det{\bf N^1_R} {\bf N^1_R}^{-1}$.

\subsection{Constraints on the results in section-{\ref{general-tori}} and further generalization}

To summarize, in this section we have given a close form expression 
for the Yukawa couplings in the magnetized brane constructions,
when in general both oblique and diagonal fluxes are present along 
the branes. However, the results of this section are somewhat 
restrictive, since the basis wavefunctions used for the computations
are well defined only when the intersection matrices satisfy  
a positivity condition given in eq.
(\ref{eq-riemann-conditions}) for arbitrary complex structure 
$\Omega$. A similar positivity criterion, for the case when 
$p_{x^ix^j}$ and $p_{y^iy^j}$ are nonzero, can be written using the
wavefunction (\ref{wavefunction-hermitian}), as well; it implies  simply the positivity of ${\bf N_R}$.

On the other hand, in realistic string model building,
 one may need intersection matrices that 
are not necessarily positive definite. The simplest examples
 correspond  simply to diagonal intersection matrices, 
having some positive and some negative elements along the diagonal.
 In such a factorized torus case, there is a 
unique prescription, to define the basis functions corresponding to
the  negative elements in the intersection matrix, as given in 
\cite{ibanez}, consisting of taking complex conjugates
of the wavefunctions  for the  positive elements. Such a prescription also works,
in the case of oblique + diagonal fluxes, when some intersection matrices
are `negative-definite' rather than being positive definite. One can then
take a complete complex conjugation over all the coordinates, in 
order to obtain a well defined wavefunction.

Such a process, however, does not work when oblique fluxes are present and 
intersection matrices have mixed eigenvalues. Note that
a diagonal flux of the type $F_{z^i\bar{z^i}}$ preserves its 
$(1, 1)$-form structure, under the interchange : $z^i \rightarrow \bar{z}^i$, 
required by supersymmetry. This is, however, no longer 
true when oblique fluxes are present, since off diagonal elements of 
a $(1, 1)$-form flux, say $F_{z^1 \bar{z^2}}$, does not remain 
of the $(1, 1)$ form when complex conjugation is taken only along 
$z^1$ or $z^2$. 

In order to cure the problem, one needs to construct new basis functions.
We present the results of our investigation in the next section, where
we first restrict to the case of a $T^4$ compactification, for simplicity. The complications arising from the
oblique nature of the fluxes are manifest in the
$T^4$ example as well, though it is possible to generalize the 
result to the full $T^6$, which  is discussed in Section \ref{T6-generalization}.

\section{Negative-chirality fermion  wavefunction}
\label{antichiral-wavefunction}

As already mentioned, the basis wavefunctions given in 
eq. (\ref{general-basis1}), used for deriving the Yukawa coupling 
expression in eq. (\ref{general-yukawa-simple2}),
are constrained by the Riemann conditions (\ref{eq-riemann-conditions}),
which imply in particular the positive-definiteness of the matrix ${\bf{N}} Im {\Omega}$.

Now, first restricting to $T^4$,
 we will show that the basis function  (\ref{general-basis1}) corresponds to the positive chirality
spinor on $T^4$.  On the other hand, to accommodate intersection matrices,
having two eigenvalues of opposite signature, one needs to find out the 
basis function corresponding to negative chirality spinor. The need 
to use such basis functions, for intersection matrices with mixed eigenvalues, can be easily seen  in the case when the $T^4$ factorizes into 
$T^2\times T^2$ and one turns on only non-oblique (diagonal) fluxes. In this case,
the intersection matrix has one positive 
diagonal element along the first $T^2$ and one negative diagonal element
along the second one. Good basis functions are then  products of 
two $T^2$ wavefunctions of opposite chiralities\cite{ibanez}, and  the total wavefunction 
on $T^4$ is  of negative chirality.

Our task therefore amounts to searching for the basis functions
corresponding to negative chirality spinors on $T^4$ with oblique fluxes.
Search for fermion wavefunctions in the presence of arbitrary fluxes
(in general oblique) has been pursued in
\cite{japan}. However, the resulting wavefunctions are presented in 
terms of diagonalized coordinates and eigenvalues of fluxes. 
Any such solution is however unsuitable for the Yukawa computation, 
both for the purpose of 
extracting the selection rules of the type given in 
eq. (\ref{integer-constraint}), 
as well as in actual evaluation, 
since the diagonalized coordinates become `stack dependent'
and  inherent nonlinearities involved in the diagonalization 
process appear in the wavefunctions, prohibiting the derivation of Yukawa couplings in a concrete form. 

In this section, we are able to write both the 
positive and negative chirality basis functions in  a `unified' 
fashion, by showing that all basis functions have a form similar
to the one given in eq. (\ref{general-basis1}). However,
the complex structure $\Omega$ appearing in eq. (\ref{general-basis1})
for a positive chirality wavefunction needs to be replaced by an
`effective' modular parameter matrix $\tilde{\Omega} = 
\hat{\Omega} \Omega$, in order to accommodate the negative chirality wavefunctions, where 
 $\hat{\Omega}$ is given in terms of the elements of the  intersection matrices (as explicitly obtained later). We also show that our results reduce to the 
ones in \cite{ibanez} for the case of diagonal fluxes.

First, in the  next subsection we present new basis functions, 
relevant for the situation when the intersection matrices
are neither positive nor negative definite. In a later subsection,
we show how the negative chirality spinor basis functions  can  be 
identified with the positive chirality ones given in 
eq. (\ref{general-basis1}),
with an effective modular parameter, defined in terms of the 
fluxes. We verify this fact
by mapping the wavefunctions into each other, as well as, by showing 
explicitly that the relevant field equations  transform into each other
through such a mapping. As a result, we are 
able to absorb the complications associated in the diagonalization process
of the modular parameter matrix, and
the final wavefunction once again has an identical form as 
given in eq. (\ref{general-basis1}), however, with a flux dependent 
modular parameter argument.

\subsection{Construction of the wavefunction}

In this subsection, as mentioned earlier, we discuss the case of 4-tori, though $T^6$ generalization
can be analyzed in a similar manner. We  first also restrict ourselves to the situation with canonical complex structure: $\Omega = i I_2$ and $\Omega = i I_3$ for $T^4$ and  $T^6$ respectively, where $I_d$ represents the $d$-dimensional identity matrix. The generalization to arbitrary $\Omega$ is given in subsections   \ref{eq-mapping} - \ref{T6-generalization}.
Now, in oder to  avoid the restriction to
the positivity condition  (\ref{eq-riemann-conditions}),
 we present an explicit solution of a wavefunction
of negative chirality satisfying both the equations of motion, as well 
as the periodicity requirements on $T^4$.

Going back to the positive chirality wavefunctions,
note that the two equations for the component $\chi_+^1$
in eq. (\ref{Dirac-equation3}) (derived from the original Dirac 
equation (\ref{Dirac-equation})) can be simultaneously solved, 
since when acting on $\chi_+^1$ with two covariant derivatives, we have:
$[D_{\bar{1}}, D_{\bar{2}}] \sim F^{ab}_{\bar{1}\bar{2}}$ and the RHS 
is zero, since all the $(0,2)$
components of the gauge fluxes are zero in order to maintain  
supersymmetry. The superscript ${ab}$ in this relation implies
that we need to take the difference of fluxes
in brane stacks $a$ and $b$ due to the combination $A^a - A^b$
that appears in eq. (\ref{Dirac-equation3}) for the bifundamental wavefunction.
Same is true for the two $\chi_+^2$ equations,
since $(2, 0)$ components of the fluxes are zero as well.
On the other hand, the relevant equations for the negative
chirality spinors are:
\beq
D_1 \chi_-^2 + D_2 \chi_-^1 = 0,
\eeq
\label{negative-chirality1}
and 
\beq
\bar{D}_2 \chi_-^2 - \bar{D}_1 \chi_-^1 = 0.
\label{negative-chirality2}
\eeq
When only one of the two components $\chi_-^{1, 2}$ is 
excited at a time, $\chi_-^{1, 2}$  satisfy: 
$\bar{D}_1 \chi_-^1 = D_2 \chi_-^1 = 0$  or
${D}_1 \chi_-^2 = \bar{D}_2 \chi_-^2 = 0$.
But none of  these sets of equations can be consistently solved
when oblique fluxes are present, since 
$[D_1, \bar{D}_2] \sim F_{1\bar{2}} \neq 0$.

The two negative chirality components $\chi_-^{1, 2}$
therefore need to be mixed up in order to obtain 
a solution of the relevant Dirac equations,
when  oblique fluxes are present. In other words, 
we need to simultaneously
excite both $\chi_-^{1, 2}$. Then, taking 
\beq
\chi_-^1 = \alpha \psi,\,\,\,
\chi_-^2 = \beta \psi, 
\label{chi-psi}
\eeq
equations (\ref{negative-chirality1})
and (\ref{negative-chirality2}) become:
\beq
( \beta \bar{D}_2 -\alpha\bar{D}_1 ) \psi = 0,
\label{negative1}
\eeq
 and 
\beq
(\beta {D}_1  + \alpha {D}_2) \psi = 0.
\label{negative2}
\eeq
In order for these two equations to have simultaneous solution,  one  
obtains the condition:
\beq
-\alpha \beta F^{ab}_{1\bar 1} - \alpha^2 F^{ab}_{2\bar{1}} + 
\beta^2 F^{ab}_{1\bar{2}} + \alpha \beta F^{ab}_{2\bar{2}} = 0,
\label{fab}
\eeq
where $F^{ab}_{i\bar{j}} \equiv {\bf{N}}_{i\bar{j}}$ is again 
the difference of fluxes in brane stacks $a$ and $b$ and 
${\bf{N}}_{i\bar{j}}$ is the same hermitian intersection matrix, 
eq. (\ref{NNRNI}), used in writing the positive chirality wavefunction and 
Yukawa couplings in eq. (\ref{general-basis1}), and other parts of 
Section \ref{general-tori}. When $p_{x^ix^j} = 0$, and $\Omega = i I_3$,
${\bf N}$ reduces to the real symmetric matrix as in eq. (\ref{def-N}).

Fortunately, equation (\ref{fab}) has arbitrary solutions of the type:
\beqa
     F^{ab} \equiv {\bf{N}} \equiv \hat{N}_{1\bar{1}}\begin{pmatrix}
 1 & -q \cr -q & q^2   \end{pmatrix} +
\tilde{N}_{2\bar{2}}\begin{pmatrix} 
q^2 & q \cr q & 1   \end{pmatrix} ,
\label{fmatrix1}
\eeqa
with $q = \frac{\beta}{\alpha} $ and $\hat{N}_{1\bar{1}}$, $\tilde{N}_{2\bar{2}}$ being arbitrary integers whose notation will become clear later (see eq. (\ref{intersection-nn}) below).
The RHS of the above relation is a general parameterization of 
a $2\times 2$ symmetric matrix, since the two terms can be 
written as 
\beqa
    F^{ab} \equiv {\bf{N}} \equiv  \hat{N}_{1\bar{1}}\begin{pmatrix}
 1 \cr -q \end{pmatrix}\begin{pmatrix} 1 & -q   \end{pmatrix} +
\tilde{N}_{2\bar{2}}
\begin{pmatrix}
  q \cr 1\end{pmatrix}\begin{pmatrix} q & 1   \end{pmatrix}. 
\label{fmatrix2}
\eeqa

After having shown the possible existence of the solution of the 
type (\ref{chi-psi}), we proceed to find  the explicit 
form of the wavefunction $\psi$ by applying the allowed 
orthogonal 
transformations on the wavefunction of the negative chirality fermion
on a $T^4$ which is factorized into $T^2\times T^2$. To obtain the 
explicit form of this orthogonal transformation, 
we start by writing the coordinate $T^4$ coordinate, $X^M = z^i,\bar{z}^i $ ($i=1,2$), in the 
spinor basis. We note, for the choice of 
Dirac Gamma matrices (in a real basis) 
given in eqs. (\ref{gammaz}), (\ref{gammabarz}) that 
\beqa
    \Gamma^{M} X_{M} = 
	\begin{pmatrix} & \bar{z}_1 & \bar{z}_2  & \cr 
	z_1 &  & & \bar{z}_2 \cr
			z_2 & &  & - \bar{z}_1 \cr 
		&z_2 & - \bar{z}_1  &  
\end{pmatrix},\;\;\;\;
\label{gammax}
\eeqa
with $z_i = x_i + i y_i$ and $\bar{z}_i = x_i - i y_i$, ($i=1, 2$), which 
factorizes into $2\times 2$ blocks providing the basis on which $SU(2)$'s 
in the Lorentz  group : $SU(2)_L \times SU(2)_R \sim  SO(1,3)$ act. 
 We get $x^i$ in the spinor basis in the form of a $2\times 2$
matrix:
 \beqa
	X_{\alpha \dot{\alpha}} = 
\begin{pmatrix} \bar{z}_1  &\bar{z}_2 \cr 
			z_2 & -z_1    \end{pmatrix} .
\label{degrees}
\eeqa

Now to understand the transformation properties of the 
fermions on $T^4$, we
consider the following transformations on $X_{\alpha \dot{\alpha}}$: \\
\beqa
\begin{pmatrix}  e^{i\theta_1} & 0  \cr 
			0 &  e^{-i\theta_1}   \end{pmatrix}
\begin{pmatrix} \bar {z_1} & \bar {z_2} \cr
                          z_2  & -z_1   \end{pmatrix}
\begin{pmatrix}  e^{-i\theta_2} & 0  \cr 
			0 &  e^{-i\theta_2}   \end{pmatrix}
= \begin{pmatrix}  e^{i(\theta_1-\theta_2)}\bar {z_1} 
&  e^{i(\theta_1+\theta_2)}\bar {z_2} \cr 
e^{-i(\theta_1+\theta_2)} z_2  & 
- e^{-i(\theta_1-\theta_2)} z_1   \end{pmatrix}
\label{Ztransformation}
\eeqa
\\
We learn from eq. (\ref{Ztransformation}) that when $T^4$ factorizes into
$T^2\times T^2$, the transformations of the positive and negative chirality
fermions on the two $T^2$'s can be read off from the transformation rules 
of $z_1$ and $z_2$ given above. Indeed, the transformation rules for the 
fermions $\psi_{\pm}^{(i)}$  on the two $T^2$'s, denoted by indices
$i=1, 2$ are:
\beqa
\psi_+^{(1)} \longrightarrow  
e^{-i \frac{(\theta_1-\theta_2)}{2}} \psi_+^{(1)}; \,\,\,\,
\psi_-^{(1)} \longrightarrow  
e^{i \frac{(\theta_1-\theta_2)}{2}} \psi_-^{(1)}, \cr
\psi_+^{(2)} \longrightarrow  
e^{-i \frac{(\theta_1+\theta_2)}{2}} \psi_+^{(2)}; \,\,\,\,
\psi_-^{(2)} \longrightarrow  
e^{i \frac{(\theta_1+\theta_2)}{2}} \psi_-^{(2)}.
\eeqa

In this case, as described in the 
section \ref{factorized-wavefunction},  the $T^4$ 
fermion wavefunctions can be written as a direct product 
of the ones on two $T^2$'s as in eq. 
(\ref{Dwavefunction-product}).
We obtain the transformation of 
$T^4$ wavefunctions (eq. (\ref{Dwavefunction-product})):
\beqa
\Psi_+^{1} \longrightarrow  e^{-i \theta_1} \Psi_+^{1}, \,\,\,
\Psi_+^{2} \longrightarrow  e^{i \theta_1} \Psi_+^{2}, \cr
\Psi_-^{1} \longrightarrow  e^{i \theta_2} \Psi_-^{1}, \,\,\,
\Psi_-^{2} \longrightarrow  e^{-i \theta_2} \Psi_-^{2}. 
\eeqa
It follows that  a left transformation ($\theta_1 \neq 0,\theta_2 = 0  $) acts 
independently on (left handed) positive chirality  wavefunctions, 
and a right transformation ($\theta_1 = 0, \theta_2 \neq 0 $)  acts on the 
negative-chirality (right handed)
wavefunctions. Now, consider the following complex  transformation 
on vectors in spinor basis:\\
\beqa
\begin{pmatrix} \bar {z_1} & \bar {z_2} \cr
                          z_2  & -z_1   \end{pmatrix} \longrightarrow
\begin{pmatrix}  a & b  \cr 
		c & d   \end{pmatrix}
\begin{pmatrix} \bar {z_1} & \bar {z_2} \cr
                          z_2  & -z_1   \end{pmatrix}
\begin{pmatrix}  e & f  \cr 
		g &  h   \end{pmatrix}
\eeqa
\\
Case-I: For $e=h=1$, $f=g=0 $, $c=-b$,  $a=d $,  i.e a left transformation   results in the following orthogonal coordinate  transformation,
\beqa
z_1 \longrightarrow a z_1 + b \bar {z_2} ; \,\,\,\,
z_2 \longrightarrow a z_2 - b \bar {z_1}.
\label{case1}
\eeqa 
Case-II: Similarly, for $a=d=1$, $c=b=0 $, $h=e$, $f=-g $, i.e a right transformation   leads to
\beqa
z_1 \longrightarrow e z_1 - f z_2 ; \,\,\,\,
z_2 \longrightarrow e z_2 + f z_1 .
\label{case2}
\eeqa 

In order to maintain the holomorphicity of the gauge fluxes, one therefore 
needs to make use of the later transformation, in order to generate 
a general wavefunction, starting with the one which corresponds to the 
diagonal (non-oblique) flux. In addition, we need to maintain the 
integrality of the fluxes, as we make such orthogonal transformations.
However, in our case, we do not make use of any specific form of the 
transformation and rather use the above analysis as a  guide  for writing down a  general solution. We then verify the equations of motion directly,
in order to confirm that  the solution  we propose is indeed the correct one.

\subsection{New wavefunction}\label{new-wavefunction}

We now  use the  transformation (\ref{case2})
to obtain the wavefunction associated with the negative chirality fermion 
bifundamentals, starting with a wavefunction associated with 
a negative chirality spinor for a diagonal flux. In the notations
of eq. (\ref{Dwavefunction}), it corresponds to exciting only 
the negative chirality component 
\beqa
 \begin{pmatrix} \Psi_-^2 \cr \Psi_-^1 \end{pmatrix} = 
\begin{pmatrix} \psi \cr 0 \end{pmatrix}.
\label{diagonal}
\eeqa 
We ignore the explicit form of $\psi$, except to note that after the 
transformation (\ref{case2}), one generates 
\beqa
\begin{pmatrix}  \psi \cr 0  \end{pmatrix} \longrightarrow
\begin{pmatrix}  \Psi^2_-  \cr \Psi^1_-   \end{pmatrix} =
\begin{pmatrix} \beta \psi  \cr \alpha \psi \end{pmatrix}, 
\label{negative}
\eeqa
while ($\Psi_+^1$, $\Psi_+^2$) remain zero.
In the gauge sector,
such wavefunctions are parameterized in the bifundamental 
representations by:
\beqa
    \Psi_{ab}  = \begin{pmatrix}C_{n_a} & \chi_{ab}\cr  & C_{n_b}
		\end{pmatrix},
\label{bi-wavefunction-text}
\eeqa
as also given in eq. (\ref{bi-wavefunction}). For  negative chirality
components, the equations to be satisfied by the various 
components  are: (see eq. (\ref{Dirac-equation2}))
\beqa
{\partial}_1 \chi_-^2 + \partial_2 \chi_-^1 + 
	(A^1 - A^2)_{{z_1}} \chi_-^2  +
	(A^1 - A^2)_{{z_2}} \chi_-^1  = 0, \cr
\bar{\partial}_2 \chi_-^2 - \bar{\partial}_1 \chi_-^1 + 
	(A^1 - A^2)_{\bar{z_2}} \chi_-^2  -
	(A^1 - A^2)_{\bar{z_1}} \chi_-^1  = 0.
\label{Dirac-equation5}
\eeqa 

We now show that the solution to eqs. (\ref{Dirac-equation5}), together with 
proper periodicity requirements on $T^4$, 
is given by the basis elements:
\beqa
\psi^{\vec{j},\hat{{\bf{N}}},{\bf{\tilde{N}}}} = 
\cn \cdot  f (z,\bar{z})  \cdot \hat{\varTheta} (z,\bar{z})
\label{anti-chiral-wavefunction}
\eeqa
where,
\beqa
f (z,\bar{z}) = 
e^{i\pi [(\hat{{\bf{N}}}_{i\bar{j}} z_i Im {z_j} )- 
({\bf{\tilde{N}_{i\bar {j}}}} \bar{z_i} Im{\bar{z_j}})] } \,\,,
\label{fzpart}
\eeqa
\beqa
\hat{ \varTheta} (z,\bar{z}) =  
\sum_{m_1, m_2 \in \inte^n} e^{\pi i(i)[ (m_i+j_i){{\bf{M}}}_{i\bar {j}}(m_j+j_j) ]} 
e^{2\pi i[ (m_i+j_i)\hat{{\bf{N}}}_{i\bar{j}} z_j }
 e^{2\pi i(m_i+j_i){\bf{\tilde{N}_{i\bar{j}}}} \bar{z_j} }\,\,,
\label{thetapart}
\eeqa
with
\beqa
 {\bf{M}}_{i\bar{j}}=\hat{{\bf{N}}}_{i\bar{j}} -{\bf{\tilde{N}_{i\bar{j}}}} 
\label{mm}
\eeqa
where  both $\hat{{\bf{N}}}$, ${\bf{\tilde{N}}}$ are real, symmetric 
matrices, given earlier in eq. (\ref{fmatrix1}), and so also is ${\bf{M}}$
(${\bf{M}}_{i\bar{j}} = {\bf{M}}_{j\bar{i}}$). We retain, however, both types of
indices: $i$ and $\bar{j}$ to incorporate real as well as complex components of
the $(1, 1)$-form fluxes $F_{i\bar{j}}$. Also, an extra factor of $i$ in 
the exponent of $\hat{ \varTheta} (z,\bar{z})$ corresponds to the fact that
we are working with the canonical complex structure : $\Omega = i I_2$ for 
the present example of the fermion wavefuncton on $T^4$.

The wavefunction  (\ref{anti-chiral-wavefunction}) satisfies the Dirac 
equations (\ref{Dirac-equation5}) for the 
following gauge potentials:
\beqa
(A^1 - A^2)_{\bar{z_1}} = (\hat{{\bf{N}}}_{1\bar1} +
{\bf{\tilde{N}_{1\bar1}}})z_1 + (\hat{{\bf{N}}}_{1\bar2} +
{\bf{\tilde{N}_{1\bar2}}} ) z_2 \cr 
(A^1 - A^2)_{\bar{z_2}} = (\hat{{\bf{N}}}_{1\bar2} +{\bf{\tilde{N}_{1\bar2}}} ) z_1 + (\hat{{\bf{N}}}_{2\bar2} +{\bf{\tilde{N}_{2\bar2}}}) z_2.
\label{gaugepotential}
\eeqa
The intersection matrix ${\bf{N}}$ is therefore given by:
\beq
{\bf{N}} = \hat{{\bf{N}}} + {\bf{\tilde{N}}}, 
\label{intersection-nn}
\eeq
as appearing previously in eqs. (\ref{fab}), (\ref{fmatrix1}).
Also, we  have imposed the following constraints, 
in order to retain the holomorphicity of gauge potentials:
\beqa
\frac{\alpha}{\beta} = \frac{-\hat{{\bf{N}}}_{1\bar1}}{\hat{{\bf{N}}}_{1\bar2}} 
= \frac{-\hat{{\bf{N}}}_{1\bar2}}{\hat{{\bf{N}}}_{2\bar2}} = 
 \frac{{\bf{\tilde{N}}}_{1\bar2}}{{\bf{\tilde{N}}}_{1\bar1}} = 
\frac{{\bf{\tilde{N}}}_{2\bar2}}{{\bf{\tilde{N}}}_{1\bar2}} = \frac{1}{q}.
\label{alphabetaconstraint}
\eeqa
Note that the ratios of the matrix elements of $ \hat{{\bf{N}}}$ 
and ${\bf{\tilde{N}}}$  are identical to those given in eq. (\ref{fmatrix1}). 
We have  therefore explicitly shown that the solution given in 
eqs. (\ref{anti-chiral-wavefunction}) - (\ref{thetapart}) satisfies the equations of motion.
The transformation properties of this wavefunction  (\ref{anti-chiral-wavefunction})  along the four 1-cycles of 
$T^4$, are given by:
\beqa
 \begin{array}{rcl} \vspace{.2cm}
\psi^{\vec{j},\hat{{\bf{N}}},{\bf{\tilde{N}}}}(\vec{z} + \vec{n})
& = &
e^{i\pi ([\hat{{\bf{N}}} \cdot \vec{n}] \cdot  \pim \vec{z}-[{\bf{\tilde{N}}} \cdot \vec{n}] \cdot  \pim \vec{\bar{z}})}
\cdot \psi^{\vec{j}, \hat{{\bf{N}}},{\bf{\tilde{N}}}}(\vec{z} ),
\label{trans1psin}
\\ 
\psi^{\vec{j},\hat{{\bf{N}}},{\bf{\tilde{N}}} }(\vec{z} + i \vec{n})
& = &
e^{-i\pi ([\hat{{\bf{N}}} \cdot \vec{n}] \cdot  Re \vec{z}  + [{\bf{\tilde{N}}} \cdot \vec{n}] \cdot  Re \vec{\bar{z}})}
\cdot \psi^{\vec{j}, \hat{{\bf{N}}},{\bf{\tilde{N}}}}(\vec{z} ),
\end{array}
\label{trans2psin}
\eeqa
provided that
\begin{itemize}
\item ${\bf{N}_{i\bar{j}} } \equiv (\hat{{\bf{N}}}+{\bf{\tilde{N}}})_{i\bar{j}} \in \inte$,  
i.e $(\hat{{\bf{N}}}+{\bf{\tilde{N}}})$ is integrally quantized, 
\item $\vec{j}$ satisfies:   $\vec{j} 
\cdot (\hat{{\bf{N}}}+{\bf{\tilde{N}}}) \in \inte^n$.     
\end{itemize}
We therefore notice that the integer quantization is imposed 
only on the intersection matrix ${\bf{N}}$ given in 
eq. (\ref{intersection-nn}) and does not necessarily hold for the matrix
${\bf{M}}$ in eq. (\ref{mm}). Explicitly, we have:
\beqa  \nonumber
{\bf{N}} = \hat{{\bf{N}}} + {\bf{\tilde{N}}} = \hat{{\bf{N}}}_{1 \bar1}
\begin{pmatrix} 1 & -q \cr
                          -q  & q^2  \end{pmatrix} + {\bf{\tilde{N}}}_{2\bar2} 
\begin{pmatrix} q^2 & q \cr
                          q  & 1  \end{pmatrix},  \\
{\bf{M}} = \hat{{\bf{N}}} - {\bf{\tilde{N}}} = \hat{{\bf{N}}}_{1 \bar1}
\begin{pmatrix} 1 & -q \cr
                          -q  & q^2  \end{pmatrix} - {\bf{\tilde{N}}}_{2\bar2} 
\begin{pmatrix} q^2 & q \cr
                          q  & 1  \end{pmatrix} \,\, ,
\label{NM}
\eeqa
where the first eq. in (\ref{NM}) is identical to the solutions in 
eq. (\ref{fmatrix1}). 

Note that the wavefunction given in 
eqs. (\ref{anti-chiral-wavefunction}), (\ref{fzpart}) 
and (\ref{thetapart}) is now well defined, as the series expansion in 
eq. (\ref{thetapart}) is now convergent. To show this, we note the following
relation:
\beq
	\det {\bf{N}} = - \det {\bf{M}} = 
\hat{{\bf{N}}}_{1 \bar1}  {\bf{\tilde{N}}}_{2\bar2} (1 + q^2)^2.
\label{det-mn}
\eeq
As a result, in the case when $\det {\bf{N}}$  is negative
( when $\bf{N}$ has two eigenvalues of opposite signatures),
$\det {\bf{M}} > 0$. So, the series  (\ref{thetapart}) is now convergent
when the two eigenvalues are of positive signature,
since it is the quadratic part, in the summation index 
in theta series, that 
dominates in the exponent of this expansion.
An overall complex conjugation will be required,
for the case when  two eigenvalues are negative rather than 
positive.

\subsection{Normalization}\label{normalization}

Now that we have found a basis of  wavefunctions, classified by the index $j_i$ in the exponent in (\ref{thetapart}), we proceed to show its orthonormality. The wavefunctions described in 
eqs. (\ref{anti-chiral-wavefunction}), (\ref{fzpart}), 
(\ref{thetapart}) can be re-written  in terms of the real 
coordinates $ \vec{x}$ and $\vec{y} $ as follows:
\beqa
\psi^{\vec{j},{\bf{N}},{\bf{M}}} = 
\cn_ {\vec{j}} \cdot e^{i\pi[\vec{x} \cdot {\bf{N}} \cdot \vec{y} 
+ i  \vec{y} \cdot  {\bf{M}} \cdot \vec{y} ] }
\sum_{\vec{m} \in \inte^n} e^{\pi i(i)[(\vec{m}+\vec{j})  
\cdot {\bf{M}}  \cdot (\vec{m}+\vec{j})]}
e^{2\pi i[ (\vec{m}+\vec{j}) \cdot  {\bf{N}}  \cdot \vec{x} + i (\vec{m}+\vec{j})  \cdot {\bf{M}}  \cdot  \vec{y}]} .
\label{xywavefunction}
\eeqa
Then the following orthonormality conditions are satisfied:
\beq
\int_{T^{4}} (\psi^{\vec{k},{\bf{N}},{\bf{M}}})^* 
\psi^{\vec{j},{\bf{N}},{\bf{M}}} = \delta_{\vec{j},\vec{k}}  .
\label{normn}
\eeq
To verify the orthogonality relation and obtain the normalization factor,
we note that,
in terms of the wavefunctions (\ref{xywavefunction}) we have:
\beqa \nonumber
&(\psi^{\vec{k},{\bf{N}},{\bf{M}}})^* \psi^{\vec{j},{\bf{N}},{\bf{M}}} & = \cn_ {\vec{k}} \cdot e^{-i\pi[\vec{x}  \cdot {\bf{N}} \cdot \vec{y} - i  \vec{y} \cdot  {\bf{M}} \cdot \vec{y} ] }
\sum_{\vec{l} \in \inte^n} e^{\pi i(i)[(\vec{l}+\vec{k})  
\cdot {\bf{M}}  \cdot (\vec{l}+\vec{k})]} \cdot
e^{-2\pi i[ (\vec{l}+\vec{k}) \cdot  {\bf{N}}  \cdot \vec{x} - i (\vec{l}+\vec{k})  \cdot {\bf{M}}  \cdot  \vec{y}]} \\  \nonumber
&& \cn_ {\vec{j}} \cdot e^{i\pi[\vec{x} \cdot {\bf{N}} 
\cdot \vec{y} + i  \vec{y} \cdot  {\bf{M}} \cdot \vec{y} ] }
\sum_{\vec{m} \in \inte^n} e^{\pi i(i)[(\vec{m}+\vec{j})  
\cdot {\bf{M}}  \cdot (\vec{m}+\vec{j})]} \cdot 
e^{2\pi i[ (\vec{m}+\vec{j}) \cdot  {\bf{N}}  
\cdot \vec{x} + i (\vec{m}+\vec{j})  \cdot {\bf{M}}  
\cdot  \vec{y}]} \\ \nonumber
&& = \cn_ {\vec{j}} \cn_ {\vec{k}} \cdot e^{-2\pi(\vec{y} 
\cdot  {\bf{M}} \cdot \vec{y} )}
\sum_{\vec{m}, \vec{l} \in \inte^n} e^{\pi i(i)[(\vec{m}+\vec{j})  
\cdot {\bf{M}}  \cdot (\vec{m}+\vec{j})]} \cdot  
e^{\pi i(i)[(\vec{l}+\vec{k})  \cdot {\bf{M}}  \cdot (\vec{l}+\vec{k})]} \\
&& e^{2\pi i[ (\vec{m}+\vec{j}) - (\vec{l}+\vec{k})] 
\cdot  {\bf{N}}  \cdot \vec{x}} \cdot
e^{2\pi i(i)[ (\vec{m}+\vec{j}) + (\vec{l}+\vec{k})] 
\cdot  {\bf{M}}  \cdot \vec{y}} .
\label{norm2}
\eeqa
The integration over $ \vec{x}$ in eq. (\ref{normn}) imposes the condition 
$\vec{j} = \vec{k} $ and equality on the summation 
indices  $\vec{m}= \vec{l}$. In particular, the 
condition $\vec{j} = \vec{k} $ gives our orthogonality
condition (\ref{normn}).  One can now obtain the normalization factor by performing the
integration:
\beqa \nonumber
 \int_0^1 d^2\vec{y} && \left[
e^{-2\pi \vec{y} \cdot  {\bf{M}} \cdot \vec{y} } \sum_{\vec{m} \in \inte^n} 
e^{-2\pi (\vec{m}+\vec{j})  \cdot {\bf{M}}  \cdot (\vec{m}+\vec{j}) } \cdot 
e^{-4\pi (\vec{m}+\vec{j})  \cdot {\bf{M}}  \cdot \vec{y} } \right]\\
&& =   \int_0^1 d^2\left(\vec{y}\right) \left[\sum_{\vec{m} \in \inte^n}  
e^{-2\pi \left((\vec{m}+\vec{j}) + 
\vec{y} \right)  \cdot {\bf{M}}  \cdot  
\left((\vec{m}+\vec{j}) + \vec{y} \right) } \right].
\label{intx}
\eeqa
One can integrate over $ \vec{y}$, using
\beqa \nonumber
\int_0^1 d^2\vec{y} \left[ \sum_{\vec{m} \in \inte^n}  
e^{-2\pi \left((\vec{m}+\vec{j}) + \vec{y} \right)  
\cdot {\bf{M}}  \cdot  \left((\vec{m}+\vec{j}) + \vec{y} \right) }\right]  & = &
\sum_{\vec{m} \in \inte^n} \int_0^1 d^2\vec{y} 
\left[ e^{-2\pi \left[(\vec{m}+\vec{j}) + \vec{y} \right]  
\cdot {\bf{M}}  \cdot  \left( (\vec{m}+\vec{j}) + \vec{y} \right) }\right]  \\
& = & \int_{-\infty}^\infty d^2\vec{y^ \prime} 
\left[  e^{-2\pi \vec{y'} \cdot {\bf{M}}  \cdot \vec{y'} } \right] 
\label{inty}
\eeqa
The integration (\ref{inty}) fixes then the normalization constant to 
\beq
\cn_{\vec{j}} = \left( 2 | {\rm det} {\bf{M}} 
|  \right)^{1/4} \cdot {\rm Vol} (T^{4})^{-1/2},\quad \forall j.
\label{normn2}
\eeq

\subsection{Eigenfunctions of the Laplace equation}\label{laplace}

The wavefunctions (\ref{anti-chiral-wavefunction}) not only represent 
zero modes of the Dirac operator, but are also  eigenfunctions of the  Laplacian.
In order to see this, we start with computing the Dirac operator in four dimensions.
In our notations:
\beqa
 \Gamma^{\mu} \partial_{\mu} = 
	\begin{pmatrix} & \bar{\partial}_1 & 
\bar{\partial}_2 & \cr \partial_1 &  & & \bar{\partial}_2 \cr
		 \partial_2 & &  & - \bar{\partial}_1 \cr 
	& \partial_2  & -\partial_1  &  \end{pmatrix},\;\;\;\;
\label{diracop}
\eeqa
which leads to
\beqa \nonumber
\left(\Dbar\right)^2 & = & 
\begin{pmatrix}  \bar{D}_1 D_1 +  \bar{D}_2 D_2 &  & & \cr 
                 & D_1 \bar{D}_1 + \bar{D}_2 D_2  & &  \cr
		 & &  D_2 \bar{D}_2 + \bar{D}_1 D_1  &  \cr
                 & &  & D_1 \bar{D}_1 + D_2 \bar{D}_2  \end{pmatrix} \\ \nonumber
\\
& = & \Delta + \begin{pmatrix}  F_{1\bar1} + F_{2\bar2} &  & & \cr 
                 & - F_{1\bar1} + F_{2\bar2} & &  \cr
		 & &  F_{1\bar1} - F_{2\bar2} &  \cr
                 & &  & -(F_{1\bar1} + F_{2\bar2})  \end{pmatrix} \,\,.
\label{diracopsq}
\eeqa
The Dirac equation  $ \Dbar \Psi = 0 $, 
with $\Psi$  given in eq. (\ref{Dwavefunction-product}), 
implies that such basis functions are also  eigenfunctions of the Laplacian  $\Delta$.
 The question whether  massless scalars exist, depends
on whether some combination of fluxes  appearing  in eq. (\ref{diracopsq})
vanish. Of course, their existence is guaranteed in the supersymmetric case.

 \subsection{Mapping of basis functions from positive to negative 
chirality}\label{mapping-positive-negative}

We now show that the basis for the negative chirality wavefunction,
given in eqs. (\ref{anti-chiral-wavefunction}), (\ref{fzpart}), 
(\ref{thetapart})  can in fact be obtained by a mapping from the 
basis of the positive chirality wavefunction given in 
eq. (\ref{general-basis1}). We also present the mapping between 
the corresponding field equations. Our mapping reduces to the ones in 
\cite{ibanez} for the case of factorized tori.

More precisely, we show that
our negative chirality wavefunction, given in eqs. 
(\ref{anti-chiral-wavefunction}), (\ref{fzpart}), 
(\ref{thetapart}), as well as (\ref{xywavefunction}) (for a trivial modular parameter matrix : $\Omega = i I_2$) is identical to the positive chirality wavefunction  (\ref{general-basis1}) for a `nontrivial'
(flux dependent) modular parameter matrix
$\Omega = i \hat{\Omega}$. Explicitly, $\hat{\Omega}$ is 
given in terms of the ratios ($q$) of flux components. This result gives
a `unified' picture of all the relevant basis functions.
Later on, in Section \ref{mapping-arbitrary}, we show that a  
similar mapping  holds for nontrivial complex structure
on $T^4$, by examining the equations of motion.


Let us write down explicitly the wavefunction (\ref{general-basis1}) for 
complex structure with arbitrary $ \Omega $ ($= i \hat{\Omega} $).
\beqa \nonumber
  \psi^{\vec{j}, {\bf{N'}}} (\vec{z}, {{\Omega}}) & = &
\cn \cdot e^{i\pi [(\vec{x} + i \hat{\Omega}\vec{y}) . {\bf{N'}} 
\hat{\Omega}^{-1}. \hat{\Omega}\vec{y} ]} \cdot 
\sum_{\vec{m} \in \inte^n} e^{i \pi [(\vec{m}+\vec{j}) . 
i {\bf{N'}} \hat{\Omega}. (\vec{m}+\vec{j}) ]} 
e^{2i \pi [(\vec{m}+\vec{j}) ({\bf{N'}} \vec{x} +  
i {\bf{N'}} \hat{\Omega}.\vec{y}) ] } \\
& \sim & e^{i\pi [\vec{x}.{\bf{N'}}.\vec{y} + 
i \hat{\Omega} \vec{y} .{\bf{N'}}.\vec{y}]}
\cdot \sum_{\vec{m} \in \inte^n} 
e^{i \pi [(\vec{m}+\vec{j}) . i {\bf{N'}} \hat{\Omega}. 
(\vec{m}+\vec{j}) ]} e^{2i \pi [(\vec{m}+\vec{j}) ({\bf{N'}} \vec{x} 
+  i {\bf{N'}} \hat{\Omega}.\vec{y}) ]}, 
\label{+vechiralwf}
\eeqa
where ${\bf{N}}$ is changed to $ {\bf{N'}}$ to show  a distinction 
between the two wavefunctions for the purpose of defining the 
mapping as given below.
Next consider the negative chirality wavefunction (\ref{xywavefunction}), 
written in terms of real coordinates $\vec{x}$ and $ \vec {y}$, 
\beqa
\psi^{\vec{j},{\bf{N}},{\bf{M}}} \sim e^{i\pi[\vec{x} 
\cdot {\bf{N}} \cdot \vec{y} + i  \vec{y} \cdot  {\bf{M}} \cdot \vec{y} ] }
\sum_{\vec{m} \in \inte^n} e^{\pi i(i)[(\vec{m}+\vec{j})  
\cdot {\bf{M}}  \cdot (\vec{m}+\vec{j})]}
e^{2\pi i[ (\vec{m}+\vec{j}) \cdot  {\bf{N}}  
\cdot \vec{x} + i (\vec{m}+\vec{j})  \cdot {\bf{M}}  \cdot  \vec{y}]}.
\label{-vechiralwf}
\eeqa
It is now easy to check that  the above equations (\ref{+vechiralwf}) 
and (\ref{-vechiralwf}) precisely match  with the following identification :
\beqa \nonumber
&&{\bf{N}} = \hat{{\bf{N}}} + {\bf{\tilde{N}}} = {\bf{N'}} ,\\ 
&&{\bf{M}} = \hat{{\bf{N}}} - {\bf{\tilde{N}}} 
= {\bf{N'}} \hat{\Omega} \Rightarrow \hat{\Omega} =
{\bf{N}}^{-1} {\bf{M}},
\label{matchwf}
\eeqa
with $\Omega = i \hat{\Omega}$, and $\hat{\Omega}$ is a real matrix.
For the ${\bf{N}}$ and ${\bf{M}}$, defined in eq. (\ref{NM}), 
${\bf{N}}^{-1} $ and $\hat{\Omega} $   are given by;
\beqa
{\bf{N}}^{-1} = \frac{1}{(1+q^2)^2 }\left[ \frac{1}{\hat{{\bf{N}}}_{1\bar1}} 
\begin{pmatrix} 1 & -q \cr
                          -q  & q^2  \end{pmatrix} +
\frac{1}{{\bf{\tilde{N}}}_{2\bar2}}
\begin{pmatrix} q^2 & q \cr
                         q  & 1  \end{pmatrix} 
\right] ,
\eeqa
\beqa
\hat{\Omega}  = \frac{1}{(1+q^2)} \begin{pmatrix} 1-q^2 & -2q \cr
                          -2q  & q^2-1  \end{pmatrix} =(\hat{\Omega})^{-1}.
\label{omega1}
\eeqa

We have therefore shown explicitly that the positive chirality 
basis wavefunction (\ref{general-basis1}), known earlier in the literature,
can be mapped to   the negative chirality 
wavefunctions that we have constructed  in eqs. (\ref{anti-chiral-wavefunction})-(\ref{thetapart}),  (\ref{xywavefunction}). 
Such a map also confirms the validity of our construction for the negative
chirality basis functions, presented using  basic principles,
such as  equations of motion as well as  periodicity requirement.
In fact, in the next subsection, the  same mapping
is also obtained through  comparison of the relevant equations of motion,
 which further confirms our results for the construction of the basis functions. 
Note that for $q=0$ or $q\rightarrow \infty$, corresponding to
the case when both matrices ${\bf N}$ and ${\bf M}$ in eq. (\ref{NM}) are 
diagonal, we have:
\beqa
\hat{\Omega}  =  \begin{pmatrix} 1 & 0 \cr
                          0  & -1  \end{pmatrix},\,\,\,\, {\rm or} \,\,\,\,
\hat{\Omega}  =  \begin{pmatrix} -1 & 0 \cr
                          0  & 1  \end{pmatrix}\, ,
\label{omega2}
\eeqa 
respectively. As a result, one reproduces the known  mapping of the 
wavefunctions between  positive and negative chirality spinors
in the case when $T^4$ is factorized into $T^2\times T^2$
\cite{ibanez}. 

\subsection{Mapping the equations of motion}\label{eq-mapping}

In order to derive a similar mapping of the equations of motion,
we show below that the covariant derivative operators appearing in 
eqs. (\ref{Dirac-equation4}) for the positive chirality wavefunction,
with a nontrivial complex structure ($i\hat{\Omega}$), are equivalent to the 
derivative operators appearing in eqs. (\ref{negative1}), (\ref{negative2}) for the negative
chirality wavefunction (with  complex structure $\Omega = i I_2$). The mapping of 
corresponding gauge potentials can also be shown in the same manner, since 
they have similar dependence on the complex structure as the derivative
operator. Note that the
complex structure appears in the wavefunctions as modular parameter 
matrices.  We therefore reconfirm the mapping between the
two wavefunctions by comparing the equations of motion as well. 

We now examine the Dirac equations for both cases.
For the first one, with arbitrary $ \Omega $($= i \hat{\Omega} $), we have
\beqa \nonumber
\vec{z} = \vec{x} + i \hat{\Omega} \vec{y}\, ; \,\,\,\,
\vec{\bar{z}} = \vec{x} - i \hat{\Omega} \vec{y}  
\quad\Rightarrow\quad  \vec{x} = \frac{\vec{z}+ \vec{\bar{z}}}{2}\, ; \,\,\,\, 
\vec{y} = (\hat{\Omega})^{-1} \left( \frac{\vec{z} - \vec{\bar{z}}}{2i}\right) ,
\label{zbarz}
\eeqa
which implies
\beqa \nonumber
\frac{\partial}{\partial z_i} = \frac{1}{2} \left( \frac{\partial}{\partial x^i} -
i (\hat{\Omega})^{-1}_{ji} \frac{\partial}{\partial y^j} \right) ,  \\
\frac{\partial}{\partial \bar{z_i}} = 
\frac{1}{2} \left( \frac{\partial}{\partial x^i} +
i (\hat{\Omega})^{-1}_{ji} \frac{\partial}{\partial y^j} \right).
\label{dz1} 
\eeqa
Then, the  Dirac equation for the positive chirality wavefunction is:
\beqa
 \bar{D}_{\bar{z}_i} 
\psi^{\vec{j}, {\bf{N'}}} (\vec{z}, {{\Omega}}) 
\equiv  \frac{1}{2} 
\left( D_{x^i} +
i (\hat{\Omega})^{-1}_{ji} 
D_{y^j} \right) 
\psi^{\vec{j}, {\bf{N'}}} (\vec{z}, {{\Omega}})  =  0 ,\,\,\,\,  i,j = 1,2 .
\label{Dirac-equation6}
\eeqa
On the other hand, for the negative chirality solution
(\ref{Dirac-equation5}), with complex structure $\Omega = i I_2$, the relevant 
derivative operators are:
\beqa
\left( \beta D_1 + \alpha D_2\right) 
\psi^{\vec{j},{\bf{N}},{\bf{M}}} = 0 ; \,\,\,\,
\left( 
\beta 
{\bar{D}}_2 - \alpha {\bar{D}}_1 \right) 
\psi^{\vec{j},{\bf{N}},{\bf{M}}} = 0 .
\label{Dirac-equation7}
\eeqa
These equations, using the definitions $z^i = x^i + i y^i$,
$\bar{z}_i = x^i - i y^i$, can be rewritten as:
\beqa \nonumber
\frac{1}{2} \left\lbrace  
D_{x^1} + 
i \left( \frac{\alpha^2 - \beta ^2 }{\alpha ^2 + \beta^2} 
D_{y^1}  - 
\frac{2 \alpha \beta}{\alpha ^2 + \beta^2} 
D_{y^2}\right) \right\rbrace  
\psi^{\vec{j},{\bf{N}},{\bf{M}}} = 0,  \\
\frac{1}{2} \left\lbrace  
D_{x^2} + 
i \left( \frac{-2\alpha\beta}{\alpha ^2 + 
\beta^2} 
D_{y^1} + \frac{\beta^2-\alpha^2}
{\alpha ^2 + \beta^2} 
D_{y^2}\right) 
\right\rbrace  \psi^{\vec{j},{\bf{N}},{\bf{M}}} = 0. 
\label{Dirac-equation8}
\eeqa
Now using $\frac{\beta}{\alpha}= q $ from eq. (\ref{alphabetaconstraint}) and  
comparing the equations (\ref{Dirac-equation6}) and 
(\ref{Dirac-equation8}), one finds that they precisely match for the 
following complex structure:
\beqa
 (\hat{\Omega})^{-1} = \frac{1}{(1+q^2)} \begin{pmatrix} 1-q^2 & -2q \cr
                          -2q  & q^2-1  \end{pmatrix}, 
\label{omega3}
\eeqa
which is exactly the same as eq. (\ref{omega1}).
Thus,  the wavefunctions as well as the Dirac equations for both 
cases match exactly. This mapping can be generalized further, 
as given in subsection \ref{T6-generalization} below.

\subsection{Mapping for arbitrary complex structure $\Omega$} 
\label{mapping-arbitrary}

In this subsection, we generalize the mapping between the equations of motion associated with the positive and negative chirality wavefunction to the case of $T^4$
compactification with arbitrary complex structure $\Omega$. Now, the negative chirality basis functions satisfy:
\beqa \nonumber
\frac{1}{2} \left\lbrace  
D_{x^1} + 
i (\Omega)^{-1}_{i1} \left( \frac{\alpha^2 - \beta ^2 }{\alpha ^2 + \beta^2} 
D_{y^i} \right)  -i (\Omega)^{-1}_{i2} \left( 
\frac{2 \alpha \beta}{\alpha ^2 + \beta^2} 
D_{y^i}\right) \right\rbrace  
\psi^{\vec{j},{\bf{N}},{\bf{M}}} = 0  \\
\frac{1}{2} \left\lbrace  
D_{x^2} + 
i (\Omega)^{-1}_{i1}\left( \frac{-2\alpha\beta}{\alpha ^2 + 
\beta^2} 
D_{y^i} \right) + i (\Omega)^{-1}_{i2}\left( \frac{\beta^2-\alpha^2}
{\alpha ^2 + \beta^2} 
D_{y^i}\right) 
\right\rbrace  \psi^{\vec{j},{\bf{N}},{\bf{M}}} = 0\, ,  
\label{Dirac-equation9}
\eeqa
which can be identified with the equations satisfied by the positive chirality 
wavefunction with $\tilde{\Omega} = \hat{\Omega} \Omega$, as can be seen through 
the  decomposition:
\beqa \nonumber
\frac{\partial}{\partial z_i} = \frac{1}{2} \left( \frac{\partial}{\partial x^i} -
i (\tilde{\Omega})^{-1}_{ji} \frac{\partial}{\partial y^j} \right) ,  \\
\frac{\partial}{\partial \bar{z_i}} = 
\frac{1}{2} \left( \frac{\partial}{\partial x^i} +
i (\tilde{\Omega})^{-1}_{ji} \frac{\partial}{\partial y^j} \right).
\label{dz2} 
\eeqa
Thus, eq. (\ref{general-basis1}) with $\tilde{\Omega} =  \hat{\Omega}\Omega$, 
with $\hat{\Omega}$ given in eq. (\ref{omega3}), provides the negative chirality 
solution for arbitrary complex structure $\Omega$, where  both `oblique' and diagonal 
fluxes are turned on.

\subsection{Generalization for the $T^6$- case}\label{T6-generalization}

In this subsection, we generalize the results obtained so far for negative chirality 
fermions on $T^4$ to the more general $T^6 $ case. We only consider the wavefunctions
that are well defined with two positive and one negative eigenvalues of the $3\times 3$
Hermitian intersection matrices, since these will complete the list of well defined 
wavefunctions, once complex conjugations are taken into account.
For the case of  $T^6$,  the relevant equations, obtained by generalization of 
eqs. (\ref{negative1}) and (\ref{negative2}) to be examined, are:
\beq
( \alpha \bar{D}_1 -\beta_{i}\bar{D}_i ) \psi = 0,
\label{Gen-negative1}
\eeq
 and 
\beq
(\alpha {D}_i  + \beta_i {D}_1) \psi = 0.
\label{Gen-negative2}
\eeq 
Note that in these equations and below, the indices $i,j = 1,2$ 
(used for the $T^4$ with wavefunctions of positive chirality).
In order for the above two equations to have simultaneous solution, one 
obtains the condition :
\beq
\alpha^2 F^{ab}_{i\bar{1}} + \alpha\beta_i  F^{ab}_{1\bar 1} - 
\alpha\beta_j  F^{ab}_{i\bar{j}}- \beta_i \beta_j F^{ab}_{1\bar{j}}   = 0,
\eeq
where $F^{ab} \equiv {\bf{N}}$ is  the difference of fluxes in brane stacks $a$ and $b$.
The general solution of this equation is of the following type:
\beqa
     F^{ab} \equiv {\bf{N}} \equiv \hat{N} \begin{pmatrix}
  1  & -(\vec{q})^T \cr -\vec{q} & 
\vec{q} (\vec{q})^T \end{pmatrix} +
\begin{pmatrix} 
 (\vec{q})^T \tilde{N} \vec{q} & \vec{q}^T \tilde{N} \cr 
\tilde{N} \vec{q} & \tilde{N}   \end{pmatrix},
\label{Gen-fmatrix1}
\eeqa
where $\tilde{N}$ is a $2 \times 2$ matrix and $\hat{N}$ is a number.
Also, $\vec{q}$ is the two-dimensional $(2d)$ vector defined as:
\beqa
\vec{q} = \begin{pmatrix}
           q_1 \cr q_2
          \end{pmatrix}
\eeqa
with $q_i = \frac{\beta_i}{\alpha} $. 

Now, after showing the possible existence of the solution by defining $F^{ab}$
in (\ref{Gen-fmatrix1}),
for the negative chirality wavefunction on $T^6$, we proceed to present a mapping 
between the equations of motion for negative chirality and positive chirality 
wavefunctions on  $T^6$.
As described before in section \ref{eq-mapping}. Here also we show that 
the covariant derivative operators appearing in 
eqs. (\ref{Dirac-equation4}), for the positive chirality wavefunction,
with a nontrivial complex structure are equivalent to the 
derivative operators appearing in eqs. (\ref{Gen-negative1}),  (\ref{Gen-negative2}) 
for the negative chirality wavefunction (with  complex structure $\Omega = i I_3$) 
and  the corresponding gauge potentials map in the same manner.

For the positive chirality  case, with arbitrary $ \Omega $($= i \hat{\Omega} $) and 
eqs. (\ref{zbarz}), (\ref{dz1}), the  Dirac equation reads:
\beqa
 \bar{D}_{\bar{z}_\mu} 
\psi^{\vec{j}, {\bf{N'}}} (\vec{z}, {{\Omega}}) 
\equiv  \frac{1}{2} 
\left( D_{x^\mu} +
i (\hat{\Omega})^{-1}_{\nu\mu} 
D_{y^\nu} \right) 
\psi^{\vec{j}, {\bf{N'}}} (\vec{z}, {{\Omega}})  =  0 ,\,\,\,\, \mu ,\nu = 1,2,3 \,\,.
\label{Dirac-equn1} 
\eeqa
On the other hand, for the negative chirality solution, 
 with complex structure $\Omega = i I_3$, the relevant 
derivative operators, given in eqs. (\ref{Gen-negative1}), (\ref{Gen-negative2}), 
take the form:
\beqa \nonumber
&&\frac{1}{2} \left\lbrace  
\left( \alpha^2 \delta_{ij} + \beta_i\beta_j\right)   D_{x^j} 
- i \left( 2 \alpha\beta_i \right)  D_{y^1}
+ i \left( \beta_i\beta_j - \alpha^2 \delta_{ij} \right) D_{y^j} 
\right\rbrace \psi^{\vec{j},{\bf{N}},{\bf{M}}} = 0 , \\
&&\frac{1}{2} \left\lbrace  \left( \beta^2_i + \alpha^2\right) D_{x^1} 
+ i \left( \alpha^2 - \beta^2_i\right) D_{y^1}
- i\left( 2\beta_i \alpha\right)  D_{y^i}  
\right\rbrace  
\psi^{\vec{j},{\bf{N}},{\bf{M}}} = 0.
\label{Dirac-equn2}
\eeqa
Now, defining new $2 \times 2$ matrices,
\beqa \nonumber
A_{ij} =  \left( \alpha^2 \delta_{ij} + \beta_i\beta_j\right),\,\,\,\,
B_{ij} =  \left( \beta_i\beta_j - \alpha^2 \delta_{ij} \right),
\eeqa
and
\beqa
P_{i}= ( 2 \alpha \beta_i ),
\eeqa
eqs. (\ref{Dirac-equn2}) can be re-written as:
\beqa \nonumber
&&\frac{1}{2} \left\lbrace 
D_{x^i} - i \left( A^{-1} P\right) _i D_{y^1}
+ i \left( A^{-1} B\right) _{ij} D_{y^j} 
\right\rbrace \psi^{\vec{j},{\bf{N}},{\bf{M}}} = 0  \\
&&\frac{1}{2} \left\lbrace D_{x^1} 
+  i \left(  \frac{\alpha^2 - \beta^2_i}{\beta^2_i 
+ \alpha^2} \right) D_{y^1}
- i \left( \frac{2 \alpha \beta_i }{\beta^2_i 
+ \alpha^2}\right) D_{y^i}  
\right\rbrace  
\psi^{\vec{j},{\bf{N}},{\bf{M}}} = 0 \,\,.
 \label{Dirac-equn3}
\eeqa
A comparison of equations (\ref{Dirac-equn1}) and (\ref{Dirac-equn3}) implies that 
they precisely match for the following complex structure:
\beqa \nonumber
(\hat{\Omega})^{-1} _{11} = 
\left(  \frac{\alpha^2 - \beta^2_i}{\beta^2_i + \alpha^2} \right), \,\,\,\,
(\hat{\Omega})^{-1} _{1i} =  \left(- A^{-1} P\right) _{i}, \\
(\hat{\Omega})^{-1} _{i1} = 
- \left( \frac{2\alpha \beta_i }{\beta^2_i + \alpha^2}\right),\,\,\,\, 
(\hat{\Omega})^{-1} _{ij}= \left( A^{-1} B\right) _{ij}.
\label{Gen-omega}
\eeqa
This expression for the complex structure generalizes  the one derived 
earlier in eq. (\ref{omega1}) for the $T^4$ case.
The results are also easily generalizable  to arbitrary complex 
structure $\Omega$ following the discussions in 
subsection \ref{mapping-arbitrary} for the special case of $T^4$ (see eq. (\ref{dz2})).

\subsection{Computation of  Yukawa couplings  }\label{yukawa-computation}

Now that we have derived both the fermionic 
and bosonic internal wavefunctions and expressed them as an orthonormal basis, 
we  compute the Yukawa couplings using 
the basis wavefunctions (\ref{xywavefunction}). 
We also point out how the results 
derived below reduce to the ones in section \ref{general-tori}.

Starting with basis functions described in eq. (\ref{xywavefunction}),  
for the case of the canonical complex structure $\Omega = i I_2$
(in the $T^4$ case), we have:
\beqa
  \psi^{\vec{i}, {\bf{N}_1},{\bf{M}_1}} (\vec{z}) \cdot
  \psi^{\vec{j}, {\bf{N}_2},{\bf{M}_2},} (\vec{z})& = &
\cn_ {\vec{i}} \cdot \cn_ {\vec{j}}  \cdot   e^{i\pi[\vec{x} \cdot ({\bf{N}_1}+{\bf{N}_2}) \cdot \vec{y} 
+ i  \vec{y} \cdot  ({\bf{M}_1} + {\bf{M}_2}) \cdot \vec{y} ] } \\ \nonumber & \cdot &
 \sum_{\vec{l_1},\vec{l_2} \in \inte^n} e^{\pi i(i)[(\vec{l_1}+\vec{i})  
\cdot {\bf{M}_1}  \cdot  (\vec{l_1}+\vec{i}) +(\vec{l_2}+\vec{j})        \cdot {\bf{M}_2}  \cdot(\vec{l_2}+\vec{j})]} \\ \nonumber & \cdot & 
e^{2\pi i[ (\vec{l_1}+\vec{i}) \cdot  {\bf{N}_1} +(\vec{l_2}+\vec{j}) \cdot  {\bf{N}_2}] \cdot \vec{x}}
e^{2 \pi i(i)[ (\vec{l_1}+\vec{i})  \cdot {\bf{M}_1} + (\vec{l_2}+\vec{j})  \cdot {\bf{M}_2} ] \cdot  \vec{y}}
\label{psi1psi2}
\eeqa 
This expression can be re-written as:
\beqa
\label{psi1-psi2}
 \psi^{\vec{i}, {\bf{N}_1},{\bf{M}_1}} (\vec{z}) \cdot
  \psi^{\vec{j}, {\bf{N}_2}{\bf{M}_2},} (\vec{z})& = &
\cn_ {\vec{i}} \cdot \cn_ {\vec{j}}  \cdot   e^{i\pi[\vec{x} \cdot ({\bf{N}_1}+{\bf{N}_2}) \cdot \vec{y} 
+ i  \vec{y} \cdot  ({\bf{M}_1} + {\bf{M}_2}) \cdot \vec{y} ] }  \\ & \cdot &
\sum_{\vec{l_1},\vec{l_2} \in \inte^n} 
e^{\pi i(i)( \vec{\textbf{l}}^{T} \cdot \hat{ \bf{Q}} \cdot\vec{\textbf{l}})} 
e^{2\pi i ( \vec{\textbf{l}}^{T} \cdot {\bf{Q}}\cdot \vec{\textbf{X}})}.
e^{2\pi i (i)( \vec{\textbf{l}}^{T} \cdot \hat{\bf{Q}}\cdot \vec{\textbf{Y}})}\,\,,
\nonumber
\eeqa
where we defined the $4d$-vectors:
\beqa
\vec{\textbf{l}} = \left( \begin{array}{c}
\vec{i} + \vec{l_1}  \\ \vec{j} + \vec{l_2} 
\end{array} \right),\,\,\,
\vec{\textbf{X}} = \left( \begin{array}{c} \vec{x} \\ \vec{x} \end{array} \right),\,\,\,
\vec{\textbf{Y}} = \left( \begin{array}{c} \vec{y} \\ \vec{y} \end{array} \right),\,\,\,
\label{4dvectors}
\eeqa
and the $4d$-matrices:
\beqa
\textbf{Q} = \left(
\begin{array}{cc} {\bf{N_1}}& 0 \\ 0 & {\bf{N_2}} \end{array} \right),\,\,\,
\hat{\textbf{Q}} = \left(
\begin{array}{cc} {\bf{M_1}}& 0 \\ 0 & {\bf{M_2}} \end{array} \right)\, .\,\,\,
\label{4dmatrices}
\eeqa

Using the transformation matrix $T$, defined in eq. (\ref{transmatrix}), and eqs. (\ref{transmatrixT})-(\ref{N1N2-identity2}), we explicitly  write   the terms appearing in the exponents in the RHS of eq. (\ref{psi1-psi2}) as:
\beqa \nonumber
( \vec{\textbf{l}})^{T} \cdot \hat{\textbf{Q}} 
\cdot (\vec{\textbf{l}}) =( \vec{\textbf{l}})^{T} \cdot  (T ^{-1} T ) \cdot \hat{\textbf{Q}} \cdot (T ^ {T} (T ^{-1})^{T} ) \cdot ( \vec{\textbf{l}}), \\ \nonumber
 ( \vec{\textbf{l}}^{T} \cdot {\bf{Q}}\cdot \vec{\textbf{X}}) =  \vec{\textbf{l}}^{T} \cdot  (T ^{-1} T ) \cdot {\bf{Q}}\cdot (T ^ {T} (T ^{-1})^{T} ) \cdot \vec{\textbf{X}}, \\ 
( \vec{\textbf{l}}^{T} \cdot \hat{\bf{Q}}\cdot \vec{\textbf{Y}}) =  \vec{\textbf{l}}^{T} \cdot  (T ^{-1} T ) \cdot \hat{\bf{Q}}\cdot (T ^ {T} (T ^{-1})^{T} ) \cdot \vec{\textbf{Y}}. 
\eeqa
Then using:
\beqa
{\textbf{Q}}^{\prime} \equiv T  \cdot \textbf{Q} \cdot T ^ {T} \hskip -0.3cm& = &\hskip -0.3cm \left(
\begin{array}{cr}
 ({\bf{N_1}} + {\bf{N_2}})  & 0  \\
0  &  \alpha {({\bf{N_1}}^{-1} + {\bf{N_2}}^{-1})  } \alpha ^ {T}
\end{array}
\right), \\ \nonumber
\hat{\textbf{Q}}^{\prime} \equiv T  \cdot \hat{\textbf{Q}} \cdot T ^ {T}\hskip -0.3cm& = &\hskip -0.3cm \left(
\begin{array}{cr}
({\bf{M_1}} + {\bf{M_2}})  &  ( {\bf{M_1}}{\bf{N_1}}^{-1}-{\bf{M_2}}{\bf{N_2}}^{-1})\alpha ^ {T} \\
\alpha( {\bf{N_1}}^{-1}{\bf{M_1}}- {\bf{N_2}}^{-1}{\bf{M_2}})  & 
 \alpha {({\bf{N_1}}^{-1}{\bf{M_1}}{\bf{N_1}}^{-1} + {\bf{N_2}}^{-1} {\bf{M_2}}{\bf{N_2}}^{-1})  } \alpha ^ {T}
\end{array}
\right),
\label{transform-qq'}
\eeqa
\beqa
( \vec{\textbf{l}})^{T} T^{-1} = 
\left( \begin{array}{c} 
(\vec{i} + \vec{l_1})({\bf{N_1}}^{-1} + {\bf{N}}_2^{-1} )^{-1} {\bf{N}}_2^{-1} + 
(\vec{j} + \vec{l_2})({\bf{N}}_1^{-1} + {\bf{N}}_2^{-1} )^{-1} {\bf{N}}_1^{-1}  
\\
\left[ (\vec{i} + \vec{l_1}) - (\vec{j} + \vec{l_2})\right]
({\bf{N}}_1^{-1} + {\bf{N}}_2^{-1})^{-1} \alpha^{-1}
\end{array} \right)^{T},
\eeqa
and
\beqa
(T^{-1})^{T} (\vec{\textbf{l}}) =\left( \begin{array}{c}
{\bf{N}}_2^{-1}({\bf{N}}_1^{-1} + {\bf{N}}_2^{-1} )^{-1}(\vec{i} + \vec{l_1}) + 
{\bf{N}}_1^{-1}({\bf{N}}_1 ^ {-1} + {\bf{N}}_2 ^ {-1} ) ^ {-1}(\vec{j} + \vec{l_2}) \\
(\alpha ^ {-1})^{T}({\bf{N}}_1 ^ {-1} + {\bf{N}}_2 ^ {-1} ) ^ {-1}[(\vec{i} + 
\vec{l_1}) - (\vec{j} + \vec{l_2})]
\end{array} \right),
\eeqa
\beqa
(T^{-1})^{T} (\vec{\textbf{X}}) =\left( \begin{array}{c}
                                         \vec{x} \\ 0
                                        \end{array} \right) ; \,\,\,\, 
(T^{-1})^{T} (\vec{\textbf{Y}}) =\left( \begin{array}{c}
                                         \vec{y} \\ 0
                                        \end{array} \right),
\eeqa
we can re-write eq. (\ref{psi1psi2}) as
\beqa
\label{trans-psi1psi2}
&&  \hskip -0.8cm
\psi^{\vec{i}, {\bf{N}_1},{\bf{M}_1}} (\vec{z}) \cdot
  \psi^{\vec{j}, {\bf{N}_2}{\bf{M}_2},} (\vec{z})= 
\cn_ {\vec{i}} \cdot \cn_ {\vec{j}}  \cdot   e^{i\pi[\vec{x} \cdot ({\bf{N}_1}+{\bf{N}_2}) \cdot \vec{y} 
+ i  \vec{y} \cdot  ({\bf{M}_1} + {\bf{M}_2}) \cdot \vec{y} ] } \times \\ \nonumber
 && \hskip -0.8cm
 \sum_{\vec{l_1},\vec{l_2} \in \inte^n} e^{\pi i(i)
 \left( \{ [(\vec{l_1}+\vec{i}){\bf{N}_1}+ (\vec{l_2}+\vec{j}){\bf{N}_2}]({\bf{N}}_1 + {\bf{N}}_2 ) ^ {-1} \} \cdot  
( {\bf{M}}_1 + {\bf{M}}_2) \cdot \{({\bf{N}}_1 + {\bf{N}}_2 ) ^ {-1} ({\bf{N}}_1(\vec{i} +\vec{l_1}) +  {\bf{N}}_2(\vec{j} + \vec{l_2})) \}
 \right)  } \times\\ \nonumber 
&& \hskip -0.8cm 
e^{2\pi i \{[ (\vec{l_1}+\vec{i}) \cdot  {\bf{N}_1} +(\vec{l_2}+\vec{j}) \cdot  {\bf{N}_2}] ({\bf{N}}_1 + {\bf{N}}_2 ) ^ {-1} \} \cdot ({\bf{N}}_1 + {\bf{N}}_2 )  \vec{x}}  \cdot
e^{2 \pi i(i) 
\{ [ (\vec{l_1}+\vec{i})  \cdot {\bf{N}_1} + (\vec{l_2}+\vec{j})  \cdot {\bf{N}_2} ] ({\bf{N}}_1 + {\bf{N}}_2 ) ^ {-1} \} \cdot ({\bf{M}}_1 + {\bf{M}}_2 ) \vec{y}} \times \\
\nonumber 
&& \hskip -0.8cm 
e^{2 \pi i(i) 
\{ [(\vec{i} + \vec{l_1}) - (\vec{j} + \vec{l_2})]
({\bf{N}}_1^{-1} + {\bf{N}}_2^{-1} ) ^ {-1} \alpha ^ {-1} \} \cdot \alpha( {\bf{N_1}}^{-1}{\bf{M_1}}- {\bf{N_2}}^{-1}{\bf{M_2}})  \cdot \vec{y} } \times \\ \nonumber
&& \hskip -0.8cm 
e^{\pi i(i)
\left( \{ [(\vec{l_1}+\vec{i}){\bf{N}_1}+ (\vec{l_2}+\vec{j}){\bf{N}_2}]({\bf{N}}_1 + {\bf{N}}_2 ) ^ {-1} \}  \cdot ( {\bf{M_1}}{\bf{N_1}}^{-1}-{\bf{M_2}}{\bf{N_2}}^{-1})\alpha ^ {T} \{(\alpha ^ {-1})^{T}{\bf{N}}_2 ({\bf{N}}_1 + {\bf{N}}_2 ) ^ {-1}{\bf{N}}_1[(\vec{i}  - \vec{j}) + 
(\vec{l_1}  - \vec{l_2})  \} \right) } \times \\ \nonumber
&& \hskip -0.8cm 
e^{\pi i(i) 
\{ [((\vec{i} - \vec{j}) + (\vec{l_1} - \vec{l_2})) 
{\bf{N}}_1 ({\bf{N}}_1 + {\bf{N}}_2 ) ^ {-1}{\bf{N}}_2  \alpha ^ {-1}] \cdot [\alpha( {\bf{N_1}}^{-1}{\bf{M_1}}- {\bf{N_2}}^{-1}{\bf{M_2}})] ({\bf{N}}_1 + {\bf{N}}_2 ) ^ {-1} ({\bf{N}}_1(\vec{i} +\vec{l_1}) +  {\bf{N}}_2(\vec{j} + \vec{l_2})) \} } \times \\ \nonumber
&& \hskip -0.8cm 
e^{\pi i(i) 
\{ [((\vec{i} - \vec{j}) + (\vec{l_1} - \vec{l_2})) 
{\bf{N}}_1 ({\bf{N}}_1 + {\bf{N}}_2 ) ^ {-1}{\bf{N}}_2  \alpha ^ {-1}] 
[ \alpha {({\bf{N_1}}^{-1}{\bf{M_1}}{\bf{N_1}}^{-1} + {\bf{N_2}}^{-1} {\bf{M_2}}{\bf{N_2}}^{-1})  } \alpha ^ {T}] [(\alpha ^ {-1})^{T}{\bf{N}}_2 ({\bf{N}}_1 + {\bf{N}}_2 ) ^ {-1}{\bf{N}}_1[(\vec{i}  - \vec{j}) + 
(\vec{l_1}  - \vec{l_2}) ]\} } 
\eeqa 

Now, in a similar exercise as the one performed earlier in 
sections \ref{proof}, \ref{yukawa-general}, \ref{summation},  
we rearrange the series in eq. (\ref{trans-psi1psi2}) in terms of 
new summation  variables 
$ \vec{l_3}, \vec{l_4} ,\vec{m} $, whose values and ranges are assigned as in 
these sections.\footnote{For details see 
sections \ref{Riemann-Theta-Function-Identity}, \ref{proof},
\ref{yukawa-general}, \ref{summation}.}
With the value of $\alpha = (\det {\bf{N}_1} \det {\bf{N}_2}) I $, defined in eq. (\ref{def-alpha}), eq. (\ref{trans-psi1psi2}) takes the form:
\beqa
\label{final-psi1psi2}
&& \hskip -0.8cm
\psi^{\vec{i}, {\bf{N}_1},{\bf{M}_1}} (\vec{z}) \cdot
\psi^{\vec{j}, {\bf{N}_2}{\bf{M}_2},} (\vec{z})= \cn_ {\vec{i}} \cdot \cn_ {\vec{j}}  \cdot e^{i\pi[\vec{x} \cdot ({\bf{N}_1}+{\bf{N}_2}) \cdot \vec{y} 
+ i  \vec{y} \cdot  ({\bf{M}_1} + {\bf{M}_2}) \cdot \vec{y} ] } \\ \nonumber
 && \hskip -0.8cm
 \sum_{\vec{l_3}, \vec{l_4}\in \inte^n}\sum_{\vec{m}} 
e^ {\pi i (i)[ (\vec{i} {\bf{N}}_1 + \vec{j} {\bf{N}}_2  + {\vec{m}}{\bf{N}}_1)
({\bf{N}}_1 + {\bf{N}}_2 ) ^ {-1} + \vec{l_3}] \cdot ( {\bf{M}}_1 + {\bf{M}}_2) \cdot
[ ({\bf{N}}_1  + {\bf{N}}_2  ) ^ {-1} ( {\bf{N}}_1\vec{i} +  {\bf{N}}_2 \vec{j}  
+{\bf{N}}_1 {\vec{m}}) + \vec{l_3}]}\times \\ \nonumber
&& \hskip -0.8cm 
e^ {2\pi i [ (\vec{i} {\bf{N}}_1 + \vec{j} {\bf{N}}_2  + {\vec{m}}{\bf{N}}_1)
({\bf{N}}_1 + {\bf{N}}_2 ) ^ {-1} + \vec{l_3}] \cdot ({\bf{N}}_1 + {\bf{N}}_2 )  \vec{x}} \cdot 
e^ {2\pi i (i)[ (\vec{i} {\bf{N}}_1 + \vec{j} {\bf{N}}_2  + {\vec{m}}{\bf{N}}_1)
({\bf{N}}_1 + {\bf{N}}_2 ) ^ {-1} + \vec{l_3}] \cdot ({\bf{M}}_1 + {\bf{M}}_2 )  \vec{y}}\times \\ \nonumber
&& \hskip -0.8cm 
e^ {2\pi i (i) [(\vec{i} - \vec{j} + \vec{m}) 
\frac{{\bf{N}}_1 ({\bf{N}}_1 + {\bf{N}}_2 ) ^ {-1}{\bf{N}}_2} 
{\det {\bf{N}_1} \det {\bf{N}_2} } + \vec{l_4}] \cdot [(\det {\bf{N}_1} \det {\bf{N}_2}) ({\bf{N_1}}^{-1}{\bf{M_1}}- {\bf{N_2}}^{-1}{\bf{M_2}})]  \cdot \vec{y} }\times \\ \nonumber
&& \hskip -0.8cm 
e^ {\pi i (i)[ (\vec{i} {\bf{N}}_1 + \vec{j} {\bf{N}}_2  + {\vec{m}}{\bf{N}}_1)
({\bf{N}}_1 + {\bf{N}}_2 ) ^ {-1} + \vec{l_3}] \cdot [(\det {\bf{N}_1} \det {\bf{N}_2}) ({\bf{M_1}}{\bf{N_1}}^{-1}-{\bf{M_2}}{\bf{N_2}}^{-1}) ] \cdot
[\frac{{\bf{N}}_2 ({\bf{N}}_1 + {\bf{N}}_2 ) ^ {-1}{\bf{N}}_1} 
{\det {\bf{N}_1} \det {\bf{N}_2} } (\vec{i} - \vec{j} + \vec{m}) + \vec{l_4}]}\times \\ \nonumber
&& \hskip -0.8cm 
e^ {\pi i (i) [(\vec{i} - \vec{j} + \vec{m}) 
\frac{{\bf{N}}_1 ({\bf{N}}_1 + {\bf{N}}_2 ) ^ {-1}{\bf{N}}_2} 
{\det {\bf{N}_1} \det {\bf{N}_2} } + \vec{l_4}] \cdot [(\det {\bf{N}_1} \det {\bf{N}_2}) ({\bf{N_1}}^{-1}{\bf{M_1}}- {\bf{N_2}}^{-1}{\bf{M_2}})]  \cdot 
[ ({\bf{N}}_1  + {\bf{N}}_2  ) ^ {-1} ( {\bf{N}}_1\vec{i} +  {\bf{N}}_2 \vec{j}  
+{\bf{N}}_1 {\vec{m}}) + \vec{l_3}]}\times \\ \nonumber
&& \hskip -0.8cm 
e^ {\pi i (i) [(\vec{i} - \vec{j} + \vec{m}) 
\frac{{\bf{N}}_1 ({\bf{N}}_1 + {\bf{N}}_2 ) ^ {-1}{\bf{N}}_2} 
{\det {\bf{N}_1} \det {\bf{N}_2} } + \vec{l_4}] [(\det {\bf{N}_1} \det {\bf{N}_2})^2 ({\bf{N_1}}^{-1}{\bf{M_1}}{\bf{N_1}}^{-1} + {\bf{N_2}}^{-1} {\bf{M_2}}{\bf{N_2}}^{-1}) ]
[\frac{{\bf{N}}_2 ({\bf{N}}_1 + {\bf{N}}_2 ) ^ {-1}{\bf{N}}_1} 
{\det {\bf{N}_1} \det {\bf{N}_2} } (\vec{i} - \vec{j} + \vec{m}) + \vec{l_4} ]} 
\eeqa
Using from eq.(\ref{xywavefunction}):
\beqa&&\hskip -3cm 
(\psi^{\vec{k},{\bf{N}_3},{\bf{M}_3}})^* = 
 \cn_ {\vec{k}} \cdot e^{-i\pi[\vec{x}  \cdot {\bf{N}_3} \cdot \vec{y} - i  \vec{y} \cdot  {\bf{M}_3} \cdot \vec{y} ] }\cr
&& \hskip 1cm
\times\sum_{\vec{l^\prime_3} \in \inte^n} e^{\pi i(i)[(\vec{l^ \prime_3}+\vec{k})  
\cdot {\bf{M}_3}  \cdot (\vec{l^ \prime_3}+\vec{k})]} \cdot
e^{-2\pi i[ (\vec{l^ \prime_3}+\vec{k}) \cdot  {\bf{N}_3}  \cdot \vec{x} - i (\vec{l^ \prime_3}+\vec{k})  \cdot {\bf{M}_3}  \cdot  \vec{y}]}, 
\eeqa
we can then proceed to calculate the Yukawa coupling:
\beqa
Y_{ijk} = \sigma_{a b c} g  \int_{T^4} dz_i d\bar{z_i} \cdot  \psi^{\vec{i}, {\bf{N}_1},{\bf{M}_1}} \cdot
\psi^{\vec{j}, {\bf{N}_2}{\bf{M}_2}} \cdot  (\psi^{\vec{k},{\bf{N}_3},{\bf{M}_3}})^* \,\,\,\,
			(i=1, 2)\,\,.
\label{yukawa-integration}
\eeqa
Consider first the integration over $\vec{x}$:
\beqa \nonumber
&&\hskip -0.8cm \int d^2 \vec{x}\, 
e^{i\pi\{\vec{x} \cdot [({\bf{N}_1}+{\bf{N}_2})-{\bf{N}_3}] \cdot \vec{y}\} }
\sum_{\vec{l_3}, \vec{l_4},\vec{l^ \prime_3} \in \inte^n}\sum_{\vec{m}} 
e^ {2\pi i [ (\vec{i} {\bf{N}}_1 + \vec{j} {\bf{N}}_2  + {\vec{m}}{\bf{N}}_1)
({\bf{N}}_1 + {\bf{N}}_2 ) ^ {-1} + \vec{l_3}] \cdot ({\bf{N}}_1 + {\bf{N}}_2 )  \vec{x}}
e^{-2\pi i (\vec{l^ \prime_3}+\vec{k}) \cdot  {\bf{N}_3}  \cdot \vec{x}}  \\
&&\hskip -0.8cm
\label{y-x-integration}
\eeqa
which implies, using $({\bf{N}_1}+{\bf{N}_2}) = {\bf{N}_3}$ ,  the following conditions:
\begin{itemize}
\item equality of the summation indices $\vec{l_3} = \vec{l^ \prime_3} $,
\item the relation $(\vec{i} {\bf{N}}_1 + \vec{j} {\bf{N}}_2  + {\vec{m}}{\bf{N}}_1)
({\bf{N}}_3) ^ {-1} = \vec{k} $ .
\end{itemize}
Note that $({\bf{N}_1}+{\bf{N}_2}) = {\bf{N}_3}$  is a valid condition in 
a triple intersection since $I_{ab} + I_{bc} = I_{ac}$, with complex conjugation 
taking care of the fact that $ I_{ac}= -I_{ca} $, which changes the signs 
of ${\bf{N}}_3 $ and ${\bf{M}}_3$. Also, as in section \ref{yukawa-general}, 
\ref{summation}, for any given solution of the above constraint equation 
for $\vec{i}, \vec{j}, \vec{k}, \vec{m}$, other solutions inside the cell
of eq. (\ref{basis-e'}) that are shifted by $\vec{m}$'s
satisfying $\vec{m}{\bf N_1 }{\bf N_3}^{-1}$ : integer are also 
allowed. In view of this,
as in eq. (\ref{replace}), we break the sum over $\vec{m}$ into two parts, one
corresponding to $\vec{\tilde{m}}$, which is a given specific solution of 
eq. (\ref{integer-constraint}) and the other ones as given by sum over 
integer variables $\vec{p}$ and $\vec{\tilde{p}}$ whose ranges are as defined in 
eq. (\ref{integer-m-in-cell}).

Imposing the constraints from the  $\vec{x}$ integration, we obtain:
\beqa
\label{negatichiral-yukawa1}
&& \hskip -0.8cm
Y_{ijk} = \sigma_{a b c} g  \cdot \cn_ {\vec{i}} \cdot \cn_ {\vec{j}} \cdot  \cn_ {\vec{k}}  \\ \nonumber
&&\int d^2 \vec{y}  \{  e^{- \pi [\vec{y} \cdot  ({\bf{M}_1} + {\bf{M}_2} + {\bf{M}_3}) \cdot \vec{y} ] }  \sum_{\vec{l_3}, \vec{l_4} \in \inte^n} 
\sum_{\vec{p}, \vec{\tilde{p}}}
e^ {\pi i (i) [\vec{k} + \vec{l_3}] \cdot ( {\bf{M}}_1 + {\bf{M}}_2) \cdot [\vec{k} + \vec{l_3}]} \times \\ \nonumber
&&\hskip -0.8cm 
e^ {\pi i (i) [\vec{k} + \vec{l_3}]    \cdot [(\det {\bf{N}_1} \det {\bf{N}_2}) ({\bf{M_1}}{\bf{N_1}}^{-1}-{\bf{M_2}}{\bf{N_2}}^{-1}) ] \cdot
[\frac{{\bf{N}}_2 ({\bf{N}}_1 + {\bf{N}}_2 ) ^ {-1}{\bf{N}}_1} 
{\det {\bf{N}_1} \det {\bf{N}_2} } (\vec{i} - \vec{j} + \vec{\tilde{m}}) 
+ \vec{l_4}]} \times\\ \nonumber
&& \hskip -0.8cm 
e^ {\pi i (i) [(\vec{i} - \vec{j} + \vec{\tilde{m}}) 
\frac{{\bf{N}}_1 ({\bf{N}}_1 + {\bf{N}}_2 ) ^ {-1}{\bf{N}}_2} 
{\det {\bf{N}_1} \det {\bf{N}_2} } + \vec{l_4}] \cdot [(\det {\bf{N}_1} \det {\bf{N}_2}) ({\bf{N_1}}^{-1}{\bf{M_1}}- {\bf{N_2}}^{-1}{\bf{M_2}})]  \cdot [\vec{k} + \vec{l_3}]}\times \\ \nonumber
&& \hskip -0.8cm 
e^ {\pi i (i) [(\vec{i} - \vec{j} + \vec{\tilde{m}}) 
\frac{{\bf{N}}_1 ({\bf{N}}_1 + {\bf{N}}_2 ) ^ {-1}{\bf{N}}_2} 
{\det {\bf{N}_1} \det {\bf{N}_2} } + \vec{l_4}] [(\det {\bf{N}_1} \det {\bf{N}_2})^2 ({\bf{N_1}}^{-1}{\bf{M_1}}{\bf{N_1}}^{-1} + {\bf{N_2}}^{-1} {\bf{M_2}}{\bf{N_2}}^{-1}) ]
[\frac{{\bf{N}}_2 ({\bf{N}}_1 + {\bf{N}}_2 ) ^ {-1}{\bf{N}}_1} 
{\det {\bf{N}_1} \det {\bf{N}_2} } (\vec{i} - \vec{j} + \vec{m}) + \vec{l_4} ]} \\ \nonumber
&& \hskip -0.8cm \times
e^ {\pi i (i) [\vec{k} + \vec{l_3}] \cdot ( {\bf{M}}_1 + {\bf{M}}_2) \cdot \vec{y}} \cdot 
e^ {\pi i (i) [(\vec{i} - \vec{j} + \vec{m}) \frac{{\bf{N}}_1 ({\bf{N}}_1 + {\bf{N}}_2 ) ^ {-1}{\bf{N}}_2} {\det {\bf{N}_1} \det {\bf{N}_2} } + \vec{l_4}]
\cdot [(\det {\bf{N}_1} \det {\bf{N}_2}) ({\bf{M_1}}{\bf{N_1}}^{-1}-{\bf{M_2}}{\bf{N_2}}^{-1}) ] \cdot \vec{y}} \},
\eeqa
where the range of the sum over $\vec{p}, \vec{\tilde{p}}$ is as used in 
eq. (\ref{integer-m-in-cell}) in section \ref{yukawa-general}.

The above expression for the Yukawa interaction can be written as following:
\beqa \nonumber
Y_{ijk} &=& \sigma_{a b c} g  \cdot (2^3)^{\frac{1}{4}} (|\det {\bf{M}_1}|.|\det {\bf{M}_2}|.|\det {\bf{M}_3}|)^{\frac{1}{4}} \left(Vol(T^{4})\right)^{- \frac{3}{2}} \\
\nonumber
&&\int d^2 \vec{y}  \{  e^{- \pi [\vec{y} 
\cdot  ({\bf{M}_1} + {\bf{M}_2} + {\bf{M}_3}) \cdot \vec{y} ] }  
\sum_{\vec{l_3}, \vec{l_4} \in \inte^n} \sum_{\vec{p}, \vec{\tilde{p}}}
e^ {\pi i (i) [\vec{\bf{K}} + \vec{\bf{L}}] \cdot \hat{ \bf{Q^\prime}}  \cdot [\vec{\bf{K}} + \vec{\bf{L}}]}  
e^ {2\pi i (i) [\vec{\bf{K}} + \vec{\bf{L}}] \cdot \vec{\bf{Y^\prime}}}\\ \nonumber
& = & \sigma_{a b c} g  \cdot (2^3)^{\frac{1}{4}} (|det {\bf{M}_1}|.|det {\bf{M}_2}|.|det {\bf{M}_3}|)^{\frac{1}{4}} \left(Vol(T^{4})\right)^{- \frac{3}{2}} \times \\
&&\sum_{\vec{p}, \vec{\tilde{p}}}
\int d^2 \vec{y}   \{  e^{- \pi [\vec{y} 
\cdot  ({\bf{M}_1} + {\bf{M}_2} + {\bf{M}_3}) \cdot \vec{y} ] } \cdot
\vt \left[
\begin{array}{c}
\vec{\bf{K}} \\ 0
\end{array}
\right] (\vec{\bf{Y^\prime}} | i \hat{ \bf{Q^\prime}} ) \}
\label{negatichiral-yukawafinal}
\eeqa
where we defined new $4d$-vectors:
\beqa
\vec{\bf{L}} =\begin{pmatrix}
 \vec{l_3}  \cr  \vec{l_4} 
 \end{pmatrix},\,\,\,
\vec{\bf{K}} = \begin{pmatrix}
\vec{k}   \cr   [(\vec{i} - \vec{j} + \vec{\tilde{m}})] 
[\frac{{\bf{N}}_1 ({\bf{N}}_1 + {\bf{N}}_2 ) ^ {-1}{\bf{N}}_2} 
{\det {\bf{N}_1} \det {\bf{N}_2}} ]
 \end{pmatrix},
\label{vec-k}
\eeqa
\beqa
\vec{\bf{Y^\prime}} =  \begin{pmatrix}
({\bf{M_1}} + {\bf{M_2}}) \vec{y} \cr  
[(\det {\bf{N}_1} \det {\bf{N}_2}) ({\bf{M_1}}{\bf{N_1}}^{-1}-{\bf{M_2}}{\bf{N_2}}^{-1}) ] \cdot \vec{y}
\end{pmatrix} 
\label{vec-y'}
\eeqa
and the $4d$-matrix:
\beqa &&\hskip -1cm 
\hat{\textbf{Q}}^{\prime} \hskip -0.2cm =\hskip -0.2cm  
\left(\hskip -0.3cm
\begin{array}{cr}
({\bf{M_1}} + {\bf{M_2}})  & 
(\det {\bf{N}_1} \det {\bf{N}_2}) ({\bf{M_1}}{\bf{N_1}}^{-1}-{\bf{M_2}}{\bf{N_2}}^{-1}) \\
(\det {\bf{N}_1} \det {\bf{N}_2})( {\bf{N_1}}^{-1}{\bf{M_1}}-{\bf{N_2}}^{-1}{\bf{M_2}})  & 
 (\det {\bf{N}_1} \det {\bf{N}_2})^2 {({\bf{N_1}}^{-1}{\bf{M_1}}{\bf{N_1}}^{-1} + {\bf{N_2}}^{-1} {\bf{M_2}}{\bf{N_2}}^{-1})  } 
\end{array} \hskip -0.3cm
\right) \cr &&
\eeqa
with $\vec{k}$ appearing in eq. (\ref{vec-k}) restricted by 
the Kronecker delta relation written above, as following from the $x$ integration, in 
eq. (\ref{y-x-integration}) and the range of the sum 
over $\vec{p}, \vec{\tilde{p}}$ is as used in 
eq. (\ref{integer-m-in-cell}) in section \ref{yukawa-general}, we skip the details
regarding them. 

In fact, the form of the result (\ref{negatichiral-yukawafinal}) is valid 
for all basis functions, whether corresponding to positive or negative chirality 
wavefunctions, since the negative chirality wavefunction (\ref{xywavefunction}),
written for the complex structure $\Omega = i I_2$ and used in obtaining the final answer
for Yukawa coupling in eq. (\ref{negatichiral-yukawafinal}),
reduces to the one for positive chirality wavefunction for the same complex structure
when ${\bf{M}}$ is set to ${\bf{N}}$ (see eq. (\ref{general-basis1}) for the general 
form of the positive chirality wavefunction).  
For such a choice: ${\bf{M_i}} =  {\bf{N_i}}$,
$\hat{\textbf{Q}}^{\prime} $ has a factorized block form and the vector
$\vec{\bf{Y^\prime}}$ in eq. (\ref{vec-y'}) now has a form:
\beqa
\vec{\bf{Y^\prime}} =  \begin{pmatrix}
({\bf{N_1}} + {\bf{N_2}})\vec{y} \cr  0
\end{pmatrix}\, .
\eeqa
The theta function in eq. (\ref{negatichiral-yukawafinal}) then
factorizes and the final answer reduces to the form given in eqs. 
(\ref{general-yukawa-series}), (\ref{general-yukawa-simple})
for the choice $\tau = i$ corresponding to the complex structure of 
our choice in the negative chirality wavefunction 
(\ref{anti-chiral-wavefunction}). 

The Yukawa coupling expression (\ref{negatichiral-yukawafinal}) can be further 
generalized to other situations. First, although the above analysis was 
very specific to the case of $T^4$ due to our choice of wavefunction in 
eq. (\ref{xywavefunction}), the generlization to the $T^6$ is staightforward.
Mapping between matrices ${\bf{N}}$ and ${\bf{M}}$ is identical and follows 
from the definition of $\hat{\Omega}$ in subsection \ref{T6-generalization}.
The final answer is identical to the one given in eq. (\ref{negatichiral-yukawafinal}).

Further generalization to the situation of arbitrary complex structure 
should also be possible, using the 
wavefunctions that emerge due to the mappings obtained in subsection 
(\ref{mapping-arbitrary}) and  scaling 
procedure presented in section (\ref{general-complex-structure})
for the positive chirality wavefunctions. One, however, also needs to examine the
symmetry property of the matrices ${\bf{N}}\hat{\Omega} \Omega$ etc., 
appearing in the definition of the wavefunction. We leave further details
for future work.

\section{Mass generation for non-chiral fermions} \label{GUT-application}

In this section, we briefly discuss one of the applications of the results 
derived in the paper, for giving mass to the 
non-chiral gauge non-singlet states of the magnetized brane model 
discussed in \cite{akp}.  This is a three generation 
$SU(5)$ supersymmetric grand unified (GUT) model in simple toroidal compactifications of type I 
string theory with magnetized $D9$ branes.
The final gauge group is just $SU(5)$ and the chiral gauge non-singlet
spectrum consists of three families with the quantum numbers of quarks and
leptons, transforming in the ${\bf 10} + {\bf\bar{5}}$ representations of $SU(5)$.
Brane stacks with oblique fluxes played a central role in this construction,
in order to stabilize all close string moduli, in a manner restricting the
chiral matter content to precisely that of $SU(5)$ GUT. Another interesting 
feature of this model is that it is free from any chiral 
exotics that often appear in such brane constructions. 
However, the model contains extra non-chiral matter that is expected to become massive
at a high scale, close to that of $SU(5)$ breaking. 

The results of the previous sections can be used for examining the issue of 
the mass generation for these non-chiral multiplets in a supersymmetric 
ground state. The aim is to analyze the D and F term conditions, and show that
a ground state allowing masses for the above matter multiplets is possible.
The exercise will further fine tune our $SU(5)$ GUT model to the ones used in 
conventional grand unification.

Although, we will not be evaluating any of the Yukawa couplings explicitly, 
which using our results of the previous sections is in principle possible to do, 
the aim of the exercise below is to show that indeed one can give masses to non-chiral matter. Our 
procedure involves the analysis of both the F and D-term supersymmetry 
conditions. In the context of our previous work \cite{akp}, 
we like to remind the reader that certain charged scalar vacuum expectation values (VEVs) were turned on 
in order to restore supersymmetry in some of the ``hidden" branes 
sector. These charged scalar VEVs gave a nontrivial solution to the 
D-term conditions, but left the F-terms identically zero in the vacuum. 
In the following, on the other hand, our aim is to find out the possibility for
a large number of scalars in various chiral multiples to acquire expectation 
values. For this, we need to examine both the F and D conditions, as already mentioned. 

The  model in \cite{akp} is described by twelve stacks of branes, namely 
$U_5, U_1$, $O_1\dots , O_8$, $A$, and $B$. The magnetic fluxes are chosen to generate the 
required spectrum, to stabilize all the geometric moduli and to satisfy the RR-tadpole 
conditions as well. The fluxes for all the stacks are summarized in Appendix \ref{Appendix-A}.
The fluxes for stacks $U_5, U_1$, $A$, $B$ are purely diagonal whereas stacks 
$O_1\dots , O_8$
carry in general both oblique and diagonal fluxes. All 36 closed string moduli are fixed in a
${\cal N}=1$ supersymmetric vacuum, apart from the dilaton, in a way that the
$T^6$-torus metric becomes diagonal with the six internal radii given in
terms of the integrally quantized magnetic fluxes.

The two brane stacks $U_5$ and $U_1$ give the particle spectrum of $SU(5)$ GUT.
We solve the condition $I_{U_5U_1} + I_{U_5U_1^*} = -3$ for the presence of three generations
of chiral fermions  transforming in $\bf{\bar 5}$ of $SU(5)$ and continue with the solution
$I_{U_5U_1} = 0,\,\,\, I_{U_5U_1^*} = -3$. The intersection of $U_5$ with $U_1$ is 
non-chiral since $I_{U_5U_1}$ vanishes.
The corresponding non-chiral massless spectrum  consists of four pairs of 
${\bf 5}+{\bf{\bar 5}}$, which we would like  to give 
mass\footnote{For details, see Section 3.7 of \cite{akp}.}.
Obviously, we would like to keep massless at least one pair of electroweak higgses
but this requires a detailed phenomenological analysis that goes beyond the scope of this paper. 
Here, we would like only to show how to use the results of the previous sections in order to give 
masses to unwanted non chiral states that often appear in intersecting brane constructions.

So, we have the following non-chiral fields where the superscript refers to 
the two stacks between which the open  string is stretched and 
the subscript denotes the charges under the respective 
$U(1)$'s :($\phi^{U_5U_1}_{+-}$,$\phi^{U_5U_1}_{-+}$, $4$), with numbers in the 
brackets denoting the corresponding multiplicities.
Similarly, the intersections of the $U_5$ stack with the two extra branes $A,B$ and their images
are non-chiral, giving rise to the extra ${\bf 5}+{\bf{\bar 5}}$ pairs:
($\phi^{U_5A}_{+-}$,$\phi^{U_5A}_{-+}$, $149$),
($\phi^{U_5A^*}_{++}$,$\phi^{U_5A^*}_{--}$, $146$),
($\phi^{U_5B}_{+-}$,$\phi^{U_5B}_{-+}$, $51$),
($\phi^{U_5B^*}_{++}$,$\phi^{U_5B^*}_{--}$, $16$).
A common feature of all these states is that they arise in non-chiral intersections,
where the two brane stacks involved have diagonal fluxes and are parallel in one of the
three tori. It is then straightforward to give masses by moving, say, the $U_5$ stack away
from the others along these tori. In the language of $D9$ branes, this amounts to turn on
corresponding open string Wilson lines.

On the other hand, analysis of the  particle spectrum on the intersections of the 
stack $U_5$ with the oblique branes ${O_a}$ and ${O^*_a}$ , satisfying the condition 
$I_{U_5 a} + I_{U_5 a^*} = 0,\,\,\,\, {\rm for}\,\,a=1,..,8\,$, leads to 
$4\times (23+14)=148$ pairs of $({\bf 5}+{\bf\bar 5})$
representations of $SU(5)$:
\begin{equation}
I_{U_5O_a} = -23\, ,\;\;\;\;I_{U_5O_a^*} = 23\, ,\,\,\,\,\,a=1,\dots ,4\, ,
\label{nonchiral-o1/o4}
\end{equation}
\begin{equation}
I_{U_5O_a} = -14\, ,\quad I_{U_5O_a^*} = 14\, ,\quad a=5,\dots ,8\, .
\label{nonchiral-o5/o8}
\end{equation}
We then have the following chiral multiplets,
($\phi^{U_5O_a}_{-+}$, $23$), ($\phi^{U_5O^*_a}_{++}$, $23$),
($\phi^{U_5O_b}_{-+}$, $14$), ($\phi^{U_5O^*_b}_{++}$, $14$) 
($ a=1,\dots ,4$, $b=5,\dots ,8$).
In order to examine the mass generation for these fields, one needs to write down the 
superpotential terms involving the above chiral multiplets, as well as those coming from 
the brane stacks $O_1,\cdots, O_8$ and their orientifold images.
The list of the later, involving purely oblique stacks, is given in Appendix \ref{Appendix-A}.

Now, using the results in Appendix \ref{Appendix-A} in 
eqs. (\ref{nonchiral-O_i}) and (\ref{nonchiral-AB}),
one can analyze the associated superpotential and D-terms and 
look for supersymmetric minima. The relevant superpotential reads: 
\beqa \hskip -1cm
W &=&  \sum_{ijk} W^{ijk}_{O_1}\; (\phi^{O_1U_5}_{+-})^i\, 
(\phi^{U_5O^*_3}_{++})^j\, (\phi^{O^*_3O_1}_{--})^k+  
\sum_{ijk} W^{ijk}_{O_2}\; (\phi^{O_2U_5}_{+-})^i\, (\phi^{U_5O^*_4}_{++})^j\, 
(\phi^{O^*_4O_2}_{--})^k \nonumber \\
\label{superpotential-su5}
&+& \sum_{ijk} W^{ijk}_{O_3}\; (\phi^{O_3U_5}_{+-})^i\, 
(\phi^{U_5O^*_8}_{++})^j\, (\phi^{O^*_8O_3}_{--})^k +
\sum_{ijk} W^{ijk}_{O_4}\; (\phi^{O_4U_5}_{+-})^i\, (\phi^{U_5O^*_7}_{++})^j\, 
(\phi^{O^*_7O_4}_{--})^k \\
&+& \sum_{ijk} W^{ijk}_{O_5}\; (\phi^{O_5U_5}_{+-})^i\, (\phi^{U_5O^*_6}_{++})^j\, 
(\phi^{O^*_6O_5}_{--})^k +
\sum_{ijk} W^{ijk}_{O_7}\; (\phi^{O_7U_5}_{+-})^i\, (\phi^{U_5O^*_8}_{++})^j\, 
(\phi^{O^*_8O_7}_{--})^k \nonumber
\eeqa
where the sum over $i,j,k$ runs over the ``flavor" indices. The couplings $W^{ijk}_{O_i}$, 
given in eq. (\ref{superpotential-su5}), can be read off from our results in the 
previous sections. In addition to the complex structure, these 
also depend on the first Chern numbers  of the branes in each triangle. 

The F-flatness conditions $\langle {\rm F}_i \rangle = \langle {\cal D}_{\phi_i}W \rangle=0 $ 
(at zero superpotential, $W=0$), imply that for each ``triangle'' at least two fields 
must have a zero VEV in order to form a supersymmetric vacuum \cite{tristan}.
In this theory, there exists indeed a supersymmetric vacuum where six charged fields remain 
unconstrained by the F-flatness conditions. Let's choose them to be
$ (\phi^{O^*_3O_1}_{--})$,
$(\phi^{O^*_4O_2}_{--})$, $(\phi^{O^*_8O_3}_{--})$, $(\phi^{O^*_7O_4}_{--})$,
$(\phi^{O^*_6O_5}_{--})$, $(\phi^{O^*_8O_7}_{--}) $ 
(they are neutral under the $U(1)$ of the $U(5)$).
The remaining fields appearing in the superpotential acquire a mass from the 
F-term potential only if these unconstrained scalars possess a non-vanishing VEV.
Indeed, their masses read:
\beqa
 \begin{array}{l}
M^2_{\phi_{u_5o_1}}\sim M^2_{\phi_{u_5o^*_3}}\sim  \langle |\phi_{o^*_3o_1}|^2 \rangle  \, , \,
M^2_{\phi_{u_5o_2}}\sim M^2_{\phi_{u_5o^*_4}}\sim  \langle |\phi_{o^*_4o_2}|^2 \rangle \, ,  \\
M^2_{\phi_{u_5o_7'}}\sim M^2_{\phi_{u_5o^*_8}}\sim  \langle |\phi_{o^*_8o_7'}|^2 \rangle  \, , \,
M^2_{\phi_{u_5o_4}}\sim M^2_{\phi_{u_5o^*_7}}\sim  \langle |\phi_{o^*_7o_4}|^2 \rangle  \, ,\\
M^2_{\phi_{u_5o_5}}\sim M^2_{\phi_{u_5o^*_6}}\sim  \langle |\phi_{o^*_6o_5}|^2 \rangle  \, , \,
\end{array}
\label{mass-generation}
\eeqa
where $\phi_{u_5o_7'}$ denotes linear combinations of $\phi_{u_5o_7}$ with $\phi_{u_5o_3}$ and
$\phi_{o^*_8o_7'}$ denotes linear combinations of $\phi_{o^*_8o_7}$ with $\phi_{o^*_8o_3}$.
Thus, 
the leftover massless states from the intersection of $U_5$ with the oblique branes 
are 60 pairs of ${\bf 5}+{\bf{\bar 5}}$:
$\phi_{u_5o^*_a}$ for $a=1,2,5$ of positive chirality together with the negative chirality states 
$\phi_{u_5o_a}$ for $a=6,7$, as well as 23 linear combinations of $\phi_{u_5o_3}$ with $\phi_{u_5o_7}$,
and 14 $\phi_{u_5o_4}$.

However, switching on non-zero VEVs for these fields, modifies the existing 
D-term conditions for the stacks of branes $ O_1,....O_8$. 
Recall that, in \cite{akp}, the 
stacks $U_5$, $O_1 \dots O_8$ satisfy the supersymmetry conditions in the absence of 
charged scalar VEVs, but VEVs for the fields $\phi^{U_1A}_{-+}$, $\phi^{U_1B^*}_{++}$ 
and  $\phi^{AB}_{+-}$  are switched on, for the same supersymmetry to be preserved by 
the stacks $U_1$, $A$ and $B$.\footnote{For details see Section 5 of \cite{akp}.}
The D-terms for each $U(1)$ factor of the eight branes $O_1,.....O_8$ read
\beqa
 \begin{array}{l}
 D_{O_1} = -|\phi^{O_1O^*_3}|^2  \, , \, D_{O_2} = -|\phi^{O_2O^*_4}|^2   \\
 D_{O_3} = -|\phi^{O_1O^*_3}|^2 - |\phi^{O_3O^*_8}|^2  \, , \, D_{O_4} = -|\phi^{O_2O^*_4}|^2 -|\phi^{O_4O^*_7}|^2  \\
 D_{O_5}=  -|\phi^{O_5O^*_6}|^2 \, , \, D_{O_6} = -|\phi^{O_5O^*_6}|^2  \\
 D_{O_7} = -|\phi^{O_4O^*_7}|^2 - |\phi^{O_7O^*_8}|^2  \, , \, D_{O_8} = -|\phi^{O_3O^*_8}|^2 -|\phi^{O_7O^*_8}|^2  
\end{array} 
\label{ex:susyD}
\eeqa

We can regain  the supersymmetry conditions $D_a=0$, $\forall a=1,\dots,8$ 
with $\xi_a(F^a,J)=0$, by switching on VEVs for the following fields:
$ (\phi^{O_1O^*_5}_{++})$, $ (\phi^{O_2O^*_7}_{++})$,  $ (\phi^{O_3O^*_7}_{++})$, 
$ (\phi^{O_3O^*_4}_{++})$, $ (\phi^{O_4O^*_8}_{++})$, $ (\phi^{O_6O^*_8}_{++})$,
provided these fields do not modify the superpotential (\ref{superpotential-su5}). 
The modified D-terms read:
\beqa \nonumber
 D_{O_1} &=& -|\phi^{O_1O^*_3}|^2 + |\phi^{O_1O^*_5}|^2  \\ \nonumber
 D_{O_2} &=& -|\phi^{O_2O^*_4}|^2 +| \phi^{O_2O^*_7}|^2  \\ \nonumber
 D_{O_3}& =& -|\phi^{O_1O^*_3}|^2 - |\phi^{O_3O^*_8}|^2 + | 
\phi^{O_3O^*_4}|^2+| \phi^{O_3O^*_7}|^2\\ \nonumber
 D_{O_4} &= &-|\phi^{O_2O^*_4}|^2 -|\phi^{O_4O^*_7}|^2 +| 
\phi^{O_3O^*_4}|^2 +| \phi^{O_4O^*_8}|^2 \\ \nonumber
 D_{O_5}&=&  -|\phi^{O_5O^*_6}|^2 +| \phi^{O_1O^*_5}|^2\\ \nonumber
 D_{O_6} &=& -|\phi^{O_5O^*_6}|^2  +| \phi^{O_6O^*_8}|^2\\ \nonumber
 D_{O_7} &=& -|\phi^{O_4O^*_7}|^2 - |\phi^{O_7O^*_8}|^2 +| 
\phi^{O_2O^*_7}|^2+|\phi^{O_3O^*_7}|^2 \\
 D_{O_8}& = &-|\phi^{O_3O^*_8}|^2 -|\phi^{O_7O^*_8}|^2 +|\phi^{O_6O^*_8}|^2 
+ |\phi^{O_4O^*_8}|^2 
\label{ex:susyDM}
\eeqa
The  supersymmetry conditions $D_a=0$, $\forall a=1,\dots,8$ with 
$\xi_a(F^a,J)=0$ can be simultaneously satisfied if and only if  the VEVs 
for all these fields appearing in the expressions (\ref{ex:susyDM}), have the same 
value, say $v^2$. Moreover we can restrict $v << 1$ in string units, as required by the 
validity of the approximation for inclusion of charged scalar fields in the D-term.

We have therefore shown the mass generation for a large set of non-chiral fields
as given in eq. (\ref{mass-generation}). It is possible, that remaining ones can also 
be made massive by incorporating non perturbative instanton contributions to the 
superpotential. However, we leave this exercise for the moment. We also do not 
give any superpotential couplings, in terms of fluxes, as given explicitly 
in the previous sections. 

\section{Discussions and Conclusions}\label{conclusion}

In this concluding section, we first comment on the case of 
magnetized branes with higher winding numbers. The form of the wrapping matrices \cite{mag7}
for $D9$ branes on $T^6$ was discussed in our earlier papers \cite{akm,akp}.
They are real $6\times 6 $ matrices giving the embedding of the brane along 
spatial internal directions.
The situation where worldvolume coordinates are identified with the spatial coordinates
corresponds to $W$ being diagonal. Then, for example, for a canonical complex structure
$\Omega = i I_3$, the spatial components of the flux matrices are of the form
given in eqs. (\ref{flux-real}), (\ref{quantization}), (\ref{b-shift}).
Taking into account the gauge indices,  one obtains a block diagonal matrix
structure for the fluxes, that reduces in the case of factorized tori to the form:
\beq
    F  = \begin{pmatrix}\frac{m^a_i}{n^a_i} I_{N^a} & \cr
	&  \frac{m^b_i}{n^b_i} I_{N^b} \end{pmatrix}, 
\label{flux-winding-n}
\eeq 
with $a$ and $b$ representing the brane-stacks and $i$ denotes the $i$'th 
$T^2$. Also $m^{a,b}_i$ are the first Chern numbers, as given in eqs. (\ref{flux-real})
and (\ref{quantization}), whereas $n^{a,b}_i$ are the product of the winding numbers
along various 1-cycles of ${(T^2)}^3 \in T^6$. Also, $N^a$ and $N^b$ are the 
number of branes in stacks $a$ and $b$ respectively and the above expression has
a straightforward generalization when many such brane stacks are involved.

In \cite{ibanez}, a gauge theoretic picture of the magnetic fluxes along brane stacks
with higher winding numbers ($> 1$) was given. For instance, consider the simplest choice
$N^a = N^b = 1$. In this case, the configuration of the brane 
stacks $a$ and $b$ with one $D$-brane each, having wrapping numbers 
$n^a, n^b$ and 1st Chern numbers $m^a, m^b$, is given by a flux matrix associated with a
$U(n^a + n^b)$ gauge group with flux having the internal (gauge) components:
\beq
    F  = \begin{pmatrix}\frac{m^a_i}{n^a_i} I_{n^a_i} & \cr
	&  \frac{m^b_i}{n^b_i} I_{n^b_i} \end{pmatrix}, 
\label{gauge-theory-flux}
\eeq 
along the $i$'th $T^2$ and $m^a_i, n^a_i$ etc. are relatively prime. 

Given the $U(n^a + n^b)$ flux in eq. (\ref{gauge-theory-flux}),
the fermion wavefunctions associated with bifundamentals were constructed
in \cite{ibanez}. The new feature is that, to have proper periodicity 
property for these fermion wavefunctions, non-abelian Wilson lines need to be 
turned on. In turn, these non-abelian Wilson lines mix up $n^a_i \times n^b_i$
components and the set of periodicity constraints only allows the bifundamentals
belonging to the representations of the gauge group:
$U(P^a_i)\times U(P^b_i)$, with $P^a_i = g.c.d. (m^a_i, n^a_i)$. In our example above
we have $P^a_i = P^b_i = 1$.

The case of oblique fluxes brings in extra complexities in the analysis due to 
the presence of six independent 1-cycles along which non-abelian
Wilson line actions need to be 
fixed. Given the action of these Wilson lines, one can then proceed to 
obtain the wavefunctions as well as the Yukawa couplings. However, unlike the 
factorized situation in \cite{ibanez}, one finds that the action of non-abelian 
Wilson lines on the wavefunction, is dependent on the particular model, or more 
precisely, on the details of the oblique fluxes that are turned on.
Further analysis along this line is, though cumbersome, possible. We now 
conclude our paper with the following remarks.

In this work, we have been able to explicitly generalize the Yukawa coupling 
expressions to the situation when the worldvolume fluxes that are responsible for 
moduli stabilization, chiral mass generation, supersymmetry breaking to $N=1$
etc., do not respect the factorization of $T^6$ into ${(T^2)}^3$.
For the factorized tori, the mappings of the Yukawa couplings, superpotentials
and K\"ahler potential between the type IIB and IIA expressions 
was discussed in \cite{ibanez}. In the IIA case, the results are obtained 
through a `diagonal' wrapping of the $D6$ branes in three $T^2$'s. 

It will be interesting to map  our IIB expressions, given in this paper
to the IIA side and find the corresponding intersecting brane picture.
As stated earlier, such a IIA construction will require putting the branes
along general $SU(3)$ rotation angles and then obtain the area of the 
triangles corresponding to the intersections of three branes giving chiral 
multiplets. 

Supersymmetry breaking is of course an important issue in model building.
Though generally, for magnetized branes, one encounters instabilities in such a 
situation, it should be however possible to obtain non-supersymmetric 
magnetized brane constructions for a rich variety of fluxes accompanied by 
orientifold planes which can possibly project out tachyons that may be generated 
during the process of supersymmetry breaking. 

The recent developments in writing the instanton induced superpotential terms
are also encouraging, for the purpose of examining the supersymmetry breaking
as well as up-quark mass generations in a GUT setting. In this context,
it has been shown that the magnetized branes too can give rise to 
interesting superpotentials through the lift of fermion zero modes when
fluxes are turned on. 

Recently, there have been interesting developments in deriving
particle models and interactions from string theory, using the non-perturbative
picture of F-theory \cite{vafa1,donagi,vafa2,vafa3,vafa4,vafa5,font,blumenhagen1}, 
with a geometric 
picture of 7-brane intersection curves inside del Pezzo surfaces giving the 
chiral spectrum as well as Yukawa interactions, including those of the 
spinors of $SO(10)$ GUT, and thus generating observable masses
for both up and down type quarks. It is also interesting to note that 
F-theory results are reproducible in a globally consistent IIB string theory, taking
into account the instanton generated superpotential terms \cite{blumenhagen2}.
It will be of interest to see the implications of these results on 
the construction of $SU(5)$ GUT in \cite{akp}, as well as on the Yukawa interactions 
discussed in this paper.

Finally, it will be interesting to explore the generalization of our results to higher-point functions
(computing couplings of higher dimensional effective operators) \cite{Abe:2009dr}
and make explicit comparisons of our results with those in \cite{mag8,mag9}, 
where the situation with diagonal intersection matrices 
${\bf N_i}$,  but non-factorized complex structure, is addressed through a 
computation of twist field correlations. However, one then needs to examine 
the effect of supersymmetry conditions (\ref{(2,0)=0}) and  (\ref{(0,2)=0})
to see if the interaction indeed remains nontrivial in a supersymmetric set up.

We hope to return to all issues above at a later stage.

\section{Acknowledgements}

We would like to thank Sachin Jain, Fernando Marchesano,
Subir Mukhopadhyay, Koushik Ray and Angel Uranga,
for fruitful discussions. We specially thank Rodolfo Russo and Stefano Sciuto
suggesting the need for improvements in our first version.
A.K. would like to thank I.P.M. in Tehran, the Theory Unit of CERN
physics department, Ecole Polytechnique in Paris and 
Galileo Galilei Institute in Florence
for their hospitalities, during the course of this work.
Work supported in part by the European Commission under the ERC Advanced Grant 226371.
B.P. acknowledges the Commission for support under this grant.

\appendix

\section{Wavefunction}\label{wavefunction}

We first present the construction of chiral fermion wavefunctions on 
tori and give their representation in terms of theta functions. For definiteness
we first discuss the case of 4-tori, though $T^6$ chiral multiplet structure 
can be analyzed in a similar manner. To be explicit, for the moment we restrict 
ourselves to the canonical complex structure: $\Omega = i I_2$ and
$\Omega = i I_3$ for $T^4$ and $T^6$ respectively, where $I_d$ represents
a $d$-dimensional identity matrix.  The general complex structure is restored 
while writing the wavefunctions as well as interaction vertices.  

To obtain the Dirac wavefunctions in $T^4$, we start by writing
four Dirac Gamma matrices (in a complex basis) :
\beqa
    \Gamma^{z_1} = \sigma^z \times \sigma^3 =
	\begin{pmatrix}0 & 2 & & \cr 0 & 0 & & \cr
			& & 0 & -2 \cr & & 0 & 0 \end{pmatrix},\;\;\;\;
\Gamma^{z_2} = I \times \sigma^z =
	\begin{pmatrix} &  & 2 & 0\cr  &  & 0 & 2\cr
			0 &0 & &  \cr 0 & 0 &  &  \end{pmatrix},
\label{gammaz}
\eeqa
where the information about the complex structure in the above expression 
is hidden in the fact that we have used the definitions: $z_i = x_i + i y_i$
in writing these Dirac matrices. Similarly,
\beqa
    \Gamma^{\bar{z}_1} = \sigma^{\bar{z}} \times \sigma^3 =
	\begin{pmatrix}0 & 0 & & \cr 2 & 0 & & \cr
			& & 0 & 0 \cr & & -2 & 0 \end{pmatrix},\;\;\;\;
\Gamma^{\bar{z}_2} = I \times \sigma^{\bar{z}} =
	\begin{pmatrix} &  & 0 & 0\cr  &  & 0 & 0\cr
			2 & 0 & &  \cr 0 & 2 &  &  \end{pmatrix}.
\label{gammabarz}
\eeqa
They satisfy the anti-commutation relations:
\beq
   \{ \Gamma^{z_i}, \Gamma^{z_j} \} = 0,\;\;
   \{ \Gamma^{\bar{z}_i}, \Gamma^{\bar{z}_j} \} = 0,\;\;
   \{ \Gamma^{z_i}, \Gamma^{\bar{z}_j} \} = 4 \delta_{ij}
\label{anticommutation}
\eeq
with $i,j = 1, 2$. In the above basis $\Gamma^5$ takes the form:
\beqa
\Gamma^5 =  \begin{pmatrix} 1 &  & & \cr  & -1 & & \cr
			& & -1 &  \cr & &  & 1 \end{pmatrix}
\label{gamma5}
\eeqa
with 4-component Dirac wavefunctions having the form:
\beqa
\Psi =  \begin{pmatrix} \Psi_+^1 \cr \Psi_-^2 \cr \Psi_-^1 \cr 
		\Psi_+^2 \end{pmatrix}\, .
\label{Dwavefunction}
\eeqa
In such a decomposition of $\psi$,  
Dirac equations for fermions in the adjoint representation are of the form:
\beqa
\bar{\partial}_1 \Psi_+^1 + \partial_2 \Psi_+^2 + 
	[A_{\bar{z_1}}, \Psi_+^1 ] +
	[A_{{z_2}}, \psi_+^2 ] = 0, \cr
\bar{\partial}_2 \Psi_+^1 - \partial_1 \Psi_+^2 + 
	[A_{\bar{z_2}}, \Psi_+^1 ] -
	[A_{{z_1}}, \Psi_+^2 ] = 0, \cr
{\partial}_1 \Psi_-^2 + \partial_2 \Psi_-^1 + 
	[A_{{z_1}}, \Psi_-^2 ] +
	[A_{{z_2}}, \Psi_-^1 ] = 0, \cr
\bar{\partial}_2 \Psi_-^2 - \bar{\partial}_1 \Psi_-^1 + 
	[A_{\bar{z_2}}, \Psi_-^2 ] -
	[A_{\bar{z_1}}, \Psi_-^1 ] = 0.
\label{Dirac-equation}
\eeqa

In a generic model, chiral fermions arise either from the string 
starting at a brane stack-$a$ and ending at another brane stack-$b$ 
(or its image $b^*$) or from strings starting at a brane stack $a$ and ending at its
image $a^*$. We already showed the correspondence between a stack of 
magnetized branes and flux quanta in supersymmetric gauge theory,
in eq. (\ref{flux-u(n)}). The correspondence is easily generalized
when several stacks of branes are present. 
Explicitly, in a construction with $P$ number of stacks of branes, 
with number of branes
being $n_i$ for the $i$'th stack, the flux (for a given 
target space component $(i\bar{j})$ ) takes a form: 
\beqa
    F_{i \bar{j}}  = \begin{pmatrix}F^1I_{n_1} & & & & \cr  
			& F^2I_{n_2} & & &\cr
			& & . &  & \cr & & & . & \cr 
			& & & & F^{n_p}I_{n_p}\, ,\end{pmatrix}
\label{flux}
\eeqa
with $I_{n_i}$ being the $n_i$-dimensional identity matrix and we 
have hidden the ${i \bar{j}}$ indices in the RHS of eq. (\ref{flux})
in constants $F^i$ that are all integrally quantized, as given earlier
explicitly in eqs. (\ref{flux-u1(n)}) and (\ref{flux-blocks}). 
The corresponding gauge potentials will also then have a
block diagonal structure:
\beqa
    A_{i}  = \begin{pmatrix}A^1_i I_{n_1} & & & & \cr  
			& A^2_i I_{n_2} & & &\cr
			& & . &  & \cr & & & . & 
			& & & & A^{n_p}_i I_{n_p}\end{pmatrix}.
\label{potential}
\eeqa

Now, in order to understand the wavefunctions associated with chiral fermion
bifundamentals, 
in such a representation of the brane stacks, we consider the flux
matrix $F_{i\bar{j}}$ in eq. (\ref{flux}) and gauge potential in 
eq.  (\ref{potential}) with only two blocks ($P=2$).
The chiral fermion bilinears between stack-$a$ and stack-$b$ are 
then represented by:
\beqa
    \Psi_{ab}  = \begin{pmatrix}C_{n_a} & \chi_{ab}\cr  & C_{n_b}
		\end{pmatrix},
\label{bi-wavefunction}
\eeqa
with $C_{n_a}$, $C_{n_b}$ being constant matrices of dimensions
$n_a$ and $n_b$ respectively. We can easily derive the equation satisfied
by the various Dirac components, as given in eq. (\ref{Dwavefunction}),
for $\chi_{ab}$ such that $\psi_{ab}$ satisfies the Dirac equation
(\ref{Dirac-equation}). We obtain:
\beqa
\bar{\partial}_1 \chi_+^1 + \partial_2 \chi_+^2 + 
	(A^1 - A^2)_{\bar{z_1}} \chi_+^1  +
	(A^1 - A^2)_{{z_2}} \chi_+^2  = 0, \cr
\bar{\partial}_2 \chi_+^1 - \partial_1 \chi_+^2 + 
	(A^1 - A^2)_{\bar{z_2}} \chi_+^1  -
	(A^1 - A^2)_{{z_1}} \chi_+^2  = 0, \cr
{\partial}_1 \chi_-^2 + \partial_2 \chi_-^1 + 
	(A^1 - A^2)_{{z_1}} \chi_-^2  +
	(A^1 - A^2)_{{z_2}} \chi_-^1  = 0, \cr
\bar{\partial}_2 \chi_-^2 - \bar{\partial}_1 \chi_-^1 + 
	(A^1 - A^2)_{\bar{z_2}} \chi_-^2  -
	(A^1 - A^2)_{\bar{z_1}} \chi_-^1  = 0,
\label{Dirac-equation2}
\eeqa
with subscript $a, b$ being dropped from $\chi_{ab}$
to make the expressions simpler.
We will, however, restore the indices at a later stage while 
evaluating the overlap of three such wave functions from different 
intersections. In particular, for the chiral components,
$\chi_+^1$ equations reduce to:
\beqa
\bar{\partial}_1 \chi_+^1  +
	(A^1 - A^2)_{\bar{z_1}} \chi_+^1  = 0, \cr
\bar{\partial}_2 \chi_+^1 + 
	(A^1 - A^2)_{\bar{z_2}} \chi_+^1  = 0.
\label{Dirac-equation3}
\eeqa

The generalization of eq. (\ref{Dirac-equation3})
to the $T^6$ case is straightforward and can be written as:
\beq
\bar{D}_i \chi^{ab}_+ \equiv 
\bar{\partial}_i \chi_+^{ab}  +
	(A^1 - A^2)_{\bar{z_i}} \chi_+^{ab}  = 0,\;\;\;\;(i=1,2,3). 
\label{Dirac-equation4}
\eeq
Eq. (\ref{Dirac-equation4}) matches with eq. (4.65) of \cite{ibanez}
for $\Omega = i I_3$, with the identification:
\beq
	(A^1 - A^2)_{\bar{z_i}} 
	\equiv \frac{\pi}{2} \left([\bf{N}.(\vec{z} + \vec{\zeta})].
		(Im \Omega)^{-1}
			\right)_i,
\label{difference-mapping}
\eeq
with $\vec{\zeta}$ being the complex constants representing the Wilson lines
and {\bf{$N$}} is the  difference of 
fluxes between the two stacks $a$ and $b$ (see eq. (\ref{def-N})), 
having constant fluxes $F^1$ and $F^2$,
giving the fermion bilinears in the representation $(n_1, \bar{n}_2)$.

Such a solution for eq. (\ref{Dirac-equation4}) and
(\ref{difference-mapping}) is given in 
\cite{ibanez} for arbitrary complex structure $\Omega$ by the basis
elements:
\beq
  \psi^{\vec{j}, {\bf{N}}} (\vec{z}, {\bf{\Omega}}) = 
\cn \cdot e^{\{i\pi [{\bf{N}}.\vec{z}]. ({\bf{N}}.Im {\bf{\Omega}})^{-1} 
Im [{\bf{N}}.\vec{z}]\}} 
\cdot 
\vt
\left[
\begin{array}{c}
{\vec{j}} \\ 0
\end{array}
\right]
({\bf{N}}.\vec{z}, {\bf{N}}. {\bf{\Omega}}), 
\label{general-basis}	
\eeq
with general definition of Riemann theta function:
\beq
\vt
\left[\begin{array}{c}
{\vec{a}} \\ {\vec{b}}
\end{array}\right]
(\vec{\nu}|{\bf{\Omega}}) = \sum_{\vec{m} \in {\bf{Z}}^n}
	e^{\pi (\vec{m} + \vec{a}) . {\bf{\Omega}} . (\vec{m} + \vec{a})}
	e^{2 \pi i (\vec{m} + \vec{a}) . (\vec{\nu} + \vec{b})}	.
\label{general-theta}
\eeq
and ${\bf{N}}$ satisfying the constraints given in eqs. (\ref{eq-riemann-conditions})
as well as:
\beq
	\vec{j}.{\bf{N}} \in {{\bf Z}}^n, 
\label{j-integer}
\eeq
implying that $\vec{j}.{\bf{N}}$ is an $n$-dimensional vector with integer entries.  
Also, the normalization factor $\cn$ in eq. (\ref{general-basis}) is given by:
\beq
 \cn = \left(2^n |\det {\bf{N}}|. \det (Im \Omega)\right)^{\frac 14}
\left(Vol(T^{2n})\right)^{- {\frac 12}}.
\label{general-normalization-constant}
\eeq
Then wavefunctions satisfy the orthonormality relations:
\beq
	\int_{T^{2n}} (\psi^{\vec{j}, \bf{N}})^* \psi^{\vec{k}, \bf{N}} 
 = \delta_{\vec{j}, \vec{k}}.
\label{normalizable-general}
\eeq
These results are useful in determining the interaction terms in 
Section \ref{general-tori}.
However, to have well-defined wavefunctions, $\bf{N}$'s must satisfy the
Riemann conditions given in eq. (\ref{eq-riemann-conditions}).


\section{More information on fluxes}\label{General-Flux}

In general, the  $(1, 1)$ form flux $F_{z^i \bar{z}^j}$ given 
by a hermitian matrix in eq.
(\ref{F(1,1)}) is constrained by two equations (\ref{(2,0)=0})
and (\ref{(0,2)=0}) which mix the matrix components 
$p_{xx}$, $p_{yy}$ and $p_{xy}$ for general $\Omega$. However, 
for a canonical complex structure, corresponding to orthogonal 
tori, the constraints simplify and are written in the matrix form:
\beq
	p_{xx} = p_{yy},\,\,\,\,p_{xy}^T = p_{xy}.
\label{constraint-tau=i}
\eeq
Fluxes of such types have been used in \cite{akp} for constructing 
an $SU(5)$ GUT with stabilized moduli and in 
Section \ref{GUT-application} we apply the Yukawa couplings computation 
results to show the mass generation for extra non-chiral states
in the model of \cite{akp}. In this case, the 
$(1, 1)$ form flux $F_{z^i \bar{z}^j}$, for $(\Omega = i I_3)$,
reduces to:
\beq
	F_{z^i \bar{z}^j} = \frac{1}{2}( p_{xy} - i p_{xx})
\label{11-tau=i}
\eeq

Explicitly, the 
hermitian flux matrix $F$ in eq. (\ref{flux}) is given as:
\beq
    F = \begin{pmatrix} p_{x^1y^1} & p_{x^1y^2} + i p_{x^1x^2} 
			& p_{x^1y^3} + i p_{x^3x^1} \cr
		p_{x^1y^2} - i p_{x^1x^2} & p_{x^2y^2} &
		p_{x^2y^3} + i p_{x^2x^3} \cr
p_{x^3y^1} - i p_{x^3x^1} & p_{x^2y^3} - i p_{x^2x^3} & p_{x^3y^3}
	\end{pmatrix}.
\label{flux-real}
\eeq
For magnetized branes in \cite{akm,akp}, we used the quantization rule for $p$'s:
\beq
	p_{x^iy^j} = \frac{m_{x^iy^j}}{n^{x^i} n^{y^j}},\;\;\;
	p_{x^ix^j} = \frac{m_{x^ix^j}}{n^{x^i} n^{x^j}},\;\;\;
	p_{x^iy^j} = \frac{m_{x^iy^j}}{n^{y^i} n^{y^j}},	
\label{quantization}
\eeq
where $m_{x^iy^j}$, $m_{x^ix^j}$, $m_{y^iy^j}$ are the first 
Chern numbers along the corresponding 2-cycles and 
$n^{x^i}$, $n^{y^i}$ etc. are the wrapping numbers along the 1-cycles
$x^i$, $y^i$.   However, for the model \cite{akp},
we have used only integral fluxes corresponding to 
$n^{x^i} = n^{y^i} = 1$. 

An additional modification comes when nonzero NS-NS $B$-field background
is turned on along some 2-cycle. In this case, the first Chern 
number along the particular 2-cycle (for $n^{x^i} = n^{y^i} = 1$)
is shifted by:
\beq
     m_{x^i y^j} \rightarrow \tilde{m}_{x^iy^j} = 
			m_{x^iy^j} + \frac{1}{2}, {\mathrm etc.}
\label{b-shift}
\eeq
In the model that we discussed in \cite{akp}, we turn on nonzero 
NS-NS $B$-field, ($B = \frac{1}{2}$), along the 2-cycles diagonally in the
three $T^2$'s. Resulting fluxes are then half-integral. However, 
as already mentioned earlier, in writing the wavefunctions of chiral 
fermions $\chi_{ab}$ in bifundamentals,
the relevant quantities are the difference of fluxes in the two stacks,
or the two diagonal blocks in the gauge theory picture. 
In addition to the $D$-branes, an orientifold model also contains
image $D$-branes with fluxes of opposite signature than the ones present in 
the original brane. In such cases, the corresponding wavefunctions
$\chi_{ab^*}$ will obey similar equations as that of $\chi_{ab}$, 
but with the addition of the gauge potentials $A^a + A^b$ 
rather than their difference as in eq. (\ref{Dirac-equation4}).
The relevant matrix $\bf{N}$ which will now be the addition of fluxes
in the two stacks, rather than their difference, will once again be integral. 

We also learnt from the second equation in (\ref{eq-riemann-conditions}) that 
$({\bf{N}}.Im {\bf{\Omega}})$ is a symmetric matrix. However, as explained in 
eqs. (\ref{F(1,1)}) in the general situation and in (\ref{11-tau=i}) for 
${\Omega = i I_3}$, fluxes are in general hermitian
when components of all types: $p_{xx}$, $p_{yy}$ and $p_{xy}$ are present.

\section{ Fluxes for the stacks $U_5, U_1$ ,$A$, $B$, $O_1,\dots ,O_8$}\label{Appendix-A}

In this Appendix, we write all the fluxes in the complex coordinate basis 
$(z,{\bar z})$ with $z = x + i y$ for our GUT model in \cite{akp} and used in 
Section \ref{GUT-application} for the non-chiral mass generation. 
\begin{eqnarray}
F^{U_5} = 
-\frac{i}{2} \begin{pmatrix} dz_1 & dz_2 &dz_3 \end{pmatrix}
\begin{pmatrix}-\frac{3}{2} & & \cr
	& -\frac{1}{2} & \cr
	& & \frac{1}{2} \end{pmatrix}
\begin{pmatrix} d\bar{z}_1 \cr d\bar{z}_2 \cr d\bar{z}_3  
\end{pmatrix} ,
\end{eqnarray}
\begin{eqnarray}
F^{U_1} = 
-\frac{i}{2} \begin{pmatrix} dz_1 & dz_2 &dz_3 \end{pmatrix}
\begin{pmatrix}-\frac{3}{2} & & \cr
	& \frac{3}{2} & \cr
	& & \frac{1}{2} \end{pmatrix}
\begin{pmatrix} d\bar{z}_1 \cr d\bar{z}_2 \cr d\bar{z}_3  
\end{pmatrix} , 
\end{eqnarray}
\begin{eqnarray}
F^{O_1} = 
-\frac{i}{2} \begin{pmatrix} dz_1 & dz_2 &dz_3 \end{pmatrix}
\begin{pmatrix}\frac{5}{2} & 4 & 3\cr
     4	& \frac{1}{2} & 1 \cr
     3	& 1 & -\frac{1}{2} \end{pmatrix}
\begin{pmatrix} d\bar{z}_1 \cr d\bar{z}_2 \cr d\bar{z}_3  
\end{pmatrix},
\label{fo1}
\end{eqnarray}
\begin{eqnarray}
F^{O_2} = 
-\frac{i}{2} \begin{pmatrix} dz_1 & dz_2 &dz_3 \end{pmatrix}
\begin{pmatrix}\frac{5}{2} & 4 & - 3\cr
     4	& \frac{1}{2} & - 1 \cr
     - 3  & - 1 & -\frac{1}{2} \end{pmatrix}
\begin{pmatrix} d\bar{z}_1 \cr d\bar{z}_2 \cr d\bar{z}_3  
\end{pmatrix},
\label{fo2}
\end{eqnarray}
\begin{eqnarray}
F^{O_3}= 
-\frac{i}{2} \begin{pmatrix} dz_1 & dz_2 &dz_3 \end{pmatrix}
\begin{pmatrix}\frac{5}{2} & - 4 & -3i\cr
     -4	& \frac{1}{2} & i \cr
     3i & -i & -\frac{1}{2} \end{pmatrix}
\begin{pmatrix} d\bar{z}_1 \cr d\bar{z}_2 \cr d\bar{z}_3  
\end{pmatrix},
\label{fo3}
\end{eqnarray}
\begin{eqnarray}
F^{O_4} = 
-\frac{i}{2} \begin{pmatrix} dz_1 & dz_2 &dz_3 \end{pmatrix}
\begin{pmatrix}\frac{5}{2} & - 4 &  3i\cr
     -4	& \frac{1}{2} & - i \cr
     -3i & i & -\frac{1}{2} \end{pmatrix}
\begin{pmatrix} d\bar{z}_1 \cr d\bar{z}_2 \cr d\bar{z}_3  
\end{pmatrix},
\label{fo4}
\end{eqnarray}
\begin{eqnarray}
F^{O_5} = 
-\frac{i}{2} \begin{pmatrix} dz_1 & dz_2 &dz_3 \end{pmatrix}
\begin{pmatrix}-\frac{25}{2} & -2i & -i\cr
     2i	& \frac{1}{2} & 1 \cr
     i & 1 & \frac{1}{2} \end{pmatrix}
\begin{pmatrix} d\bar{z}_1 \cr d\bar{z}_2 \cr d\bar{z}_3  
\end{pmatrix},
\label{fo5}
\end{eqnarray}
\begin{eqnarray}
F^{O_6} = 
-\frac{i}{2} \begin{pmatrix} dz_1 & dz_2 &dz_3 \end{pmatrix}
\begin{pmatrix} -\frac{25}{2} & -2i &  i\cr
     2i	& \frac{1}{2} & - 1 \cr
     -i & -1 & \frac{1}{2} \end{pmatrix}
\begin{pmatrix} d\bar{z}_1 \cr d\bar{z}_2 \cr d\bar{z}_3  
\end{pmatrix},
\label{fo6}
\end{eqnarray}
\begin{eqnarray}
F^{O_7} = 
-\frac{i}{2} \begin{pmatrix} dz_1 & dz_2 &dz_3 \end{pmatrix}
\begin{pmatrix}-\frac{25}{2} & 2i & - 1\cr
     -2i  & \frac{1}{2} & i \cr
     -1 & -i & \frac{1}{2} \end{pmatrix}
\begin{pmatrix} d\bar{z}_1 \cr d\bar{z}_2 \cr d\bar{z}_3  
\end{pmatrix},
\label{fo7}
\end{eqnarray}
\begin{eqnarray}
F^{O_8} = 
-\frac{i}{2} \begin{pmatrix} dz_1 & dz_2 &dz_3 \end{pmatrix}
\begin{pmatrix}-\frac{25}{2} & 2i &  1\cr
     -2i  & \frac{1}{2} & -i \cr
     1 & i & \frac{1}{2} \end{pmatrix}
\begin{pmatrix} d\bar{z}_1 \cr d\bar{z}_2 \cr d\bar{z}_3  
\end{pmatrix},
\label{fo8}
\end{eqnarray}
\begin{eqnarray}
F^{A} = 
-\frac{i}{2} \begin{pmatrix} dz_1 & dz_2 &dz_3 \end{pmatrix}
\begin{pmatrix}\frac{295}{2} & & \cr
	& \frac{1}{2} & \cr
	& & \frac{1}{2} \end{pmatrix}
\begin{pmatrix} d\bar{z}_1 \cr d\bar{z}_2 \cr d\bar{z}_3  
\end{pmatrix} , 
\end{eqnarray}
\begin{eqnarray}
F^{B} = 
-\frac{i}{2} \begin{pmatrix} dz_1 & dz_2 &dz_3 \end{pmatrix}
\begin{pmatrix}\frac{3}{2} & & \cr
	& \frac{33}{2} & \cr
	& & \frac{1}{2} \end{pmatrix}
\begin{pmatrix} d\bar{z}_1 \cr d\bar{z}_2 \cr d\bar{z}_3  
\end{pmatrix} . 
\end{eqnarray}

Using the above fluxes, one can find out the chiral multiplets in the model. This
has been done for the brane intersections involving stacks - $U_5$, $U_1$.
A computation of the chiral fermion multiplicities on the 
intersections $O_i- O_j$ and $O_i-O^*_j$ ,for $i,j= 1, \dots 8$,
implies the existence of following fields in the non-chiral spectrum of the model.
They are:\\
($\phi^{O_1O_2}_{+-}$, $\phi^{O_1O_2}_{-+}$, $40$), ($\phi^{O_1O_3}_{+-}$, $\phi^{O_1O_3}_{-+}$, $84$),
($\phi^{O_1O_4}_{+-}$, $\phi^{O_1O_4}_{-+}$, $84$),
($\phi^{O_1O_5}_{+-}$, $20$),
($\phi^{O_1O_6}_{+-}$, $\phi^{O_1O_6}_{-+}$, $49$),
($\phi^{O_1O_7}_{+-}$, $6$), ($\phi^{O_1O_8}_{+-}$, $14$),
($\phi^{O_2O_3}_{+-}$, $\phi^{O_2O_3}_{-+}$, $84$), ($\phi^{O_2O_4}_{+-}$, $\phi^{O_2O_4}_{-+}$, $84$), 
($\phi^{O_2O_5}_{+-}$, $\phi^{O_2O_5}_{-+}$, $49$),
($\phi^{O_2O_6}_{+-}$, $20$), ($\phi^{O_2O_7}_{+-}$, $14$ ), ($\phi^{O_2O_8}_{+-}$, $6$),
($\phi^{O_3O_4}_{+-}$, $\phi^{O_3O_4}_{-+}$, $40$),
($\phi^{O_3O_5}_{+-}$, $14$), ($\phi^{O_3O_6}_{+-}$, $6$), ($\phi^{O_3O_7}_{+-}$, $20$),
($\phi^{O_3O_8}_{+-}$, $\phi^{O_3O_8}_{-+}$, $49$),
($\phi^{O_4O_5}_{+-}$, $6$), ($\phi^{O_4O_6}_{+-}$, $14$),
($\phi^{O_4O_7}_{+-}$, $\phi^{O_4O_7}_{-+}$, $49$), ($\phi^{O_4O_8}_{+-}$, $20$),
($\phi^{O_5O_6}_{+-}$, $\phi^{O_5O_6}_{-+}$, $8$),
($\phi^{O_5O_7}_{+-}$, $\phi^{O_5O_7}_{-+}$, $20$), 
($\phi^{O_5O_8}_{+-}$, $\phi^{O_5O_8}_{-+}$, $20$),
($\phi^{O_6O_7}_{+-}$, $\phi^{O_6O_7}_{-+}$, $20$), 
($\phi^{O_6O_8}_{+-}$, $\phi^{O_6O_8}_{-+}$, $20$),
($\phi^{O_7O_8}_{+-}$, $\phi^{O_7O_8}_{-+}$, $8$),
($\phi^{O_1O^*_2}_{++}$, $59$), ($\phi^{O_1O^*_3}_{--}$, $33$), ($\phi^{O_1O^*_4}_{--}$, $33$),
($\phi^{O_1O^*_5}_{++}$, $86$), ($\phi^{O_1O^*_6}_{--}$, $10$), ($\phi^{O_1O^*_7}_{++}$, $24$),
($\phi^{O_1O^*_8}_{++}$, $52$), ($\phi^{O_2O^*_3}_{--}$, $33$), ($\phi^{O_2O^*_4}_{--}$, $33$),
($\phi^{O_2O^*_5}_{--}$, $10$), ($\phi^{O_2O^*_6}_{++}$, $86$), ($\phi^{O_2O^*_7}_{++}$, $52$),
($\phi^{O_2O^*_8}_{++}$, $24$), ($\phi^{O_3O^*_4}_{++}$, $59$), ($\phi^{O_3O^*_5}_{++}$, $52$),
($\phi^{O_3O^*_6}_{++}$, $24$),  ($\phi^{O_3O^*_7}_{++}$, $86$), ($\phi^{O_3O^*_8}_{--}$, $10$), 
($\phi^{O_4O^*_5}_{++}$, $24$), ($\phi^{O_4O^*_6}_{++}$, $52$),
($\phi^{O_4O^*_7}_{--}$, $10$), ($\phi^{O_4O^*_8}_{++}$, $86$), ($\phi^{O_5O^*_6}_{--}$, $41$),
($\phi^{O_5O^*_7}_{++}$, $23$), ($\phi^{O_5O^*_8}_{++}$, $23$), ($\phi^{O_6O^*_7}_{++}$, $23$),
($\phi^{O_6O^*_8}_{++}$, $23$),  ($\phi^{O_7O^*_8}_{--}$, $41$).
\vskip -1.5cm
\beq
\label{nonchiral-O_i}
\eeq

As a result of a  similar analysis for the remaining stacks $A$ and $B$, we have also the following fields:\\ 
($\phi^{U_5A}_{+-}$, $\phi^{U_5A}_{-+}$, $149$),
($\phi^{U_5A^*}_{++}$, $\phi^{U_5A^*}_{--}$, $146$),
($\phi^{U_5B}_{+-}$, $\phi^{U_5B}_{-+}$, $51$),
($\phi^{U_5B^*}_{++}$, $\phi^{U_5B^*}_{--}$, $16$),
($\phi^{U_1A}_{+-}$, $\phi^{U_1A}_{-+}$, $149$), ($\phi^{U_1B}_{+-}$, $\phi^{U_1B}_{-+}$, $45$),  
($\phi^{AB}_{+-}$, $\phi^{AB}_{-+}$, $2336$), 
($\phi^{U_1B^*}_{++}$, $\phi^{U_1B^*}_{--}$, $18$), ($\phi^{U_1A^*}_{+-}$, $292$),
($\phi^{AB^*}_{+-}$, $149 \time 17$).
\vskip -0.8cm
\beq
\label{nonchiral-AB}
\eeq

\end {document}